\documentclass[twocolumn,aps,10pt,prx,notitlepage,preprintnumbers,superscriptaddress,amsmath,amssymb,groupedaddress]{revtex4-2}
\usepackage{hyperref}
\usepackage{graphicx} 
\usepackage{xcolor}
\usepackage{soul}
\usepackage{enumitem}
\usepackage{sansmath} 

\usepackage{mathtools}

\usepackage{algorithm}
\usepackage{algpseudocode} 
\usepackage[dvipsnames]{xcolor}
\usepackage{tikz}
\usepackage{physics}
\usepackage{amsmath}
\usepackage{amssymb}
\usepackage{mathrsfs}

\usepackage{enumerate}
\usepackage{xcolor}
\usepackage{pifont}
\newcommand{\cmark}{\ding{51}}%
\newcommand{\xmark}{\ding{55}}%
\usepackage[most]{tcolorbox}
\usepackage{tabu} 
\usepackage{dsfont}
\usepackage{bbm}
\usetikzlibrary{decorations.pathreplacing} 

\newcommand{\kett}[1]{\lvert #1 \rangle\!\rangle}
\newcommand{\bbra}[1]{\langle\!\langle #1 \rvert }
\newcommand{\bbrakett}[1]{\left\langle\hspace*{-0.08cm}\left\langle #1 \right\rangle\hspace*{-0.08cm} \right\rangle}
\renewcommand{\norm}[1]{\lVert#1\rVert}

\renewcommand{\ket}[1]{\lvert #1 \rangle }
\renewcommand{\bra}[1]{\langle #1 \rvert }

\newcommand{\rrangle}{\rangle\!\rangle}
\newcommand{\llangle}{\langle\!\langle}

\newcommand{\be}{\begin{equation}}
\newcommand{\ee}{\end{equation}}
\renewcommand{\Tr}{{\rm tr}}

\newcommand{\trnorm}[1]{\norm{#1}_{\rm tr}}

\newcommand{\IA}{\langle\!\langle I_A\big\rvert}
\newcommand{\Az}{\lvert \boldsymbol{0}\rangle\!\rangle}

\newcommand{\cS}{\mathcal S}
\newcommand{\rS}{{\rm S}}
\newcommand{\cB}{\mathcal B}
\newcommand{\rB}{{\rm B}}
\newcommand{\cA}{\mathcal A}

\newcommand{\bS}{{\mathbf S}}
\newcommand{\bB}{{\mathbf B}}
\newcommand{\bA}{{\mathbf A}}

\newcommand{\bJ}{{\boldsymbol{J}}}
\newcommand{\bg}{{\boldsymbol{g}}}

 \newcommand{\DT}{\Delta t}

\begin{document}

\title{Ancilla-train quantum algorithm for  simulating non-Markovian open quantum systems}
\author{Hans Michael Christensen}
\author{Johannes Agerskov}
\author{Frederik Nathan}
\affiliation{NNF Quantum Computing Programme, Niels Bohr Institute, University of Copenhagen, Denmark}
\date{\today}

\begin{abstract}
  We present a quantum algorithm for simulating open quantum systems coupled to Gaussian environments valid for any configuration and coupling strength. The algorithm is applicable to problems with strongly coupled, or non-Markovian, environments, problems with multiple environments out of mutual equilibrium, and problems with time-dependent Hamiltonians. 
  We show that the algorithm can reproduce the true dynamics of such problems at arbitrary accuracy and, for a broad range of problems, only adds a minor resource cost relative to Trotterized time evolution; the cost is low-degree {polynomial} in the inverse target accuracy. 
  The algorithm is based on the insight that any Gaussian environment can be represented as a train of ancillary qubits that sequentially interact with the system through a time-local coupling, given by the convolution square root of the bath correlation function; this is a secondary result of our work.
  Our results open up new applications of quantum computers for efficient simulation of non-equilibrium and open quantum systems.
\end{abstract}
\maketitle

A key motivation for the development of
quantum computers is their ability to dramatically speed up simulation of physical processes governed by quantum mechanics~\cite{Feynman1982}. To harness this potential capability, a diverse range of quantum algorithms have been developed to simulate quantum systems in nature~\cite{Lloyd1996,Kassal2008,Whitfield10032011,Aspuru-Guzik2005,Bauer2020,Berry2015,Abrams1997,Lanyon2010-zu,Whitfield2010,Jones2012,Aspuru-Guzik2012}. Much effort has focused on
  {isolated} and {equilibrium} quantum systems, which can be characterized in terms of ground states, excitation spectra, or thermal ensembles~\cite{LinLin2024GSprep,Aspuru2005QCchemApplications,Ge2019GroundStatePrep,Lanyon2010-ei,Lin2020nearoptimalground,Peruzzo2014-xa,gilyen2024quantum,chen2023efficient,Matthies_2024,hahn2025,lloyd2025quantumthermalstatepreparation,lloyd2025_2,ding_2025}. 
At the same time, many important processes in nature involve {\it open and non-equilibrium} quantum systems, which feature dynamics, structured environments, time-dependent forces, or multiple baths out of mutual equilibrium. Non-equilibrium and open quantum systems emerge, e.g., in the contexts of biology~\cite{Fang2019nonEqBio}, chemistry~\cite{Rao2016nonEqReactions,10.1063/5.0268071,Dan2025-zk}, or light-harvesting~\cite{Hu2022generalquantum,Aeberhard2008nonEqPhotVolt}. {While a variety of classical methods have been developed for these  problems~\cite{redfield_theory_1965,FEYNMAN1963118,tanimura_time_1989,Imamoglu_1994,Suess_2014,deVega_2015,Breuer2016,deVega2017,kirsanskas_phenomenological_2018,nathan_topological_2019, ULE, mozgunov_completely_2020,davidovic_completely_2020, Tamascalli_2019, lacroix2025making}}, relatively simple systems  {remain} 
challenging to simulate on classical computers, e.g. due to non-Markovian effects or the breakdown of thermodynamic principles~\cite{breuer_theory_2007,deVega2017,Breuer2016}. 
This raises the question: 
{\it How can we efficiently harness quantum computers for simulating open and non-equilibrium quantum systems?}

\begin{figure}[t!]
  \centering
  \includegraphics[width=1\linewidth]{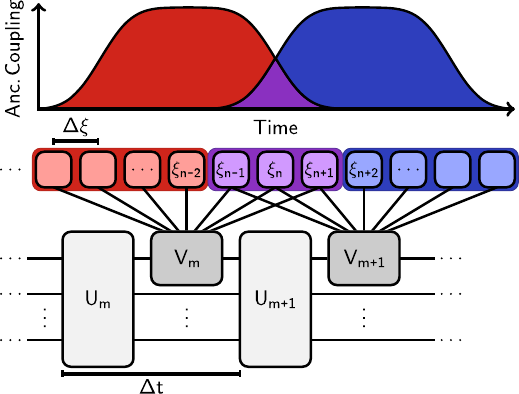}
  \caption{
  {\bf Ancilla-train algorithm (ATA):} 
     The ATA simulates the evolution of any open quantum system with Gaussian environments to arbitrary accuracy via Trotterized evolution. Each iteration---here the $m$th---propagate the system 
     through a time step $\Delta t$ via two gate sequences, $U_m$ and $V_m$, with  $U_m$  the time-evolution generated by the system Hamiltonian, while $V_m$ connects one or more system observables to a register of ancilla qubits (red, purple, and blue boxes; black lines indicate couplings). After the 
    iteration, a subset of ancillas are carried over to the next [$(m+1)$th] iteration
(purple); the remainder are discarded (red), and an identical number of new ancillas are (re)introduced in the $|0\rangle$ state (blue). 
    With a suitable  system--ancilla coupling structure obtained from the 
    jump correlator of the bath~\cite{ULE}, $g(t)$, we show that the ATA can reproduce the true dynamics of the system to arbitrary accuracy with a finite ancilla register.
  }
  \label{fig:ata_circuit}
\end{figure}

In this work, we address the question above by introducing a quantum algorithm that efficiently simulate{s} the dynamics and steady states of any open quantum system coupled to Gaussian environments. The algorithm, termed the  {a}ncilla-train algorithm (ATA), is, for instance, applicable to problems with strongly coupled, or non-Markovian, environments; problems with multiple environments out of mutual equilibrium; and problems with time-dependent Hamiltonians. 
The ATA consists of a standard Trotterized evolution where each iteration is augmented with a gate sequence  coupling system observables with a finite array---or {\it train}---of ancillary qubits (ancillas); see Fig.~\ref{fig:ata_circuit}. With a suitable but simple system{/}ancilla{-}train coupling,
we show that the ATA efficiently simulates any open quantum system with Gaussian environments to {\it arbitrary accuracy} {with respect to the true dynamics resulting from the underlying microscopic model}.

We provide rigorous error bounds and resource estimates for the ATA. {In particular}, we demonstrate that the ATA incurs only a minor cost relative to Trotterized Hamiltonian simulation for a broad class of problems: namely, a strictly or quasi-local system--ancilla coupling circuit (depending on the class of model), along with a finite ancilla register.
As another advantage, we argue that ATA can tolerate a finite hardware error rate because its output---fixed--point trajectories or steady states of open quantum systems---is perturbatively robust against generic noise.

We construct the ATA based on a single key insight: {\it any Gaussian environment can be represented by a train of qubits---or quantum binary white noise signal}---{\it sequentially interacting with the system through an appropriate time-local interaction}. Specifically, we show that such a representation is accurate when the time-local interaction is proportional to the convolution square root of the bath correlation function, or {\it jump correlator}, identified in Ref.~\cite{ULE}.
We refer to this representation 
as the {\it ancilla-train representation} of Gaussian quantum noise (ATR). \{The ATR offers a systematic  time-domain, or collision model, qubit representation of any Gaussian quantum environment, complementary to  frequency-domain, or pseudomode, bosonic representations~\cite{tanimura_time_1989, Breuer2016,Imamoglu_1994,Tamascalli_2019}; 
{As a key secondary result, we obtain rigorous bound on the error incurred by truncating the ancilla train of the ATR, enabling a systematic   approximate  representation of any Gaussian environment with a finite qubit register.}
Exploiting the natural qubit representation of Gaussian environments offered by the ATR, we  obtain the ATA simply by  Trotterizing the resulting time-evolution.

 {
Interestingly, a  construction related to the ATR  emerged recently in the open quantum systems literature~\cite{lacroix2025making},  involving a train of {h}armonic oscillators, rather than qubits, and a construction based on orthogonal polynomials that does not have the discrete time-translation symmetry of the ATR. }

An illustrative special case of the ATR 
emerges for symmetric power spectral densities, where the environment is equivalent to a classical noise signal acting on the system~\cite{breuer_theory_2007}.  Here, the ATR implies that any classical noise signal with Gaussian correlations---i.e., colored noise---can be generated as a sufficiently dense sequence of identical pulses with random signs; see Fig.~\ref{fig:classical noise}. Thus the ATR can be viewed as a quantum noise generalization of the well-known result that  a  colored  noise signal can be generated by filtering a (binary) white noise signal~\cite{brockwell_introduction_2016}; here the jump correlator provides the filtering kernel.

\begin{figure}
      \centering  \includegraphics[width=0.99\columnwidth]{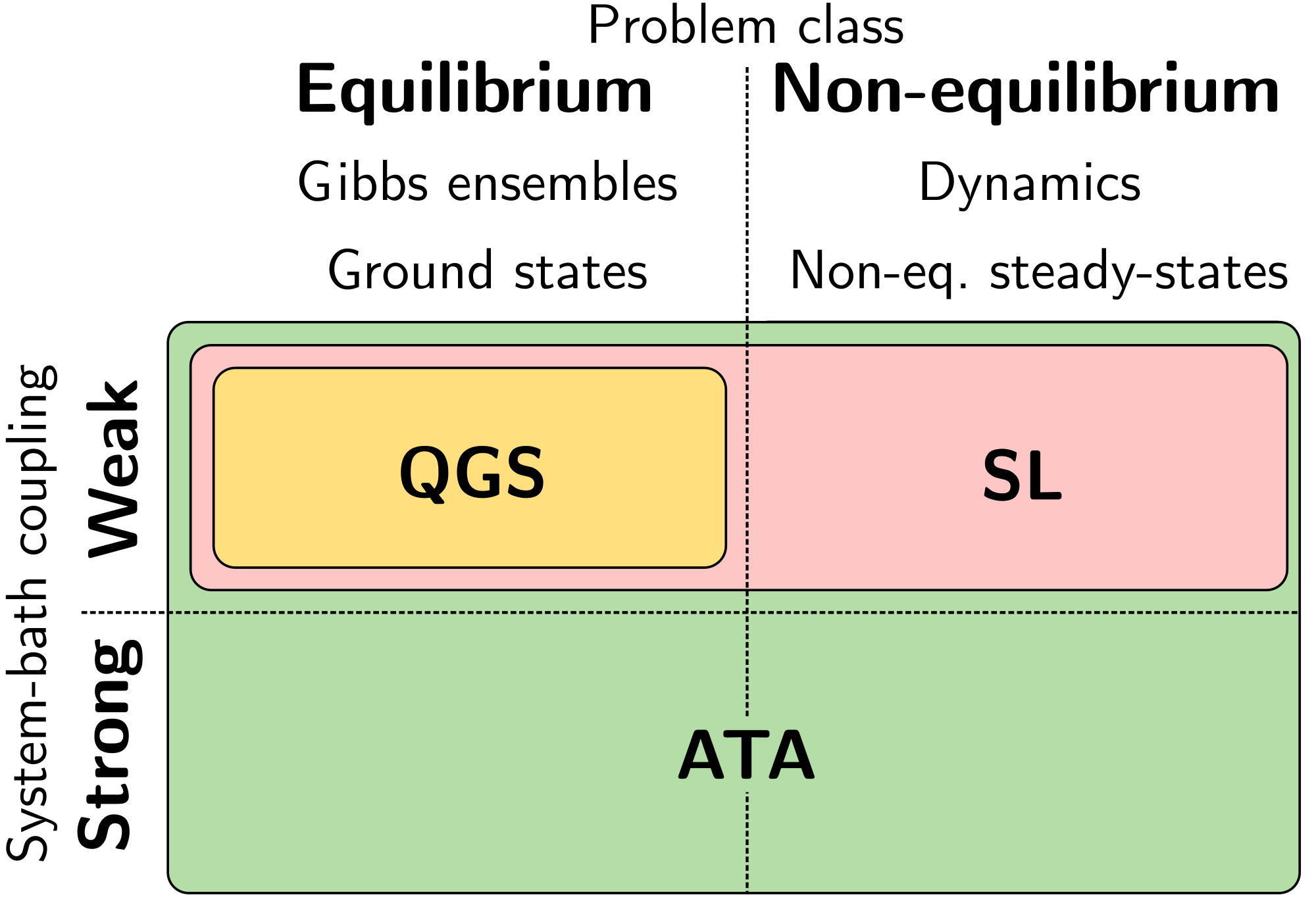} 
  \caption{\label{fig: comparison}{\bf Regimes of applicability of the ATA}, in relation to Quantum Gibbs Samplers (QGS)~\cite{chen2023efficient,LinLin2024GSprep,lloyd2025quantumthermalstatepreparation,hahn2025,Matthies_2024}, and non-Markovian collision models based on systematic, nonsecular, Lindblad equations (SL algorithms)~\cite{mozgunov_completely_2020,ULE,CICCARELLO20221,davidovic_completely_2020}; see Introduction and Sec.~\ref{sec:qgs} for details.}
  \label{fig:sttrengths}

\end{figure}

The ATA complements two other approaches that have recently emerged for systematic quantum simulation of open quantum systems:
quantum Gibbs sampling (QGS) algorithms~\cite{chen2023efficient, Matthies_2024,LinLin2024GSprep,lloyd2025quantumthermalstatepreparation,hahn2025,lloyd2025_2,ding_2025}, and Markovian collision-model algorithms based on systematic, non-secular Lindblad equations that have recently been derived rigorously~\cite{CICCARELLO20221,mozgunov_completely_2020,ULE,davidovic_completely_2020}, here termed Systematic Lindblad (SL) algorithms; see Sec~\ref{sec:qgs} for a detailed discussion
{(see also Ref.~\cite{Want_2011} for a proposal {preceding} these developments)}.
We summarize the regimes of applicability and advantages of the ATA, QGS, and SL algorithms in Fig~\ref{fig:sttrengths} and Table~\ref{tab:features}.
Compared to QGS and SL algorithms, which are generally only accurate in the limit of weak system-environment coupling~\cite{thingna_generalized_2012,nathan_quantifying_2024}, and require compilation of quasilocal jump operators and Lamb shifts, the ATA offers key advantages:
(1) the ATA  is accurate to all orders in system-environment coupling, and can thus efficiently capture non-Markovian effects. 
(2) the ATA does not  require compilation of quasilocal jump-operators  or Lamb shifts, i.e., renormalizations of the system Hamiltonian,  as required in QGS or SL algorithms, and can thus potentially be implemented with the addition of a low-depth unitary circuit to each Trotter iteration. (3)  like SL but unlike QGS algorithms, the ATA can be applied to  non-equilibrium problems, and can yield dynamics, fixed-point trajectories, along with steady states. 
The cost of the ATA relative to QGS and SL algorithms is a register of a finite number of ancilla qubits that scales linearly with system--bath coupling for a given, fixed target accuracy. 
{ Thus, the ATA offers both a broad range of applicability {\it and} an efficient implementation  in cases where circuit depth is the limiting resource. }

We expect that the advantages of the  ATA outlined above make it a promising approach for thermal state preparation and quantum simulation in a broad range of problems involving equilibrium and non-equilibrium quantum systems, and non-Markovian dynamics. Moreover, we expect that its potential robustness against finite error rates makes it a good candidate for implementation on near-term quantum hardware.

  \begin{table}
   \begin{tabular}{||c||c|c|c||}
            \hline 
        \textbf{Algorithm} &  QGS&  SL & ATA \\\hline\hline 
         {\textbf{Can be realized with}} & & & \\ 
        Quasilocal system--ancilla~coupling & \cmark & \cmark & \cmark \\ 
        {Strictly local~system--ancilla~coupling}$^1$\hspace{-0.7mm} & \xmark$^2$\hspace{-1.5mm} & \xmark & \cmark \\ 
        {No Lamb shift compilation$^{3}$} & \xmark & \xmark   & \cmark \\
        \hline\hline  \textbf{Can return} & & & \\ 
        { Thermal state}$^4$ & \cmark & (\cmark) & (\cmark) \\
        { Non-equilibrium steady state}   & \xmark & \cmark & \cmark \\ 
        { Dynamics} & \xmark & \cmark & \cmark \\     
        \hline \hline \textbf{Can simulate} & & & \\
        { Non-equilibrium baths}  & \xmark & \cmark & \cmark \\
        Time-dependent Hamiltonians  & \xmark & \cmark & \cmark \\
        Non-Markovian regimes & \xmark& \xmark& \cmark\\ 
        Non-Gaussian environments & \xmark & \xmark  & \xmark \\ \hline\hline 
         \rule{0pt}{2.5ex} \textbf{ $\#$  Ancillas required} & 1 & 1$^{5}$  & $ \sim \frac{\Gamma \tau}{\varepsilon^2},\!\frac{\Lambda\tau}{\varepsilon}$ \\[2.5pt] \hline 
        \end{tabular}
    \caption{{\bf Comparison of quantum algorithms for systematic open quantum system simulation}. We compare the features of  ATA,  Quantum Gibbs Sampling (QGS), and Systematic Lindblad (SL) algorithms; see main text for definition of these.     Here $\varepsilon$ denotes the target accuracy of the algorithm, while $\Gamma$, $\tau$, and $\Lambda$ denote the characteristic scales for system--bath coupling, correlation time, and ultraviolet cutoff for the environment defined in Eqs.~\eqref{eq: Gamma Tau def main text},~\eqref{eq:lambdadef}. {\it Explanation of notes:}
    $^1$The ATA involves a strictly local ancilla-system coupling in each  trotter iteration if observables in the simulation coupled to environments can be represented as a local observable in the system register. $^2${Refs.~\cite{lloyd2025quantumthermalstatepreparation,lloyd2025_2,hahn2025}} found  approximate QGS-like algorithms with strictly  local system--ancilla coupling and multiple ancillas, which could yield approximate (not exact) Gibbs states. $^3$Here Lamb shift refers to a quasilocal renormalization of the Hamiltonian, required for the Lindblad processes corresponding to QGS and SL algorithms. $^{4}$Physical systems only converge to Gibbs states in the limit of vanishing system--bath coupling~\cite{thingna_generalized_2012}; (\cmark) indicates that the steady state contains corrections accounting for this fact, while \cmark  indicates that the algorithms return the exact Gibbs state. $^5$E.g., for collision models based on the Lindblad equations from Refs.~\cite{ULE,davidovic_completely_2020}}
    \label{tab:features}
  \end{table}

The rest of the paper is structured as follows:
In Sec.~\ref{sec:results} we introduce the problem that we study and present our first main result: the ancilla-train representation of Gaussian environments. 
In Sec.~\ref{sec:algorithm} we use this result to present our second main result: the ancilla-train quantum algorithm, based on the ATR. 
We conclude with a discussion, including a more detailed analysis of the relationship between the ATA, QGS and SL algorithms, in Sec.~\ref{sec: Discussion}. 
Detailed error bounds, estimates and their proofs are found in the appendices.

\section{Ancilla-train representation of Gaussian environments}
\label{sec:results}
Here we introduce the class of models for open quantum systems that can be simulated with the ATA (Sec.~\ref{sec: problem introduction}) and provide our first main result, namely, that any Gaussian environment can be represented by a train of ancilla qubits that sequentially act with the system (Sec.~\ref{sec: Quantum simulation of bath}). 
We then demonstrate that this train can be truncated to have a finite number of qubits at the cost of a controllable error, that vanishes in the limit of many qubits. This result underlies the ATA. 
As a special case, we finally demonstrate that classical noise signals with Gaussian correlations can be represented by a discrete binary noise signal convolved with the jump correlator of the bath power spectral density (PSD)--defined as the inverse Fourier transform of the square root of the bath PSD.

\subsection{Introduction of problem}\label{sec: problem introduction}
We consider an arbitrary quantum system, $\cS$, coupled to one or more Gaussian environments, henceforth referred to as the bath, $\cB$.  
See below and Appendix \ref{app: Bath Constructions and Properties} for a definition of Gaussian environments.
We allow for the system Hamiltonian, $H_{\rS}(t)$, to be time-dependent. 
The Hamiltonian of the combined system we consider thus reads 
\begin{equation}\label{eq: H interaction}
  H_{\rm SB}(t) = H_{\rS}(t)+H_{\rB}+H_{\rm int},
\end{equation}
where $H_{\rm B}$ is the Hamiltonian of $\cB$, and the system--bath interaction $H_{\rm int}$ contains all terms of $H_{\rm SB}(t)$ that act non-trivially on both $\cS$ and $\cB$. 
In general, $H_{\rm int}$ can be written in the form 
\begin{equation}\label{eq: interaction multiple channels}
H_{\rm int}=\sqrt{\gamma}\sum_\alpha S_\alpha\otimes B_\alpha,
\end{equation}
with $S_\alpha, B_\alpha$ Hermitian operators (observables) of the system and bath, respectively, while $\gamma$ is an arbitrary energy scale parameterizing the system--bath coupling. We refer to each term in the sum above as a {\it noise channel}. We assume $S_\alpha$ to be bounded. Without loss of generality, we may therefore set $\norm{S_\alpha}= 1$~\footnote{Here $\norm{S}=\sup_{\norm{\psi}_2\leq 1}\norm{S\psi}_2$ denotes the usual operator norm.} and assume $\langle B_{\alpha}(t)\rangle=0$ for each $\alpha$; any non-zero expectation value can be removed by modifying $H_{\rS}(t)$. 
For simplicity, our discussion mainly considers the case where the sum above holds a single term, and hence we suppress the subscript $\alpha$ below. In Sec.~\ref{sec: multiple baths} we present our results for the general case.

Furthermore, we  assume that the system is decoupled from the bath at the beginning of the simulation, i.e. that the system--bath state is separable at some initial time $t_0$
\begin{equation}
  \rho_{\rm SB}(t_0)=\rho(t_0)\otimes \rho_{\rm B}.
\end{equation}
Our goal is to obtain the resulting evolution of the system, fully described by its reduced density matrix 
\begin{equation}\label{eq: rho(t)}
  \rho(t)\coloneq \Tr_{ \mathcal{B}}[\rho_{\rm SB}(t)].
\end{equation}
with $\rho_{\rm SB}(t)$ denoting the time-evolution of $\rho_{\rm SB}(t_0)$ generated by $H_{\rm SB}(t)$, and $\Tr_\mathcal{B}$ denoting partial trace over the environment Hilbert space.

The Gaussian property of $\cB$ 
implies that 
the correlation functions of $B$ with respect to $H_{\rm B}$ and $\rho_{\rm B}$ are fully determined from the two-point correlation function, and structured according to Wick's theorem~\cite{breuer_theory_2007}; see appendix~\ref{app: Bath Constructions and Properties} for details. 
This property emerges e.g. when $\cB$ consists of a set or continuum of bosonic modes, with $B$ a linear combination of mode creation- and annihilation operators and $\rho_{\rm B}$ a Gaussian state with respect to these modes~\cite{breuer_theory_2007}. More broadly, Gaussianity  emerges in the limit where the bath consists of many effectively decoupled subsystems, no matter their internal structure.
For this reason, Gaussian environments are ubiquitous,  emerging in a broad range of settings, including electromagnetic, phononic, electronic, vibronic, and spin environments~\cite{breuer_theory_2007,gardiner_quantum_2004}.   The universality of Gaussian environments means that they can be  represented in  many equivalent ways, e.g., in terms of configurations of discrete dissipative bosonic modes~\cite{tanimura_time_1989,deVega2017}. In the next section, we will identify one such  way of efficiently representing Gaussian environments which is particularly useful for quantum simulation.

\subsection{Ancilla{-t}rain {r}epresentation}\label{sec: Quantum simulation of bath}

We now show how the bath can be represented as a train of ancilla qubits sequentially interacting with the system. 
We first show that the bath can be represented as a train of harmonic oscillators sequentially interacting with the system, all initialized in the ground state. A bosonic construction for quantum collision models, similar to ours, was recently proposed in Ref.~\cite{lacroix2025making} for Gaussian baths with a hard UV--cutoff. Subsequently, we show that the probability for excitation to the 2nd excited state can be neglected, implying that the harmonic oscillators can be replaced by qubits {initialized in their $|0\rangle$ states}. 

First, we transform to the {interaction picture} with respect to {$H_{\rm B}$~\footnote{Specifically, the interaction picture Hamiltonian ${H}_{\rm SB}(t)$ governs the evolution of ${\rho}^{\rm (I)}_{\rm SB}(t)\coloneqq e^{iH_{\rm B} t}\rho_{\rm SB}(t)e^{-iH_{\rm B} t}$.}}, introducing $ B(t)\coloneq e^{iH_{\rB} t}B e^{-iH_{\rB}t}$. Note that the reduced density matrices of $\mathcal S$ are identical in the interaction and Schrödinger pictures. 
{In the interaction picture,} when restricting to a single term in Eq.~\eqref{eq: interaction multiple channels}, the Hamiltonian of the combined system becomes 
  \begin{equation}
    H^{\rm (I)}_{\rm SB}(t) = H_{\rS}(t)+\sqrt{\gamma} S \otimes  B(t). 
    \label{eq:hip}
  \end{equation}
The  observable $ B(t)$ can be viewed as a quantum generalization of a classical noise signal~\cite{gardiner_quantum_2004}; we henceforth refer to it as the {\it quantum noise signal} generated by the bath.

{We next exploit that $\cB$ is Gaussian. This means that, in the state $\rho_{\rm B}$, all correlation functions of $ B(t)$ are determined by the two-point correlation function, according to Wick's theorem~\cite{breuer_theory_2007}. 
Here, the two-point correlation function is given by}
\begin{equation}
  J(t-t')\coloneqq {\Tr[  B(t) B(t')\rho_{\rm B}],}
\end{equation}
The above fact means that the system evolution $\rho(t)$ can be reproduced by substituting the quantum noise signal $B(t)$ in Eq.~\eqref{eq:hip} with {\it any} observable $A(t)$ 
whose correlations 
are structured according to Wick's theorem and produce the two-point correlation function $J(t)$.
{This defines a broad equivalence class of observables that emulate the same environment as $ B(t)$.}

{We now identify {\it two} observables, which are equivalent to $ B(t)$ in the sense above {and offer efficient representations of the environment for quantum simulation}. The first, $A_{\rm osc}(t)$, is a time-dependent linear combination of creation- and annihilation operators in an array, or {\it train}, of bosonic modes. We use this to obtain the second, $A_{\rm q}(t)$, which is analogously constructed from a train of {\it qubits}, and forms the basis for the ATA in the next section.

We construct $A_{\rm q}(t)$ and $A_{\rm osc}(t)$ using the {\it jump correlator} from Ref.~\cite{ULE}, $g(t)$, defined as 
\begin{equation}
  g(t) := \frac{1}{\sqrt{2\pi}}\int_{-\infty}^\infty d\omega \sqrt{\tilde J(\omega)} e^{-i\omega t}
  \end{equation}
with $\tilde J(\omega) := \frac{1}{2\pi} \int_{-\infty}^\infty dt e^{i\omega t}J(t)$ the power spectral density of 
$B(t)$. 
Note that $\tilde J(\omega)$ is always non-negative so the square root is non-negative as well.
The jump correlator $g$ is the unique positive-semidefinite {\it convolution square root} of $J$, i.e. $J=g\ast g$ with $\ast $ denoting convolution~\footnote{For two functions $f_1$ and $f_2$ their convolution is, $[f_1\ast f_2 ](x) \coloneqq \int^\infty_{-\infty} dx' f_1(x-x')f_2(x')$ when the right-hand side is well-defined.}:
\begin{equation}\label{eq: J from g}
  J(t-t')=\int_{-\infty}^\infty ds g(t-s)g(s-t').
\end{equation}

We construct 
$A_{\rm osc}(t)$ by introducing an ancillary system consisting of {infinitely many independent harmonic oscillators, or free bosonic modes}, $\cA_{\rm osc}$, with bosonic annihilation operators $\{a_n|n\in \mathbb Z\}$. 
We initialize $\mathcal A_{\rm osc}$ in the vacuum state of all ancilla modes, $\rho_{{\rm osc}}^0:=|\boldsymbol{0}\rangle\langle \boldsymbol 0|$ with $|\boldsymbol 0\rangle := \bigotimes_{n=-\infty}^\infty |0\rangle_n $, and $|0\rangle_n$ the vacuum state of mode $n$: $a_n|0\rangle_n=0$. 
We construct $A_{\rm osc}(t)$ as a sliding linear combination of creation- and annihilation operators:
\begin{equation}\label{eq:aho}
  A_{\rm osc}(t) \coloneq \sqrt{\Delta\xi} \sum_{n=-\infty}^\infty \left(g^*(t-\xi_n)a^\dagger_n+h.c. \right),
\end{equation}
where $\xi_n=n\Delta \xi$, and $\Delta \xi$ is a timescale of our choice. 
It is straightforward to verify that, in the state $\rho_{\rm osc}^0$, $ A_{\rm osc}(t) $ satisfies Wick's theorem, and has two-point correlation function

\begin{equation}
\begin{split}
  &\Tr\left[A_{\rm osc}(t)A_{\rm osc}(t')\rho_{\rm osc}^0\right]
  =\sum_{n=-\infty}^\infty \Delta\xi g(t-\xi_n)g(\xi_n-t').
\end{split}
\end{equation}
{We identify the right-hand side above with that of Eq.~(\ref{eq: J from g}) with a discretized integral}. Thus the above expectation value converges to $J(t)$ in the limit $\Delta \xi \to 0$.
Since the correlation functions of $A_{\rm osc}(t)$ satisfy Wick's theorem, and have two-point correlation function $J(t)$, {\it $A_{\rm osc}(t)$ reproduces the same evolution as $B(t)$ in the limit $\Delta \xi \to 0$}. 
{In Appendix~\ref{app: discretization error} we we provide a rigorous proof for this fact, along with a bound for the error incurred by choosing  $\Delta\xi$ nonzero}.

{We now identify the second observable, $A_{\rm q}(t)$, that represents the bath in terms of {\it qubits}. First, note that leakage to the second excited state of any ancilla oscillator in $\cA_{\rm osc}$ is negligible} in the limit $\Delta \xi\to 0$. Even for nonzero $\Delta \xi $, the leakage error can be controlled (see Appendix \ref{app: qubit error}). This allows us to project the ancilla system into the subspace of the ground and first-excited state of each oscillator at a vanishing error cost. This effectively replaces the harmonic oscillators in $\cA_{\rm osc}$ with {\it qubits}.
Building on this result, we introduce the {\it qubit} ancilla system, $\mathcal A$, consisting of infinitely many qubits, with Pauli operators $\{(\sigma_n^x,\sigma_n^y,\sigma_n^z)|n \in \mathbb Z\}$. 
We initialize the ancilla system in the mutual $|0\rangle$ state of all qubits $\rho_{\mathcal A}^0\coloneqq \bigotimes_{n=-\infty}^\infty |0\rangle_n\langle 0|_n$ and introduce the energy-truncated ancilla-train operator. 
\begin{equation}\label{eq:aq}
  A_{\rm q}(t) \coloneq \sqrt{\Delta\xi} \sum_{n=-\infty}^\infty \left(g^*(t-\xi_n)\sigma^+_n+h.c. \right).
\end{equation}
with $\sigma^\pm_n:=\frac{1}{2}(\sigma^x_n\pm i\sigma_y^n)$ the usual ladder operators of qubit $n$. 
By the arguments above, in the limit $\Delta \xi\to 0$, this signal satisfies Wick's theorem and has the correct two-point correlation function; $J(t)$. Thus, $B(t)$ and $A_{\rm q}(t)$ result in the same system dynamics in this limit. 

The above result shows that any Gaussian environment, or quantum noise signal $B(t)$, can be represented by an array, or {\it train} of qubits that sequentially interact with the system in a time-dependent manner controlled by the jump correlator. This {\it ancilla-train representation} (ATR) of the quantum noise signal converges to the {\it exact} quantum noise signal as we take the limit $\Delta \xi\to 0$. 

\subsubsection{ATR for classical noise}

An illuminating example of the ATR emerges in  the special case  when the bath PSD is {\it even}, or, equivalently, the correlation function is real-valued~\cite{gardiner_quantum_2004}. 
In this case,  $ B(t)$ can be represented as a {\it classical}, scalar-valued, noise signal with Gaussian correlations---or colored noise signal---acting on the system through $S$.
In this case, $g(t)$ is real-valued, and Eq.~(\ref{eq: A qb}) becomes 
$
  A_{\rm q}(t) = \sqrt{\Delta\xi} \sum_{n} g(t-\xi_n)\sigma^x_n.
$ 
Note that $\sigma^x_n$ is now an integral of motion of the total time-evolution.
Since each ancilla is initialized in an even-weighted superposition of the two eigenstates of $\sigma_n^
x$, tracing out the ancilla is equivalent to replacing $\sigma^x_n$ with $1$ and $-1$ at probability $1/2$ each. 
We may therefore replace each $\sigma_n^x$ with an independent random variable $x_n$ uniformly sampled from $\{-1,1\}$. This replaces $A_{\rm q}(t)$ with the following stochastic (scalar-valued) function
\begin{equation}\label{eq: f(t)}
  A_{\rm cl}(t) = \sqrt{\Delta\xi} \sum_{n} g(t-\xi_n)x_n.
\end{equation}
Since the right-hand side above is simply a scalar-valued stochastic function, we identify it with a classical noise signal. 
This special case of the ancilla-train representation thus shows that {\it a classical Gaussian noise signal can always be generated by convolving its jump correlator with a white noise signal.} 
We provide an illustration of this result in Fig. \ref{fig:classical noise}.

The above results show how the ATR can be viewed as a quantum noise generalization of the the known result from signal processing that a colored noise signal can be generated by filtering a white noise signal~\cite{brockwell_introduction_2016}; here we identify the jump correlator  as the filtering kernel.

\subsection{Truncating ATR to finite ancilla number}\label{sec. Error scaling in tau_c and Delta xi}
{We now demonstrate that the infinite ancilla train of the ATR  can be well-approximated by an ancilla-train quantum noise signal with a finite register:  we show that a quantum noise signal $ B(t)$ can be substituted with an emulating noise signal of the form in Eq.~\eqref{eq: A qb}, with the sum truncated to a finite number of terms by picking a {\it finite} step size $\Delta \xi$, and discarding terms in the sum with $|t-\xi_n|\leq \tau_{\rm c}$, for some cutoff time $\tau_{\rm c}$}. 

We prove in Appendix~\ref{app: discretization error} that the ATR can be  truncated to a finite ancilla register as above at the cost of an arbitrary small error, if we simultaneously replace $g(t)$ with a {\it filtered} jump correlator~\footnote{The filtering is a convenient step to account for the time-discretization of the sum; for sufficiently small $\Delta \xi$, we have $g(t)\approx g_f(t)$ (with exact equality if the bath has a hard UV cutoff below $\pi/\Delta \xi$). Note that it might be possible to obtain similar proofs for unfiltered functions, for $\Delta \xi \to 0$. We leave such an investigation for future work.}, defined as  
\begin{equation}
  g_{\rm f}(t)\coloneq \int_{-\infty}^\infty ds g(s) \varphi(t-s)
\end{equation}
with $\varphi(t) = \frac{1}{2\pi}\int d\omega \tilde\varphi (\omega)e^{-i\omega t}$ a low-pass filtering kernel satisfying $\tilde \varphi(\omega)= 0$ for $|\omega| \geq \pi/\Delta \xi$, and $\tilde\varphi(\omega)=1$ for $|\omega | <\pi/2\Delta \xi$. 
In between these two ranges, $ \tilde \varphi$ should have uniformly bounded first and second derivatives (almost everywhere), implying that $g_{\rm f}(t)$ decays on approximately the same timescale as $g(t)$---in Appendix~\ref{app: discretization error} we provide a particular construction {of $\tilde{\varphi}(\omega)$} via  polynomial interpolation. 
Below, we present three operators, which faithfully reproduce the evolution generated by $ B(t)$ when choosing $\Delta\xi$ sufficiently small and choosing a sufficiently large cut-off time, $\tau_{\rm c}$:
\begin{align}\label{eq: A osc}
  A_{\rm osc,f}(t) &\coloneq \sqrt{\Delta\xi} \sum_{n=-\infty}^{\infty} \left(g_{\rm f}^*(t-\xi_n)a^\dagger_n+h.c. \right),\\
    A_{\rm q,f}(t) &\coloneq \sqrt{\Delta\xi} \sum_{n=-\infty}^{\infty} \left(g_{\rm f}^*(t-\xi_n)\sigma^+_n+h.c. \right),\label{eq: A qb}\\
      A_{\rm T}(t) &\coloneq \sqrt{\Delta\xi} \sum_{n=n_-(t)}^{n_+(t)} \left(g_{\rm f}^*(t-\xi_n)\sigma^+_n+h.c. \right),
      \end{align}
with $n_\pm(t)={\rm round}([t\pm\tau_{\rm c}]/\Delta \xi)$.
We will use the first two observables as intermediate steps to bound the error for substituting $B(t)$ with $A_{\rm T}(t)$, which only requires access of a finite ancilla register of size $\sim \tau_{\rm c}/\Delta \xi $, and serves as a basis for the ATA.
In the following, we define this error as the trace norm distance between the evolution, $\rho_{\rm T}(t)$, generated with the quantum noise signal $A_{\rm T}(t)$ with positive $\Delta \xi$ and finite $\tau_{\rm c}$, and the true evolution, $\rho(t)$, resulting from $ B(t)$.

Our error bound  depends on two quantities of the bath: a characteristic system--bath interaction rate, $\Gamma$, and the characteristic correlation time of the bath, $\tau$, defined as 
\begin{equation}\label{eq: Gamma Tau def main text}
  \Gamma = 4\gamma \left[\int_{-\infty}^\infty dt\abs{g(t)}\right]^2
  \text{ and } 
  \tau =
  \frac{\int_{-\infty}^\infty dt\abs{g(t)t}}{\int_{-\infty}^\infty dt\abs{g(t)}}.
\end{equation} 
Importantly, $\Gamma$ defines an upper bound for the rate of bath-induced evolution~\cite{ULE}.

Below, we identify values of $\Delta \xi$ and $\tau_{\rm c}$ such that the error induced by the truncation is smaller than some given target relative accuracy, $\varepsilon$. Specifically, we require that 
\begin{equation} \label{eq:rta_def}
  \norm{\rho(T)-\rho_{\rm T}(T)}_{\rm tr}=\mathcal{O}\left(\varepsilon\Gamma T\right),
\end{equation}
where $\norm{\cdot}_{\rm tr}$ denotes the trace norm. 
We employ a triangle inequality argument to bound this error by a sum of three contributions: the first error $\varepsilon_1$, arises from substituting $ B(t)$ with  
with $A_{\rm osc;f}(t)$. The second error, $\varepsilon_2$, arises from substituting the oscillator signal $A_{\rm osc;f}(t)$ with 
$A_{\rm q; f}(t)$.
The final error, $\varepsilon_3$ arises from discarding terms involving $|\xi_n-t|$ beyond the cutoff time $\tau _c$, resulting in $A_{\rm T}(t)$. 
We discuss each of these three errors separately below.

The discretization error $\varepsilon_1$ can be bounded by a quantity depending on the effective ultraviolet (UV) cutoff of the bath PSD, i.e., an energy  beyond which $\tilde J(\omega)$ effectively vanishes. 
Specifically, we find in Appendix~\ref{app: discretization error}, that 
the substitution of $B(t)$ with $A_{\rm osc;f}(t)$ induces a correction to $\rho(T)$ with trace norm of order at most $\mathcal{O}\left(\epsilon(\Omega)\Gamma T\right)$, when choosing $\Delta\xi\leq \pi/\Omega$, with 
\begin{equation}\label{eq: discretization error main text}
  \epsilon(\Omega)=\frac{4 \sqrt{\left(\int_{\abs{\omega}>\Omega}d\omega \tilde{J}(\omega)\right)\left(\int_{\abs{\omega}>\Omega}d\omega |{\tilde{J}''(\omega)}|\right)}}{\left[\int_{-\infty}^\infty dt \abs{g(t)} \right]^2}.
\end{equation}
In particular, $\epsilon(\Omega)\to 0$ as $\Omega\to \infty$ when $\tilde{J}$ and $\tilde{J}''$ are integrable. For such power spectral densities, we reach the target accuracy by choosing the following UV cut-off
\begin{equation}
  \Lambda(\varepsilon)\coloneq \inf \{\Omega\geq 1/\tau |\epsilon(\Omega)<\varepsilon\}.  \label{eq:lambdadef}
\end{equation}
The requirement that $\Omega\geq 1/\tau$ ensures that the frequency filtering does not significantly alter $\Gamma$ and $\tau$ (see Appendix~{\ref{app:Gamma tau bound proof}} for details).
We expect that $\Lambda(\varepsilon)$  can effectively be chosen to be given by the effective spectral width of the observable coupled to the bath, which is equivalent to the Trotter step in the algorithm, and hence $\Delta \xi$ can be chosen to be of the same order as the Trotter step.

We now consider the error $\varepsilon_2$ from using qubits instead of harmonic oscillators. This error is small whenever leakage to the second excited states of the harmonic oscillators is negligible. The leakage vanishes when $\Delta \xi$ goes to zero, as the strength of the interaction with each ancillary qubit becomes vanishingly small. In appendix \ref{app: qubit error} we make this intuition precise, and show that the error stays within the target accuracy when choosing $\Delta\xi\leq \frac{\varepsilon}{\Gamma}$. 

Finally, we consider the error $\varepsilon_3$ from discarding ancilla qubits whose distance from a given time $|\xi_n-t|$ exceeds the cutoff $\tau_{\rm c}$. In appendix \ref{app: cut-off error} we show that $\varepsilon_3$ is within the target accuracy when choosing $\tau_{\rm c}\geq\frac{\tau}{\varepsilon}$. 

Together, the above results imply that 
$A_{\rm T}(t)$ reproduces the correct dynamics up to a relative target accuracy $\varepsilon$ 
when choosing
\begin{equation}
  \Delta\xi \leq\min\left(\frac{\varepsilon}{\Gamma},\frac{\pi}{\Lambda(\varepsilon)}\right) ,
  \quad \tau_{\rm c}\geq\frac{\tau}{\varepsilon}.
\end{equation}

With these parameters, 
 $A_{\rm T}(t)$ only involves a {\it limited} number of ancilla qubits at any given point in time, given by 
\begin{equation}
  N = \max\left(\frac{2\Gamma\tau}{\varepsilon^2},\frac{2\Lambda(\varepsilon)\tau}{\pi\varepsilon}\right).
\end{equation}
Importantly, we note that the number of ancillas required scales as a low-degree polynomial in $\varepsilon^{-1}$ when the bath PSD decays sufficiently fast.

{Our results above show that any Gaussian environment can accurately be represented by $A_{\rm T}(t)$, even in the case of an infinite number of unbounded degrees of freedom. This is significant since $A_{\rm T}(t)$ is constructed from a finite ancillary register of $N$ qubits. 
Importantly, $A_{\rm T}(t)$ is {\it bounded}; with the choice for $\Delta \xi$ above, ${\sqrt{\gamma}}\norm{A_{\rm T}(t)}\sim \max\{{\Gamma/{\sqrt{\varepsilon}}, \sqrt{\Lambda \Gamma}}\}$~\footnote{This follows from the triangle inequality, $\norm{A}\leq \sum_n|g_{\rm f}(n\Delta \xi)|\sqrt{\Delta \xi}$, along with $\sqrt{\gamma}\sum_n|g_{\rm f}(n\Delta \xi)|\sqrt{\Delta \xi}\sim \sqrt{\Gamma/\Delta \xi}$}. This fact enables efficient Trotterization of the system evolution with the ancilla-train representation. In the next section, we exploit this fact to construct an efficient quantum algorithm for simulating dynamics of open quantum systems connected to arbitrary Gaussian environments at arbitrary coupling strength. }

\begin{figure}
  \centering
  \includegraphics[width=0.99\columnwidth]{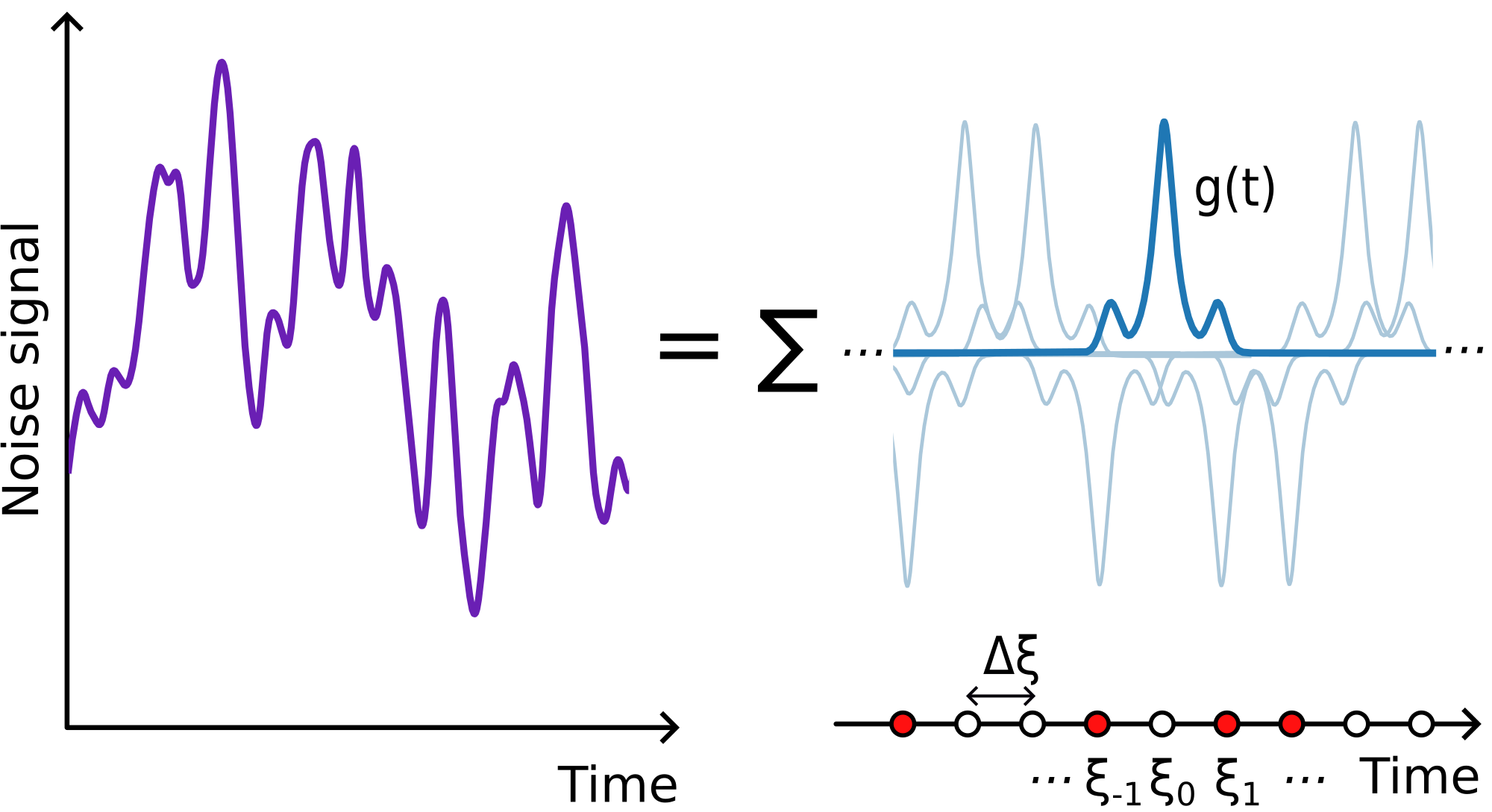}
  \caption{  {\bf Ancilla-train representation of classical colored noise:} Environments with even power spectral densities are equivalent to classical noise signals acting on the system~\cite{breuer_theory_2007}. Here our results imply that any classical noise signal with Gaussian correlations---or colored noise signals---can be generated by summing a sequence of identical pulses with random signs (red/white dots) and  spacing $\Delta \xi$, and taking the limit $\Delta \xi \to 0$. Here the is pulse given by $g(t)\sqrt{\Delta \xi}$~\cite{brockwell_introduction_2016}.
  }
 
  \label{fig:classical noise}
\end{figure}

\section{Ancilla-train algorithm}\label{sec:algorithm}
We now use the ATR to obtain our  main result: a quantum algorithm for systematically simulating 
open quantum systems coupled to any configuration of Gaussian environments at any coupling strength. The algorithm, termed the {a}ncilla-train algorithm (ATA), is obtained by Trotterizing the combined system--ancilla ($\cS\cA$) evolution in the ATR, using the truncated ancilla-train quantum noise signal $A_{\rm T}(t)$. We provide a compact pseudocode of the ATA in Fig. \ref{pseudocode1}.

The section is structured as follows: for the sake of simplicity, we first introduce the ATA for a single noise channel (Sec.~\ref{sec: Trotterizing the ATR})~\footnote{We use the term channel to signify that the system also may have multiple non-separable couplings to a single environment/bath, meaning that the interaction has tensor rank strictly greater than 1.},
and provide a resource estimate (Sec.~\ref{sec: resource estimation}). We then extend these results to the most general (multi-channel) configuration of Gaussian environments in Sec.~\ref{sec: multiple baths}). 

\subsection{Construction of ATA}\label{sec: Trotterizing the ATR}
The ATA simulates the evolution of the open quantum system of the class defined in Sec.~\ref{sec: problem introduction}: i.e. an arbitrary physical system $\cS$ with (possibly time-dependent Hamiltonian) $H_{\rm S}(t)$, connected to any configuration of Gaussian environments $\cB$, through an arbitrary system--bath coupling $H_{\rm int}$. 
For simplicity, we first consider the case where $H_{\rm int}$ factorizes into a product of a system observable $S$ and a bath observable $B$: $H_{\rm int}=\sqrt{\gamma}SB$, with $\norm{S}=1$. Here $\gamma$ parametrizes the system--bath coupling, and can be arbitrarily large. We consider the more general case in Sec.~\ref{sec: multiple baths}. 

The ATA operates on a system qubit register $\mathcal R_{\rm S}$, encoding the state of the system $\cS$, and an ancilla register, $\cA$, used to emulate the bath $\cB$ through the ATR, as described in Sec.~\ref{sec:results}. The system can be encoded in $\mathcal R_{\cS}$ through standard Hamiltonian simulation techniques~\cite{Lloyd1996,Berry2005EfficientQA,Low2017,cubitt2018universal,Watkins2022TimeDependentHS,Bosse2024EfficientAP,Berry2020timedependent}. 
Henceforth, we simply use $\cS$ in place of $\mathcal R_{\cS}$, while $S$ and $H_{\rm S}(t)$ refer to the observable and Hamiltonian in $\mathcal R_{\cS}$ representing their counterparts in the simulated system. 

The ATA takes as input the initial state of $\cS$ in the simulation, $|\psi_0\rangle_\cS$~\footnote{For mixed initial states $\rho_0$, $|\psi_0\rangle$ can, e.g., be randomly sampled from $\rho_0$}.
The ATA then simulates the evolution of $\rho(t)$ resulting from $|\psi_0\rangle$   by iteratively propagating the system forward in time through the Trotterized evolution: 
specifically, the $m$th iteration of the ATA returns a state of $\cS \cA$, $|\Psi_m\rangle $, whose corresponding reduced density matrix 
\begin{equation}
  \rho_m\coloneq \Tr_\cA[|\Psi_m\rangle \langle \Psi_m|],
\end{equation} provides a controlled approximations of $\rho(t_m)$ (see below). Here $t_m \coloneq m\DT$ and $\DT$, termed the {\it Trotter step}, is an input parameter of the simulation that controls the accuracy and output resolution.

The first step of the ATA prepares the full system--ancilla register $\cS\cA$ in the initial state $|\Psi_0\rangle \coloneq |\psi_0\rangle_{\cS}\otimes |\boldsymbol{0}\rangle_{\cA}$ with $|\boldsymbol{0}\rangle\coloneq \bigotimes_n |0\rangle_n$ the tensor product of $|0\rangle$ states in $\cA$. 
The ATA then iteratively generates the trajectory $\{|\Psi_m\rangle \}$ through Trotterized evolution with $H_{\cS\cA}(t)=H_{\rm S}(t) +\sqrt{\gamma}SA_{\rm T}(t)$. 
Specifically, $|\Psi_m\rangle$ is iteratively obtained through a sequence of  gate sequences---{\it Trotter unitaries}---$G_1,G_2,\ldots $, where the $m$th Trotter unitary, $G_m$, propagates the system forward from time $t_{m-1}$ to $t_{m}$: 
\begin{eqnarray}
  |\Psi_m\rangle = G_{m}|\Psi_{m-1}\rangle.
\end{eqnarray} 
We use a second-order Trotterization scheme, such that $G_m$ is given by 
\begin{equation}\label{eq: Trotter main}
  G_m = U_m^{1/2} V_mU_m^{1/2},
\end{equation}
where $U_m$ and $V_m$ are the Trotter Unitaries generated by $H_{\rm S}(t)$ and $\sqrt{\gamma}SA_{\rm T}(t)$: 
\begin{eqnarray}\label{eq: Trotter U}
 U_{m}&\coloneq& \mathcal{T}\exp\left[-i\int_{t_{m-1}}^{t_m} ds H_\rS(s)\right],\\\
  V_{m}& \coloneq& \mathcal{T}\exp\left[-i\int_{t_{m-1}}^{t_m} ds\sqrt{\gamma} S\otimes A_{\rm T}(s)\right],\label{eq: Trotter V 1}
\end{eqnarray}
with $\mathcal T$ denoting the time-ordering symbol, and $A_{\rm T}(t)$ the emulated bath signal from Eq.~\eqref{eq: A qb} 
\begin{equation}
  A_{\rm T}(t) \coloneq \sqrt{\Delta\xi} \sum_{n=n_-(t)}^{n_+(t)} \left(g_{\rm f}^*(t-\xi_n)\sigma^+_n+h.c. \right).
\end{equation}
Here $\xi_n\coloneq n\Delta \xi$, $n_\pm(t)={\rm round}([t\pm\tau_{\rm c}]/\Delta \xi)$, with $\Delta \xi$ and $\tau_{\rm c}$ input parameters of the simulation that, together with $\DT$, control the accuracy of the simulation; see below for more details. 

The first gate sequence of the Trotter Unitary, $U_m^{1/2}$, can be implemented with existing Hamiltonian simulation techniques \cite{Lloyd1996,Berry2005EfficientQA,Low2017,cubitt2018universal,Watkins2022TimeDependentHS,Bosse2024EfficientAP,Berry2020timedependent}. 
The bath propagator, $V_m$, is itself a product of commuting unitaries that each are generated by coupling $S$ to a single qubit in $\cA$: with 
\begin{align}
  V_{m}&=\prod_{n=n_-(t_{m-1})}^{n_+(t_m)}V_{mn},  \label{eq: Trotter V}\\  
  V_{mn}\!\!&\coloneq\! \mathcal{T} \!\exp[-i\sqrt{\gamma\Delta \xi}\!\int_{t^{(0)}_{mn}}^{t^{(1)}_{mn}} \!\!\!ds S \otimes (g_{\rm f}^*(s-\xi_n)\sigma^+_{n} \! +\! h.c.)],\notag
\end{align} 
where $t^{(0)}_{mn}\coloneq\max(t_{m-1},\xi_n-\tau_{\rm c})$ and $t^{(1)}_{mn}\coloneqq\min(t_m,\xi_n+\tau_{\rm c})$. $t^{(0/1)}_{mn}$ denote the boundaries of the integration domain, accounting both for the restriction from a finite $\tau_{\rm c}$ and from the bounds of the  time interval corresponding to the Trotter step.
Note that for local or quasilocal $S$, $V_{mn}$ can be efficiently compiled to a finite sequence of quantum gates, and may even be classically tractable.
Thus, we expect the compilation of $V_{mn}$ to generically be low-cost. 

In Appendix~\ref{app: errors} we show that the ATA returns the true system evolution to any nonzero target accuracy, $\varepsilon$, for $\Delta \xi$, $1/\tau_{\rm c}$, and $\DT$ chosen sufficiently small. 
We establish this result   by combining the discussion in Sec.~\ref{sec:results}  with an error bound for second-order Trotterization of {the} time-evolution ~\cite{Huyghebaert_1990}. Our bound assumes a time-independent system Hamiltonian, $H_{\rm S}(t)=H_\cS$~\footnote{For time-dependent $H_{\rm S}(t)$, the second-order Trotter error is significantly more complicated \cite{Huyghebaert_1990}.}, but we expect generalization to time-dependent Hamiltonians is straightforward. 
Specifically, we show that the output trajectory of the ATA is guaranteed to reproduce the true system evolution to a given (relative) target accuracy $\varepsilon$, 
\begin{equation}
  \norm{\rho_m-\rho(t_m)}_{\Tr}\leq \mathcal O(\varepsilon\Gamma t_m).
  \label{eq:target_acc_ata}
\end{equation}
provided 
\begin{equation}
  \Delta\xi \leq\min\left(\frac{\varepsilon}{\Gamma},\frac{\pi}{\Lambda(\varepsilon)}\right) ,
  \quad \tau_{\rm c}\geq\frac{\tau}{\varepsilon}, \quad \DT\leq \frac{\varepsilon}{\sqrt{\Gamma \Dot{s}}}.   \label{eq:resource requirements}
\end{equation}
where $\Dot{s}\coloneq \norm{[H_{\rm S},S]}$, while $\Gamma$, $\tau$, and $\Lambda$ are defined in Eqs.~\eqref{eq: Gamma Tau def main text}~and~\eqref{eq:lambdadef}. A more detailed requirement on $\DT$ for the bound for non-vanishing $\varepsilon$ can be found in appendix \ref{app: Trotter error}. 
The results are obtained by combining the error bounds in Sec.~\ref{sec:results} with an error bound for the Trotterization of the evolution, $\varepsilon_4$; see Appendix~\ref{app: errors} for details.

We finally emphasize that the ancilla qubits in $\cA$ need only be prepared at the first Trotter step where they are used: i.e., ancilla $n$ need only be prepared at Trotter step $m^{(\rm i)}_n\coloneq \inf\{m:n\leq  n_+(m\DT)\}$. Likewise, ancilla $n$ may be discarded after the last Trotter step where it affects the system evolution: $m^{\rm (f)}_n\coloneq \sup\{m:n\geq n_-(m\DT)\}$. 
Depending on the hardware, in particular, the resource cost of qubits, discarded ancillas may be reset and reintroduced in the register.
In this scheme, $\cA$ only needs to hold $n_+(t_m)-n_-(t_{m-1}) = \left \lceil\frac{2\tau_{\rm c}+\DT}{\Delta \xi}\right \rceil$ qubits; among these, $N_{\rm reset}\coloneq \left \lceil\DT/\Delta \xi\right \rceil$  are reset and reintroduced in the register, and the remaining qubits are preserved, at every Trotter step. It is still possible to implement the ancilla register as a continuous stream of incoming qubits in hardware, where this is more resource-efficient~\cite{slussarenko_photonic_2019,bluvstein_logical_2024}.

\subsection{Resource {c}ost}\label{sec: resource estimation}
We now estimate the additional resource cost of the ATA compared
to the standard unitary time evolution generated by $U_m$. 

We first consider the number of extra qubits required to implement the ATA---i.e. the minimal number of qubits in the ancilla register, $\cA$. 
Recall from the discussion below Eq.~\eqref{eq:resource requirements} that the ATA only requires access to $N_{\cA}=\left \lceil\frac{\DT + 2\tau_{\rm c}}{\Delta\xi}\right \rceil$ qubits in the ancilla register.
In terms of error scaling, the number of required ancillas is thus upper-bounded by
$
  \max\left(\sqrt{\frac{\Gamma}{\Dot{s}}} + \frac{2\Gamma\tau}{\varepsilon^2},\frac{\Lambda\varepsilon}{\pi\sqrt{\Gamma\Dot{s}}}+\frac{2\Lambda\tau}{\pi\varepsilon}\right).
$ 
In the limit $\varepsilon\to 0$, the required number of ancillary qubits thus scales as
\begin{equation}\label{eq: ancilla count}
  \#\, {\rm ancillas}=\mathcal{O}\left(\max\left[\frac{\Gamma\tau}{\varepsilon^2},\frac{\Lambda(\varepsilon)\tau}{\varepsilon}\right]\right).
\end{equation}

Next, we consider the gate complexity of implementing ATA relative to standard unitary evolution.
Each Trotter step $V_m$ involves a product of $N_\cA=\left \lceil\frac{2\tau_{\rm c}+\DT}{\Delta \xi}\right\rceil$ unitaries $V_{mn}$. 
As described in Sec.~\ref{sec: Trotterizing the ATR}, each $V_{mn}$ is generated by a coupling of $S$ with observables of the $n$th qubit in $\cA$. Therefore, we expect $V_{mn}$ can be efficiently compiled to a finite sequence of quantum gates. 
The additional gate count of the ATA compared to the standard Trotterized Hamiltonian simulation thus scales as $N_{\cA} T/\DT$, which leads to,
 \begin{equation}\label{eq:gate count}
   \#\, \textrm{ancilla gates }=
   \mathcal{O}\!\left(\max\left[
   \frac{\tau T \sqrt{\Gamma^3 \Dot{s}}}{\varepsilon^3}
   ,
   \frac{\Lambda(\varepsilon)\tau T \sqrt{\Gamma\Dot{s}}}{\varepsilon^2}
  \right]\right)
 \end{equation}

If we pick $\Delta\xi$ commensurate with $\DT$, i.e. $\Delta\xi/\DT=a/b$ for integers $a,b$, the gate $V_m$ is identical to $V_{m+{\rm lcm}(a,b)}$ up to a shift of the ancilla register, where $\operatorname{lcm}$ denotes the least common multiple. In this case, the gates $\{V_m\}$ may be efficiently precompiled. Given a set of unitary gates for simulating the isolated system evolution $\{U^{1/2}_m\}$, the compilation of $G_m$ follows easily. For time independent $H_{\cS}$, all $\{U^{1/2}_m\}$ are identical, and thus, the number of precompiled $G_m$ is independent of the simulation time $T$.
 
Our resource estimates for the ATA in Eqs.~\eqref{eq: ancilla count},~\eqref{eq:gate count} show that the extra resource cost of the ATA, relative to the standard Trotterized evolution, is a low-degree polynomial of the inverse target accuracy when the bath PSD decaying sufficiently fast. Importantly, the extra cost is {\it independent} of the complexity of the Hamiltonian simulation problem. 
Thus, the extra cost of replacing standard Trotterized evolution with the ATA may be negligible for a broad class of problems.

\begin{figure}
\begin{tcolorbox}[colback=gray!5!white,colframe=gray!75!black,title=Ancilla-train algorithm (2nd-order Trotterized)]
\begin{algorithmic}
\linespread{1.2}\selectfont
  \State \textsf{\textbf{Input}} $\rho(t_0)$, $\Tilde{J}(\omega)$, $H_{\rm S}$, $S$, $\gamma>0$, $T>0$, $\varepsilon>0$
  \State \textsf{\textbf{Set }}
  \begin{minipage}[t]{0.8 \textwidth}$\Lambda\geq 1/\tau$ with $\epsilon(\Lambda)<\varepsilon$\phantom{bla} \Comment{See Eq. \eqref{eq: discretization error main text}};\\$\tau_{\rm c}=\frac{\tau}{\varepsilon}$; $\Delta\xi=\min\left(\frac{\varepsilon}{\Gamma},\frac{\pi}{\Lambda}\right)$; \\ $\DT=\frac{\varepsilon}{\sqrt{\Gamma\norm{[H_{\rm S},S]}}}$; {$M=\operatorname{round}\left(\frac{T}{\DT}\right)$}
  \end{minipage}
  \\
  
  \State \textsf{\textbf{Prepare}}
  $\ket{\Psi}\sim \rho(t_0)$ \Comment{Sample from initial state}

  \For{$m\in \{1,2,3,...,M\}$} 
    \State \textsf{\textbf{Load}} new ancillas used in step $m$
    \State \textsf{\textbf{Evolve}} $\ket{\Psi} \rightarrow U_m^{1/2}V_{m}U_m^{1/2} \ket{\Psi}$ \Comment{Trotter step}
    \State \textsf{\textbf{Discard/reset}} ancillas not used in step $m+1$.
  \EndFor

  \State \textsf{\textbf{Discard}} remaining ancillas
  \State \textsf{\textbf{Return}} $\ket{\Psi}\big\vert_\cS \sim \rho(M\DT)$ to accuracy $\mathcal{O}(\varepsilon \Gamma T)$
\end{algorithmic}
\end{tcolorbox}
\caption{\label{pseudocode1} \textbf{Pseudocode for Ancilla-train algorithm}. Input parameters denote the initial system state $\rho(t_0)$, bath PSD $\Tilde{J}(\omega)$, system Hamiltonian $H_{\rm S}$, system coupling operator $S$, system--bath coupling $\gamma$, total simulation time $T$, and (relative) target accuracy $\varepsilon$. $U_m$, given in \eqref{eq: Trotter U},  denotes a Trotter interaction generated by the system Hamiltonian and $V_m$, given in \eqref{eq: Trotter V 1} and \eqref{eq: Trotter V}, denotes a Trotter iteration in the time-evolution generated by the ancilla(s)-system coupling. We write $\ket{\Psi}\sim \rho$ to indicate that the expected value of $\ket{\Psi}\bra{\Psi}$ is $\rho$.}
\end{figure}

\subsection{Definition for {m}ulti-{c}hannel {e}nvironments}
\label{sec: multiple baths}
We finally demonstrate how the ATA can be straightforwardly extended to arbitrary Gaussian environments and arbitrary system--bath coupling, $H_{\rm int}$. 

Without loss of generality~\cite{ULE}, in this case we may express $H_{\rm int}$ as \begin{equation}\label{eq: Hint dot product}
  H_{\rm int}=\sqrt{\gamma} \bS\cdot \bB.
\end{equation}
Here $\bS$ and $\bB$ represent the column vectors $(S_1, \ldots S_{N_C})$ and $\bB=(B_1 \ldots B_{N_C})$, respectively, with $S_\alpha$, $B_{\alpha}$ denoting observables on $\cS$ and $\cB$, respectively, and $N_C$ denoting the total number of noise channels. We employ this vector notation to emphasize the similarity to the single-channel case.
The bath correlation function now becomes matrix-valued, $\bJ(t)$, with $(\alpha,\beta)$ element given by 
\begin{equation}
  J_{\alpha\beta}(t-s)\coloneq \Tr\left[B_\alpha(t) B_\beta(s)\rho_{\rm B}\right],
\end{equation}
We extend the definition of the jump correlator, $\bg(t)$, to be the unique convolution matrix square root of $\bJ$ of positive type, i.e., $\bg(t)\coloneq\int d\omega\, e^{-i\omega t}\sqrt{\tilde{\bJ}(\omega)/2\pi}$ with $\sqrt{\cdot}$ denoting the matrix square root with branch cut along the negative real axis~\footnote{Note that $\tilde{\bJ}(\omega)$ is Hermitian and positive semidefinite, implying $\bg(t)$ is a positive semidefinite convolution kernel; see Appendix \ref{app: Bath Constructions and Properties} for more details.}. 

With these definitions in place, we extend the ATA from the single- to the multi-channel case as follows: First, we extending the ancilla register $\cA$ from one to $N_C$ ancilla trains, i.e., an array of qubits with Pauli operators $\{\sigma^{i}_{\lambda n}|\lambda=1,\ldots N_c,n\in \mathbb Z,i\in \{x,y,z\}\}$~\footnote{$\cA$ can be truncated to a finite ancilla register in the same way as described in Sec.~\ref{sec. Error scaling in tau_c and Delta xi}}. From these ancilla trains, we can replace each $B_\alpha(t)$ with the truncated, emulating noise signal $A_{\rm T;\alpha}(t)$, where 
\begin{equation} \label{eq: multi channel ancilla operators}
  A_{{\rm T};\alpha}(t)\coloneq 
  \sum_{\lambda=1}^{N_c}\sum_{n=n_-(t)}^{n_+(t)} \sqrt{\Delta\xi}\left(
  g_{{\rm f};\alpha\lambda}^*(t-\xi_n)\sigma^+_{\lambda n} +h.c.
  \right).
\end{equation}
where $g_{{\rm f};\alpha\lambda}(t)$ denotes the $(\alpha,\lambda)$-entry of the filtered jump correlator $\bg_{\rm f}(t)\coloneq \int ds \varphi(t-s)\bg(s)$. 
Note that each ancilla train does not necessarily emulate a single channel; instead, each channel is generally generated from combinations of all ancilla trains.

The ATA is defined for the multi-channel case as for the single-channel case, with the replacement of the system--ancilla coupling $V_m$ to account for the multiple channels, i.e., 
\begin{eqnarray} \label{eq: multi channel V}
  V_{m}& \coloneq& \mathcal{T}\exp\left[-i\int_{t_{m-1}}^{t_m} ds\sqrt{\gamma} \bS\cdot \bA_{\rm T}(s)\right].
\end{eqnarray}
This unitary can, with care, be further Trotterized by utilizing commutativity between different ancillas (see Appendix \ref{app: LRI for multiple channels}); if $\{S_\alpha\}$ mutually commute, $V_m$ directly factorizes into operators acting on individual ancillas.

Except for the substitution above, the ATA for the multi-channel case is unchanged from the single-channel case. 
Indeed, our error bound in Appendix~\ref{app: errors} was derived for the multi-channel case; it shows that the ATA for multi-channel baths is guaranteed to be $\varepsilon$-accurate in the sense of Eq.~\eqref{eq:target_acc_ata}, if $\Delta \xi$, $\tau_{\rm c}$, and $\DT$ are chosen to satisfy Eq.~\eqref{eq:resource requirements}}, with $\Gamma$, $\tau$,  and $\Lambda(\varepsilon)$, now defined by appropriate substitution of $|J(\omega)|$, $|J''(\omega)|$, and $|g(t)|$ with $\lVert\tilde\bJ(\omega)\rVert_{1,1}$, $\lVert \tilde\bJ''(\omega)\rVert_{1,1}$, and $\lVert\bg(t)\rVert_{2,1}$, respectively. Here, for any matrix $\boldsymbol{M}$, $\norm{\boldsymbol{M}}_{p,q}\coloneq   (\sum_\beta[\sum_{\alpha}|M_{\alpha\beta}|^p]^{q/p})^{1/q}$ denotes the $(p,q)$ matrix norm. 

\section{Discussion}\label{sec: Discussion}
\label{sec:discussion}
In this paper, we introduced a quantum algorithm for systematically simulating quantum systems interacting with arbitrary Gaussian environments at any coupling strength: the ancilla-train algorithm (ATA). 
The ATA can be used to simulate dynamics and steady states of equilibrium and non-equilibrium problems, and is applicable to non-Markovian regimes. 
We have provided rigorous bounds on the accuracy of the algorithm and shown that the simulation error can be made arbitrarily small with an appropriate choice of simulation resolution parameters.

The resource cost for the ATA is polynomial in the inverse target accuracy, provided modest decay properties of the bath PSD, and is controlled by a combination of four physical energy scales of the simulated model: a characteristic Heisenberg picture velocity of the system observables coupled to the environment, along with the bath correlation time, system--bath coupling strength, and effective ultraviolet cut-off of the environment. 
Importantly, resource costs are {\it independent} of the complexity of the {system}. Thus, for a broad range of problems, the ATA only requires a small amount of additional resource relative to the cost of simulating the time-evolution in the absence of external environments.

The ancilla-train representation (ATR) of quantum noise provides a secondary result of our work that underlies the ATA: Any Gaussian quantum environment can be emulated by convolving a quantum binary white noise signal---an ancilla train---with a time-local convolution kernel.  A corollary is that a classical noise signal with Gaussian correlations can be generated by convolving a white noise signal with the jump correlator.  This result is consistent with the known result in classical signal processing that a colored noise signal can be obtained from filtering a white noise signal~\cite{brockwell_introduction_2016}; here the filtering kernel is given by the jump correlator.
The ATR thus provides a generalization of this result to quantum noise, {through a non-Markovian collision model}. For quantum noise, the filtering kernel (jump correlator) is simply generalized from a real to complex valued function, and the classical (binary) white noise signal---a bit stream---is replaced with  a stream of qubits. 

\subsection{Relation to quantum Gibbs sampling and systematic Lindblad equation algorithms}
\label{sec:qgs}

We now comment on the relationship between the ATA and other algorithms for quantum simulation of open quantum systems that have recently emerged in the literature: quantum Gibbs samplers (QGS)~\cite{chen2023efficient,LinLin2024GSprep,gilyen2024quantum,lloyd2025quantumthermalstatepreparation,hahn2025,Matthies_2024} and collision models based on systematic and rigorously derived Lindblad equations (SL)~\cite{ULE,davidovic_completely_2020,mozgunov_completely_2020}.

QGS algorithms yield the exact Gibbs state for a given model Hamiltonian. This is achieved through collision models~\cite{CICCARELLO20221} based on Lindblad processes with quasilocal jump operators that exhibit KMS detailed balance~\cite{Kossakowski_1977}.
Physical systems connected to environments in thermal equilibrium are known to converge to such exact Gibbs states in the limit of weak system--bath coupling. 
However, for nonzero finite couplings---as emerge in physical systems---the steady state deviates from the Gibbs state by a correction controlled by the system--bath coupling, due to the ambiguity in defining an appropriate subsystem Hamiltonian for a partitioned quantum system~\cite{thingna_generalized_2012}. 
Thus, from a quantum simulation perspective,  QGS algorithms yield accurate physical steady states of open quantum systems only in the limit of vanishing system--bath coupling. 

In parallel with the emergence of QGS algorithms, a family of systematic rigorously-derived Lindblad equations have emerged  that circumvent the rotating-wave, or secular, approximation. These Lindblad equations have been successfully employed in classical simulation of dissipative quantum many-body systems~\cite{maimbourg_bath-induced_2021,garcia-gaitan_fate_2024,aghaee_interferometric_2025,nielsen_dynamics_2023,munk_parity--charge_2020,nathan_self-correcting_2024}. 
Similarly to  QGS Lindblad processes, these systematic Lindblad equations    feature quasilocal jump operators and can thus be efficiently implemented as quantum algorithms using collision-model type approaches~\cite{CICCARELLO20221}. We term such quantum algorithms {\it systematic Lindblad, or SL, algorithms}. Contrasting QGS algorithms, SL algorithms are systematically derived from underlying microscopic models, and extend to non-equilibrium problems, e.g. featuring driving or non-equilibrium reservoirs. 
These SL approaches yield dynamics and steady states---equilibrium or non-equilibrium---that are accurate up to corrections proportional to the system-environment coupling (relative to an intrinsic correlation timescale for the environment~\cite{nathan_quantifying_2024}).
Thus, from a quantum simulation perspective, SL algorithms yield steady states that are accurate on the {\it same level of approximation} as those returned by QGS algorithms---namely, both QGS and SL algorithms are accurate in the limit of vanishing system--bath coupling. 

The ATA complements the QGS and SL algorithms above, offering important advantages: (1) The accuracy of the ATA simulation is limited only by the available resources; it can yield the exact steady-state and dynamics of the simulated microscopic model 
to arbitrary accuracy. (2) The ATA involves only a simple, strictly local system--ancilla coupling, and thus does not necessarily require compilation of a quasi-local system--ancilla coupling. (3) The ATA does not require compilation of a quasi-local Lamb shift or generalizations thereof.
For a given, fixed target accuracy, the additional cost of the ATA is a register ancilla qubits of size scaling linearly with system--bath coupling. 
{Thus, the ATA offers both a broader range of applicability {\it and} a simpler implementation than the QGS and SL algorithms in cases where the depth of the circuit is the limiting resource.}

During the preparation of this manuscript, ancilla-train-like constructions appeared in the QGS literature in  Refs. \cite{lloyd2025_2,lloyd2025quantumthermalstatepreparation,hahn2025}. Here, the goal was to simulate a Lindblad process {with} a thermal (Gibbs) state as an approximate steady state. Ref.~\cite{lloyd2025_2,lloyd2025quantumthermalstatepreparation,hahn2025} provides rigorous performance guarantees for this construction, demonstrating that ancilla-trains are broadly applicable for QGS, where they, moreover, bring the advantage of a strictly local system ancilla coupling.

\subsection{Noise resilience}
A final potential advantage of the ATA, shared by other algorithms for open quantum systems simulation~\cite{Matthies_2024,chen2023efficient,LinLin2024GSprep,ULE,mozgunov_completely_2020,lloyd2025quantumthermalstatepreparation,hahn2025}, is a potential {\it intrinsic robustness against hardware errors}, due to its entropy-extracting nature~\cite{Schuster_2023}. 
Specifically, noise in hardware executing the ATA will map to a noise channel in the simulated microscopic model. Since the simulation already involves a mechanism for dissipation, this noise channel will compete with other channels for dissipation and noise, and thus only generate a continuous perturbation of the output (simulated steady state), if weak enough: hence for open quantum system simulators such as the ATA the, {\it noise-induced error is generically a finite-slope function of the noise strength}. Equivalently, since the dissipative channels in an ATA simulation provide a finite relaxation time for perturbations of the steady state, extrinsic noise will generate a negligible error if the characteristic rate of errors is small relative to the relaxation time. 

\subsection{Outlook}

In summary, the ATA offers a systematic and accurate approach for simulating open quantum systems at arbitrary coupling strength, and in arbitrary non-equilibrium or equilibrium configurations. 
We expect the  algorithm has intrinsic resilience against hardware error and is relatively simple to implement. 
We expect these results make the ATA  a promising candidate for quantum simulation of non-equilibrium quantum systems, and for implementation on near-term quantum hardware.

\section*{Acknowledgements}
We gratefully acknowledge insightful discussions with Michael Kastoryano, Chi-Fang Chen, Matthew S. Teynor, Gemma C. Solomon, and Daniel Malz. 
This work is supported by the Novo Nordisk Foundation, Grant number NNF22SA0081175, NNF Quantum Computing Programme, and the Danish E-Infrastructure Consortium, grant number 4317-00014B.

\counterwithin*{equation}{section}
\renewcommand\theequation{\thesection\arabic{equation}}

\section*{Appendices}

In the appendices, we provide proofs of our claims from the main text. In Appendix \ref{app: Bath Constructions and Properties} we
demonstrate that the evolution of 
a system coupled to a Gaussian environment  
only depends on the environment through its bath correlation function.
{In Appendix \ref{app: errors} we give rigorous error bounds for our ancilla-train representation (ATR) as well as for the Trotter approximation required to obtain the ancilla-train algorithm (ATA). 
In Appendix \ref{app:Gamma tau bound proof}, we prove that {the} jump correlator can be passed through a suitable low-pass filter without significantly increasing 
its characteristic interaction rate and correlation timescales.
In Appendix \ref{app: LRI for multiple channels}, we give a formula for $V_m$ from Eq.~\eqref{eq: Trotter V 1} in the case where $\{S_\alpha\}$ are mutually commuting operators. Finally, in Appendix \ref{app: Common Relations and Notation} we derive some important relations that are frequently used throughout the appendices.

Throughout these appendices, we consider the general case of multiple quantum noise channels {[i.e., with multiple terms in the sum in Eq. \eqref{eq: interaction multiple channels}]}, where the bath correlation function $\bJ(t)$ is matrix-valued. The special case considered in the main text of the article is equivalent to the case where $\bJ(t)$ is a $1\times 1$ matrix.
As a help to the reader, in Table \ref{tab: jump corr etc} we provide an overview of the relevant functions of the bath {for} 
multiple quantum noise channels. 
\begin{table}[h!]
    \centering
    \bgroup
    \def\arraystretch{1.6}
    \begin{tabular}{|c|c|c|}
        \hline
        Name & Notation & Definition  \\
        \hline
        Bath Correlation Function & $\bJ(t-s)$ & $\Tr\left[\bB(t) \bB(s)^\dagger\rho_{\rm B}\right]$ 
        \\
        \hline
        Power Spectral Density (PSD) & $\tilde{\bJ}(\omega)$ & $\frac{1}{2\pi}\int_{-\infty}^\infty dt\bJ(t)e^{i\omega t}$ 
        \\
        \hline
        Square-root PSD & $\tilde{\bg}(\omega)$ & $\sqrt{\tilde{\bJ}(\omega)/2\pi}$
        \\
        \hline
        Jump Correlator & $\bg(t)$ & $\int_{-\infty}^\infty d\omega \tilde{\bg}(\omega)e^{-i\omega t}$
        \\
        \hline
    \end{tabular}
    \egroup
    \caption{Overview of the relationship between bath correlation function, power spectral density, square root of the power spectral density, and the jump correlator.}
    \label{tab: jump corr etc}
\end{table}

\appendix

\section{Properties of Gaussian baths}\label{app: Bath Constructions and Properties}

In this appendix, we show that the effect on a given quantum system  from a Gaussian environment solely depends on the two-point correlation function of the environment. 

To recap, we consider the evolution of the reduced density matrix of the system, $\rho(t)$, generated {by} the full system--bath Hamiltonian {$H_{\rm SB}(t) = H_{\rS}(t)+H_{\rB}+\sqrt{\gamma}\sum_\alpha S_\alpha\otimes B_\alpha$ [see also Eq. \eqref{eq: H interaction} of the main text].

We first go to the 
interaction picture  
with respect to the  system--bath interaction, 
$\sqrt{\gamma} \mathbf{S}\cdot \mathbf{B}$, obtaining
\begin{equation}
    \rho(t)=U_{\rm S}(t)\Tr_{\mathcal{B}}\left[U^{\rm (C)}_{\rm SB}(t)\rho(t_0)\otimes \rho_{\rm B} U^{{\rm (C)}\dagger}_{\rm SB}(t)\right]U^\dagger_{\rm S}(t),
    \label{eqa:a1}
\end{equation}
with $U_{\rm S}(t) = \mathcal{T}e^{-i\int_{t_0}^{t'} ds H_{\rm S}(s)}$, $U^{\rm (C)}_{\rm SB}(t)=\mathcal{T}e^{-i\int_{t_0}^t dt' \sqrt{\gamma} \bS{(t')}\cdot \bB(t') }$, and $\bS(t
)=U_{\rm S}(t)\bS U_{\rm S}^\dagger(t)$. Here we use the dot product notation that was introduced in Eq. \eqref{eq: Hint dot product} of the main text. Next, we expand $U_{\rm SB}^{\rm (C)}$ in terms of its Dyson series:\begin{equation}
\begin{aligned}
        &U^{\rm (C)}_{\rm SB}(t,t_0)=\\&1\!+\!\sum_{n=1}^{\infty}(-i)^n \!\!\int_{t_0}^{t}\!\!dt_1\int_{t_0}^{t_1}\!\!dt_2... \int_{t_0}^{t_{n-1}}\!\!\!dt_n\! \prod_{i=1}^n (\sqrt{\gamma} \bS(t_i)\!\cdot\! \bB(t_i)).
\end{aligned}
\end{equation}
We refer to the series expansion of $\rho(t)$ obtained by inserting the above expression into Eq.~\eqref{eqa:a1} as the {\it Dyson series for $\rho(t)$}.

Importantly, each term in the Dyson series for $\rho(t)$ depends {solely} on  $H_{\rm S}(t)$,  $\{S_\alpha\}$, and the $N$-point correlation functions of the bath (for $N\in \mathbb N$),

\begin{equation}\label{eq: cumulants}
    \left\langle B_1...B_N\right\rangle\coloneqq 
    \Tr\left[B_1...B_N\rho_{\rm B}\right],
\end{equation}
where $B_j$ is short-hand for
$
    B_{\alpha_j}(t_j)\coloneq e^{iH_{\rB} t_j}B_{\alpha_j}e^{-iH_{\rB}t_j}.
$ 
A defining fact about Gaussian baths used in this work is that they obey Wick's theorem \cite{Itzykson:1980rh}
\begin{equation}\label{eq: Wick's theorem}
    \left\langle B_1...B_N\right\rangle
    =
    \sum_{j=2}^N\left\langle B_1B_j\right\rangle
    \left\langle B_2...B_{j-1}B_{j+1}...B_N\right\rangle.
\end{equation}

Repeated use of Wick's theorem together with our assumption (without loss of generality) that $\langle B(t)\rangle=0$ shows that all terms in the Dyson series for $\rho(t)$ are {\it completely determined} by the system Hamiltonian $H_{\rm S}(t)$, the local coupling operators $\{S_{\alpha}\}$, and the matrix-valued two-point correlation function, defined as the matrix, $\bJ(t)$, whose $(\alpha,\beta)$-entry is given by
\begin{equation}
    J_{\alpha\beta}(t-t')\coloneqq \langle B_{\alpha}(t)B_{\beta}(t')\rangle.
\end{equation}

We now briefly argue that the above bath properties ensures that the dynamics are fixed by the two-point correlation function of the environment. Assuming $\Gamma<\infty$, $\norm{\bJ}_{1,1}$ is integrable ~\footnote{This follows from \eqref{eq: jump correlator Cauchy Schwarz}.}. This implies that the Dyson series for $\rho(t)$ is absolutely convergent. Therefore, the effects of Gaussian baths on the system dynamics are also {\it completely determined} by $H_{\rm S}(t)$, $\{S_\alpha\}$, and $J(t)$ \cite{FEYNMAN1963118,breuer_theory_2007}. For a more detailed proof of this fact, see \cite{tamascelli2018nonperturbative,park2408quasi}.

\section{Error bounds of the ATR and ATA}\label{app: errors}

In this appendix, we bound the error of the ATA, i.e., deviation of the evolution generated by the ATA and the true evolution of the simulated microscopic model.

We obtain the ATA through 4 successive approximations, outlined in Fig. \ref{Fig: error flowchart}, and described in Appendices \ref{app: discretization error}--\ref{app: Trotter error} below. We establish a rigorous bound on the error induced by each of these approximations; the error of the ATA is bounded by the sum of these  bounds. Likewise, the error of the ATR---i.e. the deviation of the system dynamics obtained from the  ATR  with respect to the true dynamics---is bounded by the {sum of the} errors induced by the first 3 of these approximations. 

Throughout this appendix, we make use of superoperator notation. We provide an introduction to this notation in the sub-appendix \ref{app: superoperator notation}. Throughout this Appendix, we  make frequent reference to a set of relations associated with matrix norms or  Dyson expansions; we provide proofs for these relations in
Appendix \ref{app: Common Relations and Notation}

\definecolor{blueish}{HTML}{2f97b1}

\definecolor{lightblue}{HTML}{b8ceea}

\definecolor{orangeish}{HTML}{ffa366}

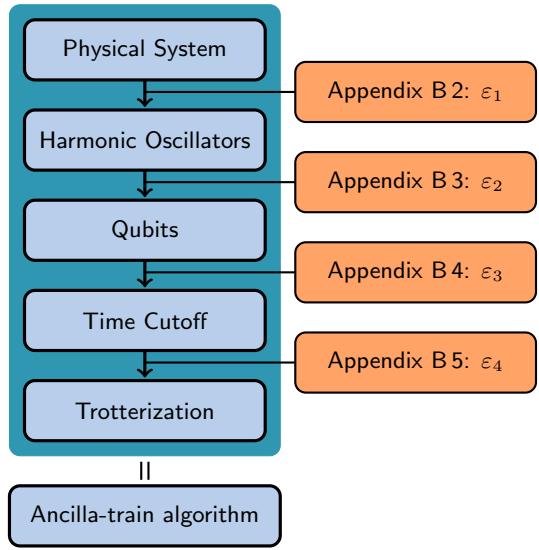
\begin{figure}[h!]
    \centering
\begin{tikzpicture}[scale = 0.8, every text node part/.style={align=center},font=\sffamily]

\filldraw[ rounded corners, color = blueish] (-2.25,0.25) rectangle (2.25,-7.25); 

\filldraw[rounded corners, color = lightblue] (-2,0) rectangle (2,-1);
\draw[rounded corners, very thick] (-2,0) rectangle (2,-1);
\node[] at (0,-0.5) {Physical System};

\draw[very thick, ->] (0,-1)--(0,-1.45);

\draw[thick] (0,-1.2)--(2.5,-1.2);

\filldraw[rounded corners, color = orangeish](2.5,-0.7) rectangle (6.5,-1.7);
\draw[rounded corners,thick] (2.5,-0.7) rectangle (6.5,-1.7);

\node[] at (4.5,-1.2) {Appendix \ref{app: discretization error}: $\varepsilon_1$};

\filldraw[rounded corners, color = lightblue] (-2,-1.5) rectangle (2,-2.5);
\draw[rounded corners, very thick] (-2,-1.5) rectangle (2,-2.5);
\node[] at (0,-2) {Harmonic Oscillators};

\draw[very thick, ->] (0,-2.5)--(0,-2.95);

\draw[thick] (0,-2.7)--(2.5,-2.7);

\filldraw[rounded corners, color = orangeish](2.5,-2.2) rectangle (6.5,-3.2);
\draw[rounded corners,thick] (2.5,-2.2) rectangle (6.5,-3.2);

\node[] at (4.5,-2.7) {Appendix \ref{app: qubit error}: $\varepsilon_2$};

\filldraw[rounded corners, color = lightblue] (-2,-3) rectangle (2,-4);
\draw[rounded corners, very thick] (-2,-3) rectangle (2,-4);
\node[] at (0,-3.5) {Qubits};

\draw[very thick, ->] (0,-4)--(0,-4.45);

\draw[thick] (0,-4.2)--(2.5,-4.2);

\filldraw[rounded corners, color = orangeish](2.5,-3.7) rectangle (6.5,-4.7);
\draw[rounded corners,thick] (2.5,-3.7) rectangle (6.5,-4.7);

\node[] at (4.5,-4.2) {Appendix \ref{app: cut-off error}: $\varepsilon_3$};

\filldraw[rounded corners, color = lightblue] (-2,-4.5) rectangle (2,-5.5);
\draw[rounded corners, very thick] (-2,-4.5) rectangle (2,-5.5);
\node[] at (0,-5) {Time Cutoff};

\draw[very thick, ->] (0,-5.5)--(0,-5.95);

\draw[thick] (0,-5.7)--(2.5,-5.7);

\filldraw[rounded corners, color = orangeish](2.5,-5.2) rectangle (6.5,-6.2);
\draw[rounded corners,thick] (2.5,-5.2) rectangle (6.5,-6.2);

\node[] at (4.5,-5.7) {Appendix \ref{app: Trotter error}: $\varepsilon_4$};

\filldraw[rounded corners, color = lightblue] (-2,-6) rectangle (2,-7);
\draw[rounded corners, very thick] (-2,-6) rectangle (2,-7);
\node[] at (0,-6.5) {Trotterization};

\draw[thick] (-0.06,-7.35)--(-0.06,-7.65);
\draw[thick] (0.06,-7.35)--(0.06,-7.65);

\filldraw[rounded corners, color = lightblue] (-2.25,-7.75) rectangle (2.25,-8.75);
\draw[rounded corners, very thick] (-2.25,-7.75) rectangle (2.25,-8.75);
\node[] at (0,-8.25) {Ancilla-train algorithm};

\end{tikzpicture}
\caption{\label{Fig: error flowchart}
    Overview of approximations leading from the physical system to the ATA. Orange boxes indicate the sub-appendices where the approximations are introduced and bounded.}
\end{figure}

\subsection{Superoperator notation}\label{app: superoperator notation}
We first introduce a convenient notation for {superoperators} that will be used extensively in this and in the following appendices. Superoperators are defined as linear operators acting on the {set of linear}  operators on {a given} Hilbert space $\mathscr{H}$~\footnote{\label{footnote: Hilbert-Schmidt}Technically, we mean Hilbert-Schmidt operators on $\mathscr{H}$.}. {Importantly, this set of operators itself forms a Hilbert space, termed {\it operator space}; superoperators can  be represented as matrices acting on this operator space.}
To highlight the{ir} Hilbert space nature, we will, in the following, use double brackets, $|\cdot\rrangle$, to indicate operators acting on $\mathscr{H}$, and use $\llangle \cdot|\cdot\rrangle $ to indicate the Hilbert-Schmidt inner product between two such operators~\footnote{When $\rho$ is traceclass, we will often abuse notation and write $\bbrakett{I\vert \rho}$ to mean $\Tr\left(\rho\right)$ even when $I$ is not Hilbert-Schmidt.}:
\begin{equation}
\llangle A|B\rrangle \coloneqq \Tr[A^\dagger B]. 
\end{equation}
A (bounded) operator on 
$\mathscr{H}$, $O$, induces two natural superoperators, namely left- and right-multiplication by $O$. We denote these induced superoperators  by \begin{equation}
O^L\kett{A}\coloneqq\kett{OA}\textnormal{ and } O^R\kett{A}\coloneqq\kett{A O}.
\end{equation}
Note that $[A^L,B^R]=0$ for any operators $A,B$.
As an important example, the  von-Neumann equation, $\frac{\partial}{\partial t}\rho(t) =-i[H,\rho(t)]$, may be written in the superoperator notation as \begin{equation}
    \frac{\partial}{\partial t}\kett{\rho(t)}=-i\nu_d H^d \kett{\rho(t)},
\end{equation}
{where  $\nu_L=-\nu_R=1$,} and we use the Einstein summation convention
\begin{equation}\label{eq: Eistein notation}
    \nu_d H^d \coloneq \sum_{d\in\{L,R\}}\nu_dH^d,
\end{equation}
We use {the above summation convention} for the indices  $\{L,R\}$ {when they appear in the following}.
We will {moreover}
in general use calligraphic symbols to indicate superoperators, when they are \emph{not} given by left- or right-multiplication by bounded operators on $\mathscr{H}$. In particular, we reserve $\mathcal{U}$ for time evolution superoperators. Given a Hamiltonian, $H(t)$, we define
\begin{equation}
    \mathcal{U}(t,t_0)
    \coloneq
    \mathcal{T}\exp\left(-i\nu_d\int_{t_0}^tds H^d(s)\right).    
\end{equation}
We refer to $\mathcal{U}(t,t_0)$ as \emph{the time evolution superoperator generated by} $H(t)$.
\subsection{Approximating the bath with a discrete train of harmonic oscillators}\label{app: discretization error}

We  now proceed to the proofs {of our error bounds}. We first {bound} the error that comes from representing the bath as a discrete train of harmonic oscillators {of the form described in Eq.~\eqref{eq: A osc} of the main text}~\footnote{Note that a similar construction was recently proposed in Ref.~\cite{lacroix2025making}}.

In the following, we let $\rho(t)$ denote the exact time-evolved system state as in Eq.~\eqref{eq: rho(t)} and by $\rho_{\rm osc}(t)$ the time-evolved system state resulting from an interaction with a discrete train of harmonic oscillators in place of the bath, defined via  Eqs. \eqref{eq: Hamiltonian osc} and \eqref{eq: rho osc} below.
Our goal is to bound the trace norm distance between these two evolutions, 
\begin{equation}
    \varepsilon_1 \coloneq \trnorm{\rho_{\rm osc}(t)-\rho(t)}.
\end{equation}
{We {call} this error} the \emph{discretization error}, since the system of harmonic oscillators can be seen as a discretization of the bath. 
Our derivation proceeds as follows: we first  show that the bath $\mathcal B$ is  equivalent to  a  combination of {\it two} specific environments, $\mathcal B_1$ and $\mathcal B_2$, in the sense that they produce the same evolution of the reduced density matrix of the system, $\rho(t)$. Here $\mathcal B_1$  can be taken to have a compactly supported PSD and $\mathcal B_2$ can be neglected at the cost of a small error [Sec.~\ref{seca:split}]. Subsequently, we show, via Poisson's summation formula that   $\mathcal B_1$ is equivalent to the  harmonic oscillator train Hamiltonian that generates the evolution of $\rho_{\rm osc}(t)$ [Sec.~\ref{app: B1 harmonic osc equivalence}].

\subsubsection{Dividing the bath in two parts}
\label{seca:split} 
To construct $\mathcal B_1$ and $\mathcal B_2$, we  express the bath PSD $\tilde{\bJ}(\omega)$ as a sum of two power spectral densities, $\tilde{\bJ}(\omega)= \tilde{\bJ}_1(\omega)+ \tilde{\bJ}_2(\omega)$, where $\tilde{\bJ}_1(\omega)$ and $\tilde{\bJ}_2(\omega)$ are positive semi-definite, and have support only for $|\omega|<\Omega$ and $\abs{\omega} >\Omega/2$ respectively, with
$\Omega$ some arbitrary effective frequency cutoff we introduce as a free parameter. In the following, we show that, when choosing $\Omega$ large enough,
$\tilde{\bJ}_2(\omega)$ can be neglected at the cost of a small bounded error [we give {an} explicit expression below in Eq.~\eqref{eq: J2 bound}], resulting in a bath whose PSD [$\tilde{\bJ}_1(\omega)$] has compact support. 

We construct the bath PSDs $\tilde{\bJ}_1(\omega)$ and $\tilde{\bJ}_2(\omega)$ {as follows:} 
\begin{equation}
    \begin{aligned}
        \tilde{\bJ}_1(\omega) &\coloneqq (\tilde{\varphi}(\omega))^2 \tilde{\bJ}(\omega),\\
    \tilde{\bJ}_2(\omega) &\coloneqq (1-(\tilde{\varphi}(\omega))^2) \tilde{\bJ}(\omega), 
    \end{aligned}
\end{equation}
{Here $\varphi$ can be any function satisfying} $0\leq \tilde{\varphi}(\omega)\leq 1$ for all $\omega\in \mathbb{R}$, $\tilde{\varphi}(\omega)=1$ for $\abs{\omega}\leq \Omega/2$ and $\tilde{\varphi}(\omega)=0$ for $\abs{\omega}>\Omega$.
 To avoid introducing long time-correlations in the bath described by $\tilde{\bJ}_1(\omega)$, we shall be careful in {choosing $\varphi(\omega)$ to be {sufficiently} smooth}. {To this end,}
we pick a second-order polynomial interpolation for $\tilde{\varphi}(\omega)$, ensuring that it is twice differentiable almost everywhere:
\begin{equation}\label{eq: phi bump function}
    \tilde{\varphi}(\omega)=\begin{cases}
        1&\textnormal{for }\abs{\omega}<\frac{\Omega}{2}\\
        1-\frac{1}{2}\left(\frac{4}{\Omega}\right)^2(\abs{\omega}-\Omega/2)^2&\textnormal{for }\frac{\Omega}{2}\leq \abs{\omega}<\frac{3\Omega}{4}\\
        \frac{1}{2}\left(\frac{4}{\Omega}\right)^2(\abs{\omega}-\Omega)^2&\textnormal{for }\frac{3\Omega}{4}\leq \abs{\omega}<\Omega\\
        0& \textnormal{for }\abs{\omega}\geq \Omega
    \end{cases}.
\end{equation}
By this choice, the characteristic  timescales $\Gamma_{\rm f}$ and $\tau_{\rm f}$ of $\mathcal{B}_1$ are of similar magnitude as the corresponding timescales $\Gamma$ and $\tau$ of the   original bath $\mathcal{B}$ (see Appendix~\ref{app:Gamma tau bound proof}).

Importantly, since $ \tilde{\bJ}(\omega) = \tilde{\bJ}_1(\omega)+ \tilde{\bJ}_2(\omega)$, and the effects of a Gaussian environment is completely determined  {by} its two-point correlation function (see Appendix~\ref{app: Bath Constructions and Properties} and Ref.~\cite{FEYNMAN1963118}), the  physical environment we wish to simulate can be exactly represented
as a combination of 
two uncorrelated environments, $\mathcal{B}_1$ and $\mathcal{B}_2$, with PDSs $  \tilde{\bJ}_1(\omega)$ and $ \tilde{\bJ}_2(\omega)$ such that the system--bath Hamiltonian takes the form~\cite{FEYNMAN1963118}\begin{equation}
    H_{12}(t)=H_{\rm S}(t)+H_{\mathcal{B}_1}+H_{\mathcal{B}_2}+\sqrt{\gamma}\mathbf{S}\cdot (\mathbf{B}_1+\mathbf{B}_2),
\end{equation}
with $\mathbf{B}_{1,2}$  observables of Gaussian environments with PSDs
\begin{align}
    \bJ_1(t)&=\langle \mathbf{B}_1(t) \left(\mathbf{B}_1(0)\right)^\dagger\rangle_{\mathcal{B}_1},\\
    \bJ_2(t)&=\langle \mathbf{B}_2(t) \left(\mathbf{B}_2(0)\right)^\dagger\rangle_{\mathcal{B}_2},\\
    \langle \mathbf{B}_1(t) \left(\mathbf{B}_2(0)\right)^\dagger\rangle_\mathcal{B}&=\langle \mathbf{B}_2(t)\left(\mathbf{B}_1(0)\right)^\dagger \rangle_\mathcal{B}=0.
\end{align}
Specifically, the corresponding equation of motion for the reduced density of the system described by $H_{12}(t)$ is the same as that arising from the system--bath Hamiltonian given in Eq.~\eqref{eq: H interaction} of the main text:
\begin{equation} \label{eq: rho exact}
\begin{aligned}
        \kett{\rho_{}(t)}\coloneqq&\bbra{I}_{\mathbf{B}}\mathcal{U}_{}(t,t_0)\kett{\rho_0}\kett{\rho_{\mathbf{B}}}\\=&\bbra{I}_{\mathbf{B}}\mathcal{U}_{12}(t,t_0)\kett{\rho_0}\kett{\rho_\mathbf{B}}.
\end{aligned}
\end{equation}
Here $\mathcal{U}_{}$  denotes the time evolution superoperator of $H_{\rm SB}$ from Eq. \eqref{eq: H interaction} and $\mathcal{U}_{12}$   the time evolution superoperator generated by $H_{12}$. 
As the combination of the two baths $\mathcal{B}_1 \otimes \mathcal{B}_2$ is equivalent to the original bath $\mathcal{B}$, we use $\mathcal{B}$ also to denote the combined baths $\mathcal{B}_1 \otimes \mathcal{B}_2$.

We now show, that the  bath $\mathcal{B}_2$ can be neglected at the cost of a small error, when $\Omega$ is sufficiently large. To formalize this statement, we define $H_1(t)=H_{\rm S}(t)+H_{\bB_1}+H_{\bB_2}+\sqrt{\gamma}\mathbf{S}\cdot\mathbf{B}_1$, and the time evolution superoperator of $H_1$ by $\mathcal{U}_{1}$. We also denote the time-evolved reduced density matrix of the system arising from $\mathcal{U}_1$ by  
\begin{equation}\label{eq: rho 1}
\begin{aligned}
\kett{\rho_1(t)}\coloneqq\bbra{I}_{\mathbf{B}}\mathcal{U}_{1}(t,t_0)\kett{\rho_0}\kett{\rho_{\mathbf{B}}}.
\end{aligned}
\end{equation}
Since the time-evolution of $\rho(t)$  generated by $H_{\rm SB}(t)$ and $H_{\rm 12}(t)$ are identical, bounding the distance $\norm{\rho_{1}(t)-\rho(t)}_{\Tr}$ is reduced to bounding the effect of decoupling the split system from $\mathcal{B}_2$.

To proceed, we transform the system described by $H_{12}(t)$ to the interaction picture with respect to $ \sqrt{\gamma} \mathbf{S}\cdot\mathbf{B}_2$, finding \begin{equation}
    \kett{{\rho}_{}(t)}=\bbra{I}_{B}\mathcal{U}_1(t,t_0)\mathcal{U}_2(t,t_0)\kett{\rho_0}\kett{\rho_{\rm B}}
\end{equation} with ${\mathcal{U}}_2(t,t_0)=\mathcal{T}e^{-i\int_{t_0}^{t}ds{\mathcal{H}}_2(s)}$ and ${\mathcal{H}}_2(s)=\sum_{d\in \{L,R\}}\nu_{d}\sqrt{\gamma}\mathbf{S}^d(s)\cdot\mathbf{B}_2^d(s)$. Here we defined the interaction picture operators; $\mathbf{S}^d(s)=e^{iH_1 s}\mathbf{S}^de^{-iH_1 s}$ and $\mathbf{B}_2(s)=e^{iH_{\bB_2} s}\mathbf{B}_2e^{-iH_{\bB_2} s}$. 
Using a simple Dyson expansion on $\mathcal{U}_2$   [see Eq.~\eqref{Simple Dyson}], along with Eq.~\eqref{eq: rho 1}, we conclude 
\begin{widetext}
    \begin{equation}\label{eq: rho dyson}
    \begin{aligned}
        \kett{{\rho}_{}(t)}&=\kett{\rho_{1}(t)}-i\int_{t_0}^tds \bbra{I}_{B}\mathcal{U}_1(t,t_0){\mathcal{H}}_2(s){\mathcal{U}}_2(s,t_0)\!\kett{\rho_0}\!\!\kett{\rho_{\rm B}}.
    \end{aligned}
\end{equation}
 Next, we write $\mathcal{U}_2$ as a Trotterized product with infinitesimal Trotter step: \begin{equation}
    \mathcal{U}_2(s,t_0)=\lim_{M\to\infty}\prod_{m=0}^M\left[1-i \frac{s-t_0}{M} \sum_{d\in \{L,R\}}\nu_d \sqrt{\gamma} \mathbf{S}^d\left(\frac{m(s-t_0)}{M}\right)\cdot \mathbf{B}^d\left(\frac{m(s-t_0)}{M}\right)\right].
\end{equation} Upon direct substitution of the above into Eq.~\eqref{eq: rho dyson}, and using Wick's theorem \eqref{eq: Wick's theorem} for $\mathcal{B}_2$, it follows that 
  \begin{equation}
    \begin{aligned}
        |\rho(t)\rrangle =|\rho_1(t)\rrangle  -i\gamma \sum_{\alpha,\beta}\nu_{d}\nu_{e}\!\!\int_{t_0}^t\!\!ds\int_{t_0}^s\!\! ds' J^{de}_{2,{\alpha\beta}}(s-s') \bbra{I}_{\!B}\mathcal{U}_1(t,t_0){\mathbf{S}}_\alpha^{d}(s){\mathcal{U}}_2(s,s'){\mathbf{S}}^{e}_\beta(s') {\mathcal{U}}_2(s',t_0)\!\kett{\rho_0}\!\kett{\rho_{\rm B}},
    \end{aligned}
\end{equation}
where, for $d=L,R$, $J^{de}_{2,{\alpha\beta}}$ denotes the $(\alpha,\beta)$-entry of the matrix $\bJ_2$ when $e=L$ and the $(\alpha,\beta)$-entry of the matrix $\bJ_2^\dagger$ for $e=R$. Furthermore, we used the Einstein summation convention from \eqref{eq: Eistein notation}.
To bound the integrand above we use that 
$|\Tr[X\rho Y]|\leq\norm{X}\norm{Y}$; this auxiliary result is  established in Appendix~\ref{app: Common Relations and Notation} [Eq.~\eqref{Bound on trace norm of generic error term}]. Inserting this above implies \begin{equation}
\begin{aligned}
        \label{eqa: B19} \norm{\!\bbra{I}_{\!B}\mathcal{U}_1(t,t_0){\mathbf{S}}_\alpha^{d}(s){\mathcal{U}}_2(s,s'){\mathbf{S}}^{e}_\beta(s') {\mathcal{U}}_2(s',t_0)\!\kett{\rho_0}\!\kett{\rho_{\rm B}}\!}_{\rm tr}\leq 1.
\end{aligned}
\end{equation}
Thus,
\begin{equation}
\begin{aligned}
        &\norm{\rho_{1}(t)-\rho_{}(t)}_{\Tr}\leq 4\gamma \int_{t_0}^{t}ds\int_{t_0}^{s}ds' \norm{\bJ_{2}(s-s')}_{1,1}.
    \end{aligned}
\end{equation}
Splitting the integration domain in $s'>s_0$ and $s'<s_0$, where $s_0$ is some free parameter to be chosen below, we find 
\begin{equation}
\begin{aligned}\label{eq: J2 bound}
        \norm{\rho_{1}(t)-\rho_{}(t)}_{\Tr}&\leq 4\gamma \int_{t_0}^{t}ds\left(\int_{-s_0}^{s_0}ds' \norm{\bJ_{2}(s')}_{1,1}+2\int_{s_0}^{\infty}ds' \frac{1}{(s')^2} \norm{(s')^2 \bJ_{2}(s')}_{1,1}\right).
    \end{aligned}
\end{equation}
\end{widetext}
{Using} pointwise bounds $\abs{J_{2,{\alpha\beta}}(s)}\leq \int_{-\infty}^{\infty} d\omega \tilde{J}_{2,\alpha\beta}(\omega) $ and $s^2\abs{J_{2,{\alpha\beta}}(s)}\leq \int_{-\infty}^{\infty} d\omega \abs{\tilde{J}''_{2,\alpha\beta}(\omega)} $, we find 
\begin{equation}
\begin{aligned}
        \norm{\rho_{1}(t)-\rho_{}(t)}_{\Tr}\leq 4\gamma\left(s_0\epsilon_1(\Omega)+\frac{1}{s_0}\epsilon_2(\Omega)\right)T.
    \end{aligned}
\end{equation}
with 
\begin{eqnarray} 
\epsilon_1(\Omega)&=&2\int_{\abs{\omega}>\Omega}d\omega \norm{\tilde{\bJ}(\omega)}_{1,1}\\ 
\epsilon_2(\Omega)&=&2\int_{\abs{\omega}>\Omega}d\omega \norm{\tilde{\bJ}''(\omega)}_{1,1}.
\end{eqnarray}
It follows that, for $t<T$, 
\begin{equation}
\begin{aligned}
        \norm{\rho_{1}(t)-\rho_{}(t)}_{\Tr}\leq \epsilon(\Omega)\Gamma T
    \end{aligned},
\end{equation}
with  \begin{equation}\label{eq: discretization epsilon function}
    \epsilon(\Omega)\coloneqq\frac{2\sqrt{\epsilon_1(\Omega)\epsilon_2(\Omega)}}{\left[\int_{-\infty}^\infty dt \norm{\bg(t)}_{2,1} \right]^2}.
\end{equation} For $\lVert\tilde{\bJ}\rVert _{1,1}, {\lVert}\tilde{\bJ}''{\rVert}_{1,1}\in L^1(\mathbb{R})$ we note that $\epsilon(\Omega)\to 0$ as $\Omega \to \infty$.
Hence,  the full bath can be closely approximated by the interaction with a bath where the PSD has compact support. In particular, fixing $\Omega= \Lambda(\varepsilon)$, where
\begin{equation}
  \Lambda(\varepsilon)\coloneq \inf \{\Omega\geq 1/\tau |\epsilon(\Omega)<\varepsilon\},
\end{equation}
the trace distance between $\rho(t)$ and $\rho_1(t)$ stays within the target accuracy, i.e.
\begin{equation}
    \norm{\rho_{1}(t)-\rho_{}(t)}_{\Tr}\leq \varepsilon\Gamma \tau. 
    \label{eqa:epsilon_1_bound}
\end{equation}

\subsubsection{Equivalence between the bath \texorpdfstring{$\mathcal B_1$}{B1}  and a train of harmonic oscillators.}\label{app: B1 harmonic osc equivalence}
We now show that the dynamics of a system coupled to  the bath $\mathcal{B}_1$ are identical to those generated by the a train of harmonic oscillators, as in Eq.~\eqref{eq: A osc}. 
Specifically, exploiting the compact support of $\tilde{\boldsymbol{J}}_1(\omega)$, we show that the evolution of $|\rho_1(t)\rrangle$ is equivalent to the reduced density matrix  evolution 
generated by the harmonic oscillator train Hamiltonian
\begin{equation}\label{eq: Hamiltonian osc}
    H_{\rm osc}(t)=H_{\rm S}(t)+\sqrt{\gamma} {\mathbf S}\cdot {\mathbf A}_{\rm osc}(t),
\end{equation}
where $\mathbf A_{\rm osc}(t)$ is the column vector whose $\alpha$th entry is given by 
\begin{equation}
{A}_{ {\rm osc},\alpha}(t)=\!\!\!\sum_{n=-\infty}^{\infty}\sum_{\lambda=1}^{N_C}\sqrt{\Delta\xi}\left(g_{{\rm f};\alpha\lambda}^\ast (t-\xi_n){a}_{\lambda n}^\dagger+{h.c.}\right).
\end{equation}
Here, $\Delta \xi$ can take any value smaller than, or equal to, $2\pi /\Omega$, $  \{a_{\lambda n}|\lambda=1,\ldots N_C,n\in \mathbb Z\}$ are a set of distinct bosonic annihilation operators, and $\bg_{\rm f}(t)$ denotes the jump correlator of the filtered environment:
\begin{equation}
    \bg_{\rm f}(t)=\int d\omega e^{-i\omega t} \sqrt{\tilde{\bJ}_1(\omega)/2\pi},
\end{equation}
with $\sqrt{\cdot}$ denoting the positive-semidefinite matrix square-root. 
We define the harmonic oscillator train evolution of the system as
\begin{equation}\label{eq: rho osc}
    \rho_{\rm osc}(t)\coloneqq \llangle I_{\rm osc}|\mathcal U_{\rm osc}(t,t_0)|\rho_0\rangle |\boldsymbol {0}\rangle 
\end{equation}
with $\mathcal U_{\rm osc}(t,t_0)$ the unitary time-evolution superoperator generated by $H_{\rm osc}(t)$, and $\ket{\boldsymbol{0}}=\ket{0,0,\ldots, 0}$ the mutual vacuum states of all bosonic modes corresponding to $\{a_{\lambda n}\}$. 
We note that $|\boldsymbol{0}\rangle\langle \boldsymbol{0}|$ is  a Gaussian state; hence the $n$-point correlation functions of $\mathbf{A}_{\rm osc}$ in this state satisfy Wick's theorem. 
Thus, due to the equivalence among Gaussian environments described in Appendix \ref{app: Bath Constructions and Properties}, the evolution of $\rho_{\rm osc}(t)$ is identical to that of any open quantum system coupled to a Gaussian environment with system Hamiltonian $H_{\rm S}(t)$, system--bath coupling $\sqrt{\gamma}\bS\cdot \bB'$, with $\bB'$ a bath observable with  two-point correlation function given by 
\begin{equation}
    \langle \boldsymbol{0}\vert \mathbf{A}_{\rm osc}(t)\left(\mathbf{A}_{\rm osc}(s)\right)^\dagger \vert \boldsymbol{0}\rangle=\sum_n \bg_{\rm f}(t-\xi_n)\bg_{\rm f}(\xi_n-s)\Delta\xi,  
    \label{eqa:Aoscdef}
    \end{equation}
We now show that the right-hand side above is equal to $\bJ_1(t-s)$, implying that $\rho_{\rm osc}(t)=\rho_1(t)$. 

To establish that the right-hand-side of Eq.~\eqref{eqa:Aoscdef} is identical to $\bJ_1(t-s)$, we use Poisson's summation formula \cite{benedetto1997sampling} which states 
\begin{equation}
\begin{aligned}
        \sum_{n=-\infty}^{\infty} &\bg_{\rm f}(t-\xi_n)\bg_{\rm f}(\xi_n-s)\Delta\xi\\
        &=\sum_{k=-\infty}^{\infty} \int_{-\infty}^\infty d\xi \bg_{\rm f}(t-\xi)\bg_{\rm f}(\xi-s)e^{-i\frac{2\pi \xi k}{\Delta\xi}}.
\end{aligned}
\end{equation}
Inserting $\bg_{\rm f}(t)=\int_{-\infty}^{\infty} d\omega {\sqrt{\tilde{\bJ}_1(\omega)/2\pi}} e^{-i\omega t}$, it is straightforward to show that
\begin{equation}\label{EqDiscretization1}
\begin{aligned}
        &\int_{-\infty}^\infty d\xi \bg_{\rm f}(t-\xi)\bg_{\rm f}(\xi-s)e^{-i\frac{2\pi \xi k}{\Delta\xi}}.\\
        &=\int_{-\infty}^{\infty}d\omega {\sqrt{\tilde{\bJ}_1\left(2\pi k/\Delta\xi+\omega\right)}\sqrt{\tilde{\bJ}_1(\omega)}}e^{-i\omega (t-s)}e^{-i\frac{2\pi kt}{\Delta\xi}}.
\end{aligned}
\end{equation}
Thus, since ${\sqrt{\tilde{\bJ}_1(\omega)}}=0$ for $\abs{\omega}\geq \pi/\Delta\xi$, we find, for $k\neq 0$,
\begin{equation}
\begin{aligned}
        &\int_{-\infty}^\infty d\xi \bg_{\rm f}(t-\xi)\bg_{\rm f}(\xi-s)e^{-i\frac{2\pi \xi k}{\Delta\xi}}=0.
\end{aligned}
\end{equation}
It follows that
\begin{equation}
\begin{aligned}
        \sum_{n=-\infty}^{\infty} &\bg_{\rm f}(t-\xi_n)\bg_{\rm f}(\xi_n-s)\Delta\xi\\
        &= \int_{-\infty}^\infty d\xi \bg_{\rm f}(t-\xi)\bg_{\rm f}(\xi-s)=\bJ_1(t-s).
\end{aligned}
\end{equation}
This is what we wanted to show; thus \begin{equation}\rho_1(t)=\rho_{\rm osc}(t).\end{equation} 

Combining the above result with the bound on  $\norm{\rho_1(t)-\rho(t)}_{\rm tr}$ in Eq.~\eqref{eqa:epsilon_1_bound}, and recalling that $\Delta \xi = 2\pi /\Omega$ we conclude that 
\begin{equation}
    \varepsilon_1\coloneq \norm{\rho_{\rm osc}(t)-\rho_{}(t)}_{\Tr}\leq \varepsilon\Gamma T. 
\end{equation}
when $\rho_{\rm osc}$ is constructed with $\Delta \xi= 2\pi/\Omega$ and $\Omega =\Lambda(\varepsilon)$.
This establishes our bound on $\varepsilon_1$.

\subsection{Error from using qubits in place of harmonic oscillators in the ancilla train}\label{app: qubit error}

In this section we bound the error incurred by using qubits in place of harmonic oscillators in the ancilla train. 
Specifically, we let $\rho_{\rm q}(t)$ denote the time evolution of the reduced density matrix of the system generated by $H_{\rm q}(t)$, where
\begin{equation}\label{eq: Hamiltonian q}
    H_{\rm q}(t)=H_{\rm S}(t)+\sqrt{\gamma} {\mathbf S}\cdot {\mathbf A}_{\rm q}(t).
\end{equation}
The qubit error is defined as the trace norm distance between these states:
\begin{equation}
    \varepsilon_2\coloneq \norm{\rho_{\rm osc}(t)-\rho_{\rm q}(t)}_{\rm tr},
\end{equation}
In this subsection, we obtain the following bound for this error
\begin{equation}
    \varepsilon_2
    \leq
    25.96 \Delta \xi \left(T+\Delta \xi\right)  \Gamma^2,
    \label{eqa:e2_bound}
\end{equation}
where $T$ is the total simulation time.

To establish Eq.~\eqref{eqa:e2_bound}, it is convenient to consider the quantum channels for the two time evolutions $\rho_{\rm q}$ and $\rho_{\rm osc}$; $\mathcal{C}_{\rm q}$ and $\mathcal{C}_{\rm osc}$, {respectively}, such that $\rho_{\rm osc}(t)=\mathcal{C}_{\rm osc}(t,t_0)[\rho(t_0)]$ and $\rho_{\rm q}(t)=\mathcal{C}_{\rm q}(t,t_0)[\rho(t_0)]$. We quantify the error in terms of the difference between the two quantum channels, defining
\begin{equation}
    \mathcal{E}_{\rm q}(t,t_0)\coloneq \mathcal{C}_{\rm q}(t,t_0)-\mathcal{C}_{\rm osc}(t,t_0).
\end{equation}
We term $\mathcal{E}_{\rm q}(t,t_0)$ the error map. Importantly, $\varepsilon_2 $ is upper bounded by the superoperator norm of the error map
\begin{equation}
    \varepsilon_2\leq\norm{\mathcal{E}_{\rm q}(t,t_0)}_s,
\end{equation}
where $\norm{\cdot}_{\rm s}$ denotes the superoperator norm induced by the trace norm:
\begin{equation}
    \norm{\mathcal{E}_{\rm q}(t,t_0)}_s
    \coloneq
    \sup_{\rho: \norm{\rho}_{\rm tr}\leq 1}
    \norm{\mathcal{E}_{\rm q}(t,t_0)[\rho]}_{\rm tr}.
\label{eq:qchannelnorm}\end{equation}
The rest of this appendix is dedicated to bounding $\norm{\mathcal{E}_{\rm q}(t,t_0)[\rho]}_{\rm tr}.$ To this end, we first obtain a convenient explicit expression for $\mathcal{E}_{\rm q}$ in subsection \ref{sec: qubit construct error map}. Subsequently, {in subsection \ref{sec: qubit expand error map}}, we expand this expression in a collection of terms that we bound individually through a Dyson expansion. In subsection \ref{sec: qubit combine errors}, we finally combine these bounds through triangle inequality to obtain a bound on $\norm{\mathcal{E}_{\rm q}(t,t_0)}_s$, and thereby $\varepsilon_2$.

\subsubsection{Constructing the error map}\label{sec: qubit construct error map}
To obtain an  explicit expression for the error map $\mathcal{E}_{\rm q}(t,t_0)$
we seek to express the quantum channel $\mathcal C_{\rm q}$  in terms of $\mathcal C_{\rm osc}$. Letting $\mathcal{U}_{\rm osc}(t,t_0)$ denote the time-evolution superoperator generated by $H_{\rm osc}(t)$ [Eq.~\eqref{eq: Hamiltonian osc}], the quantum channel $\mathcal{C}_{\rm osc}$, can be explicitly written as 
\begin{equation}\label{eq: quantum channel osc}
    \mathcal{C}_{\rm osc}(t,t_0)=\bbra{I_A}\mathcal{U}_{\rm osc}(t,t_0)\kett{\boldsymbol{0}},
\end{equation}
with $|I_A\rrangle$ the identity operator on the Hilbert space of the ancilla oscillators. Since the harmonic oscillators are initiated in the vacuum state and since $H_{\rm osc}(t)$ is linear in creation operators acting on the harmonic oscillators, we can obtain $\mathcal C_{\rm q}$ by removing  terms in $H_{\rm osc}$ that couple the first and second excited level in each ancillary oscillator. We introduce the following notation: Given an operator defined on the Hilbert space of a single harmonic oscillator, $X$, we denote by $(X)_{\lambda n}$ the operator which acts with $X$ on oscillator $(\lambda, n)$ and acts as the identity on all other oscillators\footnote{Formally, we define $(X)_{\lambda n}\coloneq \left(\bigotimes_{(\lambda',n') \prec (\lambda,n)}{\rm I_{\lambda'n'}}\right)\otimes X_{\lambda n} \otimes \left( \bigotimes_{ (\lambda,n)\prec (\lambda'',n'')}{\rm I_{\lambda''n''}}\right)$, where $\prec$ denotes some total ordering of the oscillators. The choice of ordering is not important as long as it is used consistently.}. This allows us to arrive at an equivalent expression for $H_{\rm q}$ in terms of $H_{\rm osc}$ by removing the terms in the individual oscillators which mix the first and second excited states~\footnote{The two operators in \eqref{eq: Hamiltonian q} and \eqref{eqa:hqdef} are equivalent in the sense that \eqref{eq: Hamiltonian q} is isospectral to \eqref{eqa:hqdef} when the latter is restricted to the invariant subspace of the oscillator space, where all oscillators are in the span of the vacuum state and first excited state. As the initial state of the oscillators is taken to belong to this subspace, so are the entire dynamics. Therefore, \eqref{eq: Hamiltonian q} and \eqref{eqa:hqdef} generate the same dynamics, with our choice of initial state for the ancillas/oscillators.}:
\begin{equation} 
    H_{\rm q}(t) = H_{\rm osc}(t)-\sum_{\lambda}\sum_{n=-\infty}^\infty K_{\lambda n;2}(t)
    \label{eqa:hqdef}
\end{equation}
with
\begin{equation}\label{eq: definition of K's for qb error}
\begin{split}
    K_{\lambda n;j}(t)
    \coloneq
    &\sum_\alpha \sqrt{j \gamma\Delta \xi}S_\alpha \\&\left[g_{\rm f\it;\alpha\lambda}^*(t-\xi_n)(\ketbra{j}{j-1})_{\lambda n} +h.c.\right],
\end{split}
\end{equation}
where $\ket{j}$ denotes the $j$th harmonic oscillator Fock state. Specifically, denoting the time evolution superoperator generated by $H_{\rm q}(t)$ by $\mathcal{U}_{\rm q}(t,t_0)$,
we can express $\mathcal{C}_{\rm q}$ in terms of $\mathcal{U}_{\rm q}$ via
\begin{equation}\label{eq: Quantum chanel q}
\mathcal{C}_{\rm q}(t,t_0)=\bbra{I_A}\mathcal{U}_{\rm q}(t,t_0)\kett{\boldsymbol{0}}.
\end{equation}
The error map can now be found as the difference between the two quantum channels.
Through Eq.~\eqref{eqa:hqdef}, we can now express $\mathcal{U}_{\rm q}$ in terms of $\mathcal{U}_{\rm osc}$ via a Dyson expansion allowing us to obtain a compact expression for $\mathcal E_{\rm q}$. To this end, we make extensive use of the following two Dyson-type identities in the remainder of this Appendix: Given a superoperator $\mathcal{U}(t,t_0)$ generated by a Hamiltonian of the form $H(t)=H_1(t)+H_2(t)$, the following two relations hold:
\begin{align}
\label{eq: Dyson split expansion in text 1}
        \mathcal{U}(t,t_0)&=&
        \mathcal{U}_1(t,t_0)
        -i\nu_d 
        \int_{t_0}^tds 
        \mathcal{U}_1(t,s)H^d_2(s)\mathcal{U}(s,t_0)\\
\label{eq: Dyson split expansion in text 2}
        \mathcal{U}(t,t_0)&=&
        \mathcal{U}_1(t,t_0)
        -i\nu_d 
        \int_{t_0}^tds 
        \mathcal{U}(t,s)H^d_2(s)\mathcal{U}_1(s,t_0),
\end{align}
where  $\mathcal{U}_1(t,t_0)$ is the time evolution superoperator generated by $H_1(t)$.
Due to our extensive use of these identities, for shorthand we refer to the substitutions in Eqs.~\eqref{eq: Dyson split expansion in text 1},\eqref{eq: Dyson split expansion in text 2} as a left and right $H_2(t)$--Dyson substitution of $\mathcal{U}(t,t_0)$ respectively. We moreover refer to $\mathcal{U}_1(t,t_0)$ as the leading order term, and the second terms above as the residual term. The effect of a $H_2(t)$--Dyson substitution is to isolate the dynamics generated by $H_2(t)$ to the residual term. Using suitable choices of $H_2$, this allows us {to} eliminate leading-order terms in our analysis below. We provide  proofs of {the} Dyson substitutions above in Appendix \ref{app: Dyson Expansion}.

To express $\mathcal{U}_{\rm q}$ in terms of $\mathcal{U}_{\rm osc}$ via the above Dyson substitutions, we first perform a $\sum_\lambda\sum_{n=-\infty}^\infty K_{\lambda n;2}(t)$--Dyson substitution of $\mathcal{U}_{\rm osc}(t,t_0)$, obtaining
\begin{equation}
\begin{aligned}
    \mathcal{U}_{\rm osc}(t,t_0) =& \mathcal{U}_{\rm q}(t,t_0)
    \\
    -&i\sum_{\lambda,n}\nu_d\int_{t_0}^t ds \mathcal{U}_{\rm osc}(t,s) K_{\lambda n;2}^d(s) \mathcal{U}_{\rm q}(s,t_0).
\end{aligned}
\end{equation}
From the expressions for the  quantum channels [Eqs.~\eqref{eq: Quantum chanel q} and \eqref{eq: quantum channel osc}], we thus identify the error map as
\begin{equation}\label{eq: error map}
    \mathcal{E}_{\rm q}(t,t_0)\!=\!\sum_{\lambda,n}
    \IA
    i\nu_d\!\!\int_{t_0}^t ds \mathcal{U}_{\rm osc}(t,s) K_{\lambda n;2}^d(s) \mathcal{U}_{\rm q}(s,t_0)
    \Az.
\end{equation}
This expression serves as a starting point for obtaining our bound on the error map.

\subsubsection{Expanding the error map}\label{sec: qubit expand error map}
In this section, we use Dyson substitution to expand the error map in Eq.~\eqref{eq: error map}  into several maps whose superoperator norm can be bounded individually. To this end, we seek to bound each individual term in the sum in Eq. ~\eqref{eq: error map} 
\begin{equation}\label{Error before Dyson exp}
\begin{aligned}
    &\mathcal{E}_{\rm q\it ;\lambda n}(t,t_0)
    \coloneq
    \\&
    \IA
    i\nu_d\int_{t_0}^t ds \mathcal{U}_{\rm osc}(t,s) K_{\lambda n;2}^d(s) \mathcal{U}_{\rm q}(s,t_0)
    \Az.
\end{aligned}
\end{equation}
Our strategy is to further expand $\mathcal{E}_{\rm q\it ;\lambda n}(t,t_0)$ in powers of $K_{\lambda n;j}(t)$ using Dyson substitutions in such a way that the leading order terms vanish when we trace out the harmonic oscillators. Each time we perform such a Dyson substitution, the resulting expression gets suppressed by a factor $\sqrt{\gamma \Delta \xi}$. We are able to perform $4$ such Dyson substitutions, allowing us to eventually achieve a bound for the norm of $\mathcal{E}_{\rm q\it ;\lambda n}(t,t_0)$ which is of order $(\gamma\Delta \xi)^2$.
To establish {that} the leading-order terms {are indeed vanishing}, we make extensive use of arguments based on the support of quantities in the operator spaces of the individual ancillas. For simplicity, we introduce the following terminology: Below, we say that $\kett{A}$ has its $(\lambda,n)$--support confined to the subspace spanned by $\ketbra{i_1}{j_1},...,\ketbra{i_m}{j_m}$ if $\Tr_{\lambda n}\left[A(\ketbra{\phi}{\psi})_{\lambda n}\right]=0$, for any $(\ketbra{\phi}{\psi})_{\lambda n}$ orthogonal to the subspace spanned by $(\ketbra{i_1}{j_1})_{\lambda n},...,(\ketbra{i_m}{j_m})_{\lambda n}$. Here $\Tr_{\lambda n}$ denotes the partial trace over oscillator $(\lambda,n)$, and orthogonality is defined with respect to the Hilbert-Schmidt inner product. 

To identify the first vanishing leading order term in $\mathcal{E}_{\rm q\it ;\lambda n}(t,t_0)$, we perform a $K_{\lambda n;1}$--Dyson substitution of $\mathcal{U}_{\rm q}$ in Eq.~\eqref{Error before Dyson exp} to obtain
\begin{equation}
\begin{aligned}
        \mathcal{U}_{\rm q}(s,t_0)
    =
    &\mathcal{U}_{\rm q}^{\lambda n;1}(s,t_0)
    \\& +i\nu_e \int_{t_0}^sdu\mathcal{U}_{\rm q}(s,u)K^e_{\lambda n;1}\mathcal{U}_{\rm q}^{\lambda n;1}(u,t_0),
\end{aligned}
\end{equation}
{where we use $\mathcal{U}_{\rm q}^{\lambda n;j}$ to denote the time evolution superoperator generated by $H_{\rm q}-K_{\lambda n;j}$. Crucially, $\mathcal{U}_{\rm q}^{\lambda n;1}$ cannot couple the $0$th Fock  state of oscillator $(\lambda,n)$ to  {any} other Fock state. 
({note that }the first Fock state is {already} decoupled from all higher Fock states{, by the very definition of $H_{\rm q}$}). 
Hence, the operator $\mathcal{U}_{\rm q}^{\lambda n;1}\kett{\boldsymbol{0}}$ has its $(\lambda,n)$--support confined to the subspace spanned by $\ketbra{0}$. Reinserting the leading order term of the above Dyson substitution into Eq.~\eqref{Error before Dyson exp} results in a map, which is proportional to the operator $K_{\lambda n;2}\mathcal{U}_{\rm q}^{\lambda n;1}\kett{\boldsymbol{0}}$. Recall from the definition in Eq.~\eqref{eq: definition of K's for qb error} that $K_{\lambda n;2}$ annihilates any operator with $(\lambda,n)$--support confined to the subspace spanned by $\ketbra{0}{0}$. $K_{\lambda n;2}\mathcal{U}_{\rm q}^{\lambda n;1}\kett{\boldsymbol{0}}$ is therefore the zero operator, which in turn means that the map resulting from the leading order of the Dyson substitution is the zero map. We can therefore replace $\mathcal{U}_{\rm q}$ in Eq.~\eqref{Error before Dyson exp} with the residual of the above Dyson substitution, resulting in
\begin{equation}\label{eq: error map qb missing leading order}
    \begin{aligned}
        \mathcal{E}_{\rm q\it ;\lambda n}(t,t_0)
        =
        \IA
        &
        \nu_d\nu_e\int_{t_0}^t ds \int_{t_0}^s du 
        \\
        &\times\mathcal{U}_{\rm osc}(t,s) 
        \\
        &\times K_{\lambda n;2}^d(s) \mathcal{U}_{\rm q}(s,u)
        \\
        &\times K^e_{\lambda n;1}(u)\mathcal{U}_{\rm q}^{\lambda n;1}(u,t_0)
        \Az.
    \end{aligned}
\end{equation}
We now further subdivide the error map by performing a $K_{\lambda n;1}$--Dyson substitution of $\mathcal{U}_{\rm q}(s,u)$ in Eq.~\eqref{eq: error map qb missing leading order}
\begin{equation}
    \mathcal{E}_{\rm q\it ;\lambda n}(t,t_0)=\mathcal{E}^{(1)}_{\rm q\it ;\lambda n}(t,t_0)+\mathcal{E}^{(2)}_{\rm q\it ;\lambda n}(t,t_0),
\end{equation}
where $\mathcal{E}^{(1)}_{\rm q\it ;\lambda n}(t,t_0)$ and $\mathcal{E}^{(2)}_{\rm q\it ;\lambda n}(t,t_0)$ result from the leading order term and residual term of the Dyson substitution respectively:
\begin{equation}\label{eq: error map qb 1}
    \begin{aligned}
        \mathcal{E}^{(1)}_{\rm q\it ;\lambda n}(t,t_0)
        =
        \IA
        &
        \nu_d\nu_e\int_{t_0}^t ds \int_{t_0}^s du 
        \\
        &\times\mathcal{U}_{\rm osc}(t,s) 
        \\
        &\times K_{\lambda n;2}^d(s) \mathcal{U}_{\rm q}^{\lambda n;1}(s,u)
        \\
        &\times K^e_{\lambda n;1}(u)\mathcal{U}_{\rm q}^{\lambda  n;1}(u,t_0)
        \Az
    \end{aligned}
\end{equation}
and
\begin{equation}\label{eq: error map qb 2}
    \begin{aligned}
        \mathcal{E}^{(2)}_{\rm q\it ;n}(t,t_0)
        =
        -i\IA
        &
        \nu_d\nu_e \nu_f \int_{t_0}^t ds \int_{t_0}^s du \int_u^s dv
        \\
        &\times\mathcal{U}_{\rm osc}(t,s) 
        \\
        &\times K_{\lambda n;2}^d(s) \mathcal{U}_{\rm q}(s,v)
        \\
        &\times K_{\lambda n;1}^f(v) \mathcal{U}_{\rm q}^{\lambda n;1}(v,u)
        \\
        &\times K^e_{\lambda n;1}(u)\mathcal{U}_{\rm q}^{\lambda  n;1}(u,t_0)
        \Az.
    \end{aligned}
\end{equation}
We bound the norm of these error maps individually below.

To bound the norm of $\mathcal{E}^{(2)}_{\rm q\it ;\lambda n}(t,t_0)$, we analyze the structure of the $(\lambda,n)$--support of the operators making up the integrand in $\mathcal{E}^{(2)}_{\rm q\it ;\lambda n}(t,t_0)$. For brevity, the time dependencies are implicit in the rest of this paragraph. We first note that $\Az$ has its $(\lambda,n)$--support confined to the span of $\ketbra{0}{0}$. Since $\mathcal{U}^{\lambda  n;1}_{\rm q}$ cannot couple the vacuum state of oscillator $(\lambda,n)$ {with} higher Fock states, the operator $\kett{\mathcal{O}_1^{(2)}}\coloneq\mathcal{U}_{\rm q}^{\lambda  n;1}\Az$ also has its $(\lambda,n)$--support confined to the span of $\ketbra{0}{0}$. Recall from the definition in Eq.~\eqref{eq: definition of K's for qb error} that $K_{\lambda n;1}^e$ couples the vacuum and first excited state of oscillator $(\lambda,n)$. Thus, the operator $\kett{\mathcal{O}_2^{(2);e}}\coloneq K_{\lambda n;1}^e\kett{\mathcal{O}_1^{(2)}}$ has its $(\lambda,n)$--support confined to the subspace spanned by $\ketbra{0}{1},\ketbra{1}{0}$. We observe that $\mathcal{U}_{\rm q}^{\lambda n;1}$, $ K_{\lambda n;1}^f$, and $\mathcal{U}_{\rm q}$ at most couple the vacuum and first excited state of oscillator $(\lambda,n)$. Therefore, the operator $\kett{\mathcal{O}_5^{(2);ef}}\coloneq \mathcal{U}_{\rm q} K_{\lambda n;1}^f\mathcal{U}_{\rm q}^{\lambda n;1}\kett{\mathcal{O}_2^{(2);e}}$ has its $(\lambda,n)$--support confined to the subspace spanned by $\ketbra{0}{0},\ketbra{0}{1},\ketbra{1}{0},\ketbra{1}$. Finally, since $K_{\lambda n;2}^d$ couples the first and second Fock state of oscillator $(\lambda,n)$ and annihilates any other Fock state, we conclude that the operator to the right of $\mathcal{U}_{\rm osc}$ in Eq.~\eqref{eq: error map qb 2}, $\kett{\mathcal{O}_6^{(2);def}}\coloneq K_{\lambda n;2}^d\kett{\mathcal{O}_5^{(2);ef}}$, has its $(\lambda,n)$--support confined to the subspace spanned by $\ketbra{0}{2},\ketbra{2}{0},\ketbra{1}{2},\ketbra{2}{1}$. 

We now perform a $K_{\lambda n;2}$--Dyson substitution on $\mathcal{U}_{\rm osc}$ in Eq.~\eqref{eq: error map qb 2}. {The leading order term resulting from this consists of an integral with integrand $\llangle I_A\kett{\mathcal{O}^{(2);def}_7}$ with $\kett{\mathcal{O}^{(2);def}_7}\coloneq\mathcal{U}_{\rm osc}^{\lambda n;2}{\kett{\mathcal{O}_6^{(2);def}}}$ and $\mathcal{U}_{\rm osc}^{\lambda n;j}$ denoting the time evolution superoperator generated by $H_{\rm osc}-K_{\lambda n;j}$.} {Crucially, } $\mathcal{U}_{\rm osc}^{\lambda n;j}$ cannot couple {Fock states of oscillator $(\lambda,n)$ with index below $j$ to Fock states with  index $j$ or above.}  Hence $\mathcal{U}_{\rm osc}^{\lambda n;2}$} is unable to couple the second Fock state of oscillator $(\lambda, n)$ with the zeroth and first Fock state. {Since we found above that $\kett{\mathcal{O}_6^{(2);def}}$ has its $(\lambda,n)$--support confined to the subspace spanned by $\ketbra{0}{2},\ketbra{2}{0},\ketbra{1}{2},\ketbra{2}{1}$, we conclude that} $\kett{\mathcal{O}^{(2);def}_7}$ has its $(\lambda,n)$--support confined to the span of operators on the form $\ketbra{0}{j},\ketbra{j}{0},\ketbra{1}{j},\ketbra{j}{1}$ with $j\geq 2$.  Since the $(\lambda,n)$--support of $\kett{\mathcal{O}^{(2);def}_7}$ is fully off-diagonal, we thus conclude that $\llangle I_A\kett{\mathcal{O}^{(2);def}_7}=0$. This leaves us  with the map resulting from the residual term of the Dyson substitution
\begin{equation}\label{eq: e2qlm}
    \begin{aligned}
        \mathcal{E}^{(2)}_{\rm q\it ;\lambda n}(t,t_0)&\\
        =
        -\IA
        &
        \nu_d\nu_e \nu_f\nu_g \int_{t_0}^t ds \int_{t_0}^s du \int_u^s dv\int_{s}^t dw
        \\
        &\times\mathcal{U}_{\rm osc}(t,w) 
        \\
        &\times K_{\lambda n;2}^g(w)\mathcal{U}_{\rm osc}^{\lambda n;2}(w,s) 
        \\
        &\times K_{\lambda n;2}^d(s) \mathcal{U}_{\rm q}(s,v)
        \\
        &\times K_{\lambda n;1}^f(v) \mathcal{U}_{\rm q}^{\lambda n;1}(v,u)
        \\
        &\times K^e_{\lambda n;1}(u)\mathcal{U}^{\lambda  n;1}_{\rm q}(u,t_0)
        \Az.
    \end{aligned}
\end{equation}

To bound the right-hand side above, we now use that, for $\rho$ trace class and $X$ and $Y$ bounded operators, we have 
\begin{equation}\label{eq: tracenorm of partial trace bound}
    \norm{\Tr_A[X\rho Y]}\leq \norm{X}\norm{Y}\norm{\rho}_{\Tr}
\end{equation}
where $\Tr_A$ is a partial trace, {\it i.e.}, $\Tr_{A}[\mathcal O]=\llangle I_A|\mathcal O\rrangle$ [see Appendix \ref{app: Inequality for the trace norm of a partial trace} for the proof]. Using the above result, along with the triangle inequality and the sub-multiplicative property of the operator norm, we arrive at the following bound
\begin{equation}
\begin{aligned}
    \norm{\mathcal{E}_{\rm q; \it \lambda n}^{(2)}(t,t_0)} 
    \leq  16 \int_{t_0}^t ds& \int_{t_0}^s du \int_u^s dv \int_{s}^t dw (\gamma\Delta \xi)^2
    \\ &\times \norm{K_{\lambda n;2}(s)}\norm{K_{\lambda n;1}(u)}
    \\&\times\norm{K_{\lambda n;1}(v)}\norm{K_{\lambda n;2}(w)}.\label{eqa: bound 1}
\end{aligned}
\end{equation}
We next note that
\begin{equation}\label{eq: Bound on K_qb}
    \norm{K_{\lambda n;j}}\leq \sqrt{j \gamma\Delta \xi}\sum_\alpha \abs{g_{\rm f\it;\alpha\lambda}(t-\xi_n)}.
\end{equation}
Using this bound in Eq.~\eqref{eqa: bound 1}, and exploiting the symmetry of the resulting integral under exchange of integration variables, we find
\begin{equation}\label{eq: qubit bound 2}
\begin{aligned}
    &\norm{\mathcal{E}_{\rm q; \it \lambda n}^{(2)}(t,t_0)} 
    \leq
    \frac{32(\gamma\Delta \xi)^2}{4!}\left[\int_{t_0}^t ds \sum_\alpha\abs{g_{\rm f\it;\alpha\lambda}(s-\xi_n)}\right]^4\!\!.
\end{aligned}
\end{equation}
Here, the factor of $32$ arises from counting over the indices governing right and left multiplication and the two factors of $\sqrt{2}$ from $K_{n;2}$.

We now bound the norm of $\mathcal{E}^{(1)}_{\rm q\it ;\lambda n}(t,t_0)$. To this end, we observe in Eq.~\eqref{eq: error map qb 1} that $\kett{\mathcal{O}^{(1);e}_3}\coloneq \mathcal{U}_{\rm q}^{\lambda n;1}K^e_{\lambda n;1}\mathcal{U}_{\rm q}^{\lambda  n;1}\Az$ has its $(\lambda,n)$--support confined to the subspace spanned by $\ketbra{0}{1}, \ketbra{1}{0}$. Recall that $K_{\lambda n;2}^d$ couples the first and second Fock state and annihilates the vacuum state of oscillator $(\lambda,n)$. Therefore, the operator $\kett{\mathcal{O}^{(1);de}_4}\coloneq K_{\lambda n;2}^d\kett{\mathcal{O}^{(1);e}_3}$ has its $(\lambda,n)$--support confined to the subspace spanned by $\ketbra{0}{2},\ketbra{2}{0}$. We further note that $\kett{\mathcal{O}^{(1);de}_4}$ is non-zero only when $d=e$.
Performing a $(K_{\lambda n;1}+K_{\lambda n;2})$--Dyson substitution on $\mathcal{U}_{\rm osc}$ in Eq.~\eqref{eq: error map qb 1} results in a leading order term, $\mathcal{U}_{\rm osc}^{\lambda n;1,2}$ generated by $H_{\rm osc}-K_{\lambda n;1}-K_{\lambda n;2}$. 
$\mathcal{U}_{\rm osc}^{\lambda n;1,2}$ is unable to couple the vacuum, first, and second Fock state of oscillator $(\lambda,n)$. The operator, $\kett{\mathcal{O}^{(1);de}_5}\coloneq \mathcal{U}_{\rm osc}^{\lambda n;1,2} \kett{\mathcal{O}^{(1);de}_4}$, therefore has its $(\lambda, n)$--support confined to the span of operators on the form $\ketbra{0}{j},\ketbra{j}{0}$ with $j\geq 2$. The map resulting from the leading order term of the Dyson substitution consists of sums and integrals of terms on the form $\IA\mathcal{O}^{(1);de}_5\rrangle$. Since the $(\lambda,n)$--support of $\kett{\mathcal{O}^{(1);de}_5}$ is purely off diagonal, this map vanishes in the partial trace over the oscillators, leaving us with two separate maps resulting from the individual Hamiltonians in the residual term of the Dyson expansion
\begin{equation}
    \mathcal{E}^{(1)}_{\rm q\it ;\lambda n}(t,t_0)
    =
    \mathcal{E}^{(1a)}_{\rm q\it ;\lambda n}(t,t_0)
    +
    \mathcal{E}^{(1b)}_{\rm q\it ;\lambda n}(t,t_0)
\end{equation}
where
\begin{equation}\label{eq: error map def 1,1 qb}
    \begin{aligned}
        \mathcal{E}^{(1a)}_{\rm q\it ;\lambda n}(t,t_0)
        =
        -i\IA
        &
        \nu_d\nu_e\nu_f\int_{t_0}^t ds \int_{t_0}^s du \int_s^t dv
        \\
        &\times\mathcal{U}_{\rm osc}(t,v) 
        \\
        &\times K_{\lambda n;1}^{f}(v)\mathcal{U}_{\rm osc}^{\lambda n;1,2}(v,s) 
        \\
        &\times K_{\lambda n;2}^d(s) \mathcal{U}_{\rm q}^{\lambda n;1}(s,u)
        \\
        &\times K^e_{\lambda n;1}(u)\mathcal{U}^{\lambda  n;1}_{\rm q}(u,t_0)
        \Az 
    \end{aligned}
\end{equation}
and 
\begin{equation}\label{eq: error map def 1,2 qb}
    \begin{aligned}
        \mathcal{E}^{(1b)}_{\rm q\it ;\lambda n}(t,t_0)
        =
        -i\IA
        &
        \nu_d\nu_e \nu_f\int_{t_0}^t ds \int_{t_0}^s du \int_s^t dv
        \\
        &\times\mathcal{U}_{\rm osc}(t,v) 
        \\
        &\times K_{\lambda n;2}^{f}(v)\mathcal{U}_{\rm osc}^{\lambda n;1,2}(v,s) 
        \\
        &\times K_{\lambda n;2}^d(s) \mathcal{U}_{\rm q}^{\lambda n;1}(s,u)
        \\
        &\times K^e_{\lambda n;1}(u)\mathcal{U}^{\lambda  n;1}_{\rm q}(u,t_0)
        \Az \Bigg.
    \end{aligned}
\end{equation}
We consider the two residual terms individually, starting with $\mathcal{E}^{(1a)}_{\rm q\it ;\lambda n}$. 
To this end, we note that the operator $\kett{\mathcal{O}^{(1a);def}_6}\coloneq  K_{\lambda n;1}^{f}\kett{\mathcal{O}^{(1);de}_5}$ has its $(\lambda,n)$--support confined to the span of operators of the form $\ketbra{1}{j},\ketbra{j}{1}$ with $j\geq 2$. We further note that $\kett{\mathcal{O}^{(1a);def}_6}$ is non-zero only when $d=e\neq f$. We now perform a $K_{\lambda n; 2}$--Dyson substitution on $\mathcal{U}_{\rm osc}$ in Eq.~\eqref{eq: error map def 1,1 qb} with leading order term $\mathcal{U}_{\rm osc}^{\lambda n;2}$. The $(\lambda,n)$--support of the operator $\kett{\mathcal{O}^{(1a);def}_7}\coloneq \mathcal{U}_{\rm osc}^{\lambda n;2}\kett{\mathcal{O}^{(1a);def}_6}$ is confined to the span of $\ketbra{0}{j},\ketbra{j}{0},\ketbra{1}{j},\ketbra{j}{1}$ with $j\geq 2$. The map resulting from the leading order term of the Dyson substitution consists of sums and integrals of terms of the form $\IA \mathcal{O}^{(1a);def}_7\rrangle$. Once again this map vanishes in the partial trace, leaving us with the map resulting from the residual 
\begin{equation}
    \begin{aligned}
        \mathcal{E}^{(1a)}_{\rm q\it ;\lambda n}(t,t_0)&\\
        =
        -\IA&
        \nu_d\nu_e\nu_f\nu_g\int_{t_0}^t ds \int_{t_0}^s du \int_s^t dv\int_v^tdw
        \\
        &\times\mathcal{U}_{\rm osc}(t,w) 
        \\
        &\times K_{\lambda n;2}^g(w)\mathcal{U}_{\rm osc}^{\lambda n;2}(w,v) 
        \\
        &\times K_{\lambda n;1}^f(v)\mathcal{U}_{\rm osc}^{\lambda n;1,2}(v,s) 
        \\
        &\times K_{\lambda n;2}^d(s) \mathcal{U}_{\rm q}^{\lambda n;1}(s,u)
        \\
        &\times K^e_{\lambda n;1}(u)\mathcal{U}^{ \lambda n;1}_{\rm q}(u,t_0)
        \Az.
    \end{aligned}
\end{equation}
By the same arguments as for the bound on the norm of $\mathcal{E}_{\rm q \it ; \lambda n}^{(2)}$, we arrive at the following bound
\begin{equation}\label{eq: qubit bound 1a}
\begin{aligned}
    &\norm{\mathcal{E}^{(1a)}_{\rm q\it ;\lambda n}(t,t_0)} 
    \leq 
    \frac{8(\gamma\Delta \xi)^2}{4!}\left[\int_{t_0}^t ds \sum_\alpha\abs{g_{\rm f\it;\alpha\lambda}(s-\xi_n)}\right]^4.
\end{aligned}
\end{equation}
The pre-factor is smaller than before since we showed that only terms where $d=e\neq f$ provide non-zero contributions to the bound.

By a line of arguments similar to the ones we made below Eq.~\eqref{eq: error map def 1,2 qb} we conclude that the $(\lambda, n)$--support of $\kett{\mathcal{O}_6^{(1b);def}}\coloneq K_{\lambda n;2}^f\kett{\mathcal{O}^{(1);de}_5}$ is confined to the span of $\ketbra{0}{1},\ketbra{1}{0}$ and is moreover only nonzero when $d=e=f$. Performing a $K_{\lambda n;1}$--Dyson substitution on $\mathcal{U}_{\rm osc}$ in Eq.~\eqref{eq: error map def 1,2 qb}, we thus see that the leading order term vanishes, leaving us with
\begin{equation}
    \begin{aligned}
        \mathcal{E}^{(1b)}_{\rm q\it ;\lambda n}(t,t_0)&
        \\
        =
        -i\IA
        &
        \nu_d\nu_e\nu_f\nu_g \int_{t_0}^t ds \int_{t_0}^s du \int_s^t dv \int_v^tdw
        \\
        &\times\mathcal{U}_{\rm osc}(t,w) 
        \\
        &\times K_{\lambda n;1}^g(w)\mathcal{U}_{\rm osc}^{\lambda n;1}(w,v) 
        \\
        &\times K_{\lambda n;2}^f(v)\mathcal{U}_{\rm osc}^{\lambda n;1,2}(v,s) 
        \\
        &\times K_{\lambda n;2}^d(s) \mathcal{U}_{\rm q}^{\lambda n;1}(s,u)
        \\
        &\times K^e_{\lambda n;1}(u)\mathcal{U}^{\lambda  n;1}_{\rm q}(u,t_0)
        \Az.
    \end{aligned}
\end{equation}
By the same arguments as for the bound on the norm of $\mathcal{E}_{\rm q \it ; \lambda n}^{(2)}$, we arrive at the following bound
\begin{equation}\label{eq: qubit bound 1b}
\begin{aligned}
    \norm{\mathcal{E}^{(1b)}_{\rm q\it ;\lambda n}(t,t_0)} \leq
    \frac{8(\gamma\Delta \xi)^2}{4!}\left[\int_{t_0}^t ds \sum_\alpha\abs{g_{\rm f\it;\alpha\lambda}(s-\xi_n)}\right]^4.
\end{aligned}
\end{equation}
The small pre-factor results from the fact that only terms where $d=e=f$ provide non-zero contributions to the bound. We have now bounded the norm of all terms that make up the error map.

\subsubsection{Combining the error bounds}\label{sec: qubit combine errors}

We finally combine the bounds on the norm of the subdivisions of the error map in order to arrive at a bound on the norm of the entire error map. We start by combining the three bounds from the last section. Using the triangle inequality, we find \begin{equation}
    \begin{aligned}
        &\lVert\mathcal{E}_{\rm q\it;\lambda n}(t,t_0)\rVert\leq\\
        &\lVert{\mathcal{E}^{(2)}_{\rm q\it;\lambda n}(t,t_0)}\rVert+\lVert\mathcal{E}^{(1a)}_{\rm q\it;\lambda n}(t,t_0)\rVert +\lVert\mathcal{E}^{(1b)}_{\rm q\it;\lambda n}(t,t_0)\rVert.
    \end{aligned}
\end{equation}
Substituting in the bounds for the terms from Eqs.~\eqref{eq: qubit bound 2}, \eqref{eq: qubit bound 1a}, and \eqref{eq: qubit bound 1b} and
changing the integration variable results in the bound
\begin{equation}\label{eq: error qb n bound}
\begin{aligned}
    \norm{\mathcal{E}_{\rm q\it ;\lambda n}(t,t_0)}
    \leq & 
    2(\gamma\Delta \xi)^2\left[\int_{t_0-\xi_n}^{t-\xi_n} ds \sum_\alpha\abs{g_{\rm f\it;\alpha\lambda}(s)}\right]^4.
\end{aligned}
\end{equation}
Using the triangle inequality once more, we can bound the norm of the entire error map by summing up the bounds for the individual oscillators 
\begin{equation}\label{eq: first bound on entire qb error map}
    \norm{\mathcal{E}_{\rm q}(t,t_0)}\leq
    \sum_{\lambda, n}
    2(\gamma\Delta \xi)^2\left[\int_{t_0-\xi_n}^{t-\xi_n} ds \sum_\alpha\abs{g_{\rm f\it;\alpha\lambda}(s)}\right]^4.
\end{equation}
To simplify this bound, we use the inequality, which states that for any function $f(t)$,
\begin{equation}
    \sum_{n=-\infty}^\infty
    \int_{t_0- \xi_n}^{t-\xi_n}dt \abs{f(t)}
    \leq
    \left(\frac{T}{\Delta \xi}+1 \right)
    \int_{-\infty}^\infty dt \abs{f(t)},
    \label{eqa:f result}
\end{equation}
where $T=t-t_0$ and $\xi_n=n\Delta\xi$ (see Appendix \ref{app: bound on sums of integrals} for  proof). Extending the domain of integration to the entire real line for three of the integrals in Eq.~\eqref{eq: first bound on entire qb error map} and using the above inequality on the fourth integral allows to get rid of the sum over $n$ and provides the following bound
\begin{equation}
\begin{aligned}
    &\norm{\mathcal{E}_{\rm q}(t,t_0)}\leq 
    \\&2(\gamma \Delta \xi)^2 \left(\frac{T}{\Delta\xi}+1\right)
    \sum_{\lambda}\left[\int_{-\infty}^\infty ds \sum_\alpha\abs{g_{\rm f \it;\alpha\lambda}(s)}\right]^4.
\end{aligned}
\end{equation}
 since $\sum_n |a_n|^2 \leq (\sum_n |a_n|)^2$ for any set of numbers $\{a_n\}$, we obtain
 \begin{equation}
     \begin{aligned}
         &\norm{\mathcal{E}_{\rm q}(t,t_0)}\leq 
         \\&
         2\Delta \xi \left(T+\Delta \xi\right)
    \gamma^2\left( \sum_{\lambda}\left[\int_{-\infty}^\infty ds \sum_\alpha\abs{g_{\rm f\it;\alpha\lambda}(s)}\right]^2\right)^2.
     \end{aligned}
 \end{equation}
Minkowski's 
inequality, $\sum_\lambda \abs{\int_{-\infty}^\infty ds \sum_\alpha  f_{\alpha\lambda}(s)}^p\leq \left[\int_{-\infty}^\infty ds \sum_\alpha \left(\sum_\lambda \abs{f_{\alpha \lambda}(s)}^p\right)^{1/p}\right]^p$,
allows us to further bound the error by 
 \begin{equation}
     \begin{aligned}
         &\norm{\mathcal{E}_{\rm q}(t,t_0)}\leq 
         \\&
         2\Delta \xi \left(T+\Delta \xi\right)\gamma^2
    \left( \int_{-\infty}^\infty ds \sum_{\alpha}\sqrt{\sum_\lambda\abs{g_{\rm f\it;\alpha\lambda}(s)}^2}\right)^4.
     \end{aligned}
 \end{equation}
 Using $\abs{g_{\alpha\lambda}(s)}=\abs{g_{\lambda\alpha}(-s)}$, and relabeling the integration variable $s$ to $-s$, we obtain
  \begin{equation}
     \begin{aligned}
         &\norm{\mathcal{E}_{\rm q}(t,t_0)}\leq 
         \\&
         2\Delta \xi \left(T+\Delta \xi\right)\gamma^2
    \left( \int_{-\infty}^\infty ds \sum_{\lambda}\sqrt{\sum_\alpha\abs{g_{\rm f\it;\alpha\lambda}(s)}^2}\right)^4.
     \end{aligned}
 \end{equation}
 We recognize the integrand above as the $2,1$--norm of $\bg$. {Defining} {$\Gamma_{\rm f} \coloneq 4\gamma \left[\int_{-\infty}^\infty dt \norm{\bg_{\rm f}(t)}_{2,1} \right]^2$} analogously to $\Gamma$, we
 arrive at the following error bound
\begin{equation}
    \norm{\mathcal{E}_{\rm q}(t,t_0)}\leq 
    \frac{ \Delta \xi}{8} \left(T+\Delta \xi\right) \Gamma_{\rm f}^2,
\end{equation}

If we choose $\Delta\xi\leq \frac{\varepsilon}{\Gamma}$ and $\varepsilon\leq \Gamma T$, we get the following error scaling
\begin{equation}
    \norm{\mathcal{E}_{\rm q}(t,t_0)}=  \mathcal{O}\left( \varepsilon\Gamma_{\rm f}^2 T/\Gamma\right).
\end{equation}
Importantly, with the choice of filtering function $\tilde \varphi$ made in Eq.~\eqref{eq: phi bump function}, we have $\Gamma_{\rm f}\leq 14.41\Gamma$---see Eq.~\eqref{eq: Gamma physical vs Gamma filtered} for {the} proof. Thereby, we obtain 
\begin{equation}
    \varepsilon_2\leq 25.96 \Delta \xi \left(T+\Delta \xi\right)  \Gamma^2
\end{equation}
This tells us that a train of qubits can emulate the effect of a train of harmonic oscillators in the limit of a small discretization step.

\subsection{Error from restricting interactions to a finite number of qubits}\label{app: cut-off error}
In this section, we bound the error from truncating the  qubit register connected to the system at each point in time. Specifically, we introduce a cutoff time, $\tau_{\rm c}$, such that the $(\lambda,n)$th ancilla qubit only interacts with the system at times $t$ for which $\abs{t-\xi_n}\leq \tau_{\rm c}$. 
\begin{equation}\label{eq: H ATR}
    H_{\rm T}=H_{\rm S}+\sqrt{\gamma} \sum_\alpha S_\alpha\otimes A_{{\rm T};\alpha}(t),
\end{equation} with $A_{{\rm T};\alpha}(t)$ given in Eq.~\eqref{eq: multi channel ancilla operators}.
We define $\rho_{\rm T}(t)$ to be the reduced density matrix of the system time evolved under $H_{\rm T}$} from the the initial state $\rho(t_0)\otimes|\boldsymbol{0}\rangle\langle\boldsymbol{0}|_{\mathcal A}$. Similar to before, our goal is to quantify the error by considering the trace distance between the time evolution with and without a time cutoff 
\begin{equation}
    \varepsilon_3\coloneq \norm{\rho_{\rm T}(t)-\rho_{\rm q}(t)}_\Tr.
\end{equation}
In this appendix, we show that this error can be upper bounded by
\begin{equation}
    \varepsilon_3\leq \Gamma_{\rm f}(T+\Delta\xi) \frac{\tau_{\rm f}}{\tau_{\rm c}},
\end{equation}
where $\tau_{\rm f}$ and $\Gamma_{\rm f}$ are related to the characteristic {interaction }rate and correlation timescale of the bath. 

Similarly to the previous subsection, we arrive at the bound above by defining a quantum channel for the time truncated dynamics, $\mathcal{C}_{\rm T}$, such that $\rho_{\rm T}(t)=\mathcal{C}_{\rm T}(t,t_0)[\rho_0]$. As in the last appendix, we define the error map as the difference between the two quantum channels and arrive at the bound by bounding the {super}operator norm of the following error map
\begin{equation}
    \mathcal{E}_{\rm T}(t,t_0) \coloneq 
    \mathcal{C}_{\rm T}(t,t_0)[\rho(t_0)]
    -
    \mathcal{C}_{\rm q}(t,t_0)[\rho(t_0)],
\end{equation}
where $\mathcal{C}_{\rm q}(t,t_0)[\rho(t_0)]$ is the quantum channel for the non-truncated evolution. The rest of this appendix is dedicated to constructing the error map in subsection \ref{sec: time cut error map} and subsequently splitting and bounding the norm of the terms in the error map in subsection \ref{sec: bounding error map time cut}.

\subsubsection{Constructing the error map}\label{sec: time cut error map}

To obtain an expression for the error map $\mathcal{E}_{\rm T}$, we define $\chi_n(t)$ to be the indicator function of an interval of width $2\tau_{\rm c}$ centered  at $\xi_n$:
\begin{equation}
    \chi_n(t)\coloneq \begin{cases}
        1, &|t-\xi_n|\leq \tau_{\rm c} 
        \\
        0, &|t-\xi_n|>\tau_{\rm c}
    \end{cases}.
\end{equation}
We can then construct a Hamiltonian which is equivalent to $H_{\rm T}$ by explicitly subtracting interactions with ancillary qubits, which fall outside outside of the cut-off interval
\begin{equation}
    H_{\rm T}(t) = H_{\rm q}(t)-\sum_{\lambda}\sum_{n=-\infty}^\infty K_{\lambda n}(t)\left[1-\chi_n(t)\right],
\end{equation}
where
\begin{equation}
\begin{aligned}
    K_{\lambda n}(t)
    =
    \sum_\alpha &S_\alpha 
    \sqrt{\gamma\Delta\xi}
    &\left(g_{\rm f\it;\alpha\lambda}^*(t-\xi_n)\sigma^+_{\lambda n} + h.c.\right)
\end{aligned}
\end{equation}
is the time dependent coupling to qubit $(\lambda,n)$. Note that $K_{\lambda n}$ is simply a restriction of $K_{\lambda n;1}$ from the previous appendix to the subspace spanned by the $0$th and $1$st Fock states, which is equivalent to the Hilbert space of the ancilla  qubits. We can now construct the quantum channel for the time truncated dynamics $\mathcal{C}_{\rm T}(t,t_0)=\IA \mathcal{U}_{\rm T}(t,t_0)\Az$, where $\mathcal{U}_{\rm T}$ is the time evolution superoperator generated by $H_{\rm T}$. Recall that the quantum channel for the qubit evolution is $\mathcal{C}_{\rm q}(t,t_0)=\IA \mathcal{U}_{\rm q}(t,t_0)\Az$, where $\mathcal{U}_{\rm q}$ is the time evolution superoperator generated by $H_{\rm q}$. The error map can now be found as the difference between the two channels. Similarly to before, we relate the two time evolution superoperators using a Dyson substitution to obtain a compact expression for the error map [see paragraph above Eq.~\eqref{eq: Dyson split expansion in text 1}]. Specifically, we perform a $\sum_{\lambda}\sum_{n=-\infty}^\infty K_{\lambda n}(t)\left[1-\chi_n(t)\right]$--Dyson substitution on $\mathcal{U}_{\rm q}$, allowing us to write 
\begin{equation}
\begin{aligned}
    &\mathcal{U}_{\rm q}(t,t_0)
    =
    \mathcal{U}_{\rm T}(t,t_0)
    -i\sum_\lambda\sum_{n=-\infty}^\infty\Bigg[
    \\
    &\nu_d
    \int_{t_0}^tds
    \mathcal{U}_{\rm q}(t,s)K^d_{\lambda n}(s)\left[1-\chi_n(t)\right]\mathcal{U}_{\rm T}(s,t_0)\Bigg].
\end{aligned}
\end{equation}
This allows us to identify
\begin{equation}
    \begin{aligned}
        &\mathcal{E}_{\rm T}(t,t_0)=\\
    &\sum_{\lambda, n}i\nu_d
    \int_{t_0}^tds
    \IA\mathcal{U}_{\rm q}(t,s)K^d_{\lambda n}(s)\left[1-\chi_n(s)\right]\mathcal{U}_{\rm T}(s,t_0)\Az.
    \end{aligned}
\end{equation}
We now set out to expand and bound the error map by methods similar to those used in the previous subsection.

\subsubsection{Expanding and bounding the error map}\label{sec: bounding error map time cut}

In this section, we expand the error map into a sum of terms that can be conveniently bounded individually. We first define the error map associated with each individual ancilla qubit $(\lambda,n)$ as
\begin{equation}\label{eq: Error map time cut-off}
    \begin{aligned}
        &\mathcal{E}_{\rm T \it;\lambda n}(t,t_0)
        \\=&
        i\nu_d
    \int_{t_0}^tds
    \IA\mathcal{U}_{\rm q}(t,s)K^d_{\lambda n}(s)\left[1-\chi_n(s)\right]\mathcal{U}_{\rm T}(s,t_0)\Az.
    \end{aligned}
\end{equation}
The total error map is the sum over the error maps above.
We now use Dyson substitutions on the unitary superoperators above to express $\mathcal{E}_{\rm T \it;\lambda n}(t,t_0)$ in such a way that it can be conveniently bounded. As in the previous subsection, our strategy is to choose Dyson substitutions where the leading order terms vanish based on considerations about $(\lambda,n)$--support (see the second paragraph under Eq.~\eqref{Error before Dyson exp} for a definition of this term{inology}).
To aid this analysis, we introduce the following notation: we let $\mathcal{U}_{\rm q}^{\lambda n}(t)$ denote the time-evolution superoperator generated by $H_{\rm q}(t)-K_{\lambda n}(t)$ and let $\mathcal{U}_{\rm T}^{\lambda n}(t)$ denote the time-evolution superoperator generated by $H_{\rm T}(t)-K_{\lambda n}(t)\chi_n(t)$. Crucially, both  these superoperators act trivially on the operator space of ancilla qubit $(\lambda,n)$, implying that they do not modify the $(\lambda,n)$--support of the operator they act on.

Having laid out the strategy and notation, we are ready to expand and bound  $\mathcal{E}_{\rm T \it;\lambda n}(t,t_0)$. As a first step, we perform a $K_{\lambda n}\chi_n$--Dyson substitution on $\mathcal{U}_{\rm T}$ in Eq.~\eqref{eq: Error map time cut-off}, resulting in
\begin{equation}
    \begin{aligned}
        &\mathcal{E}_{\rm T \it ;\lambda n}
        =
        \mathcal{E}^{(1)}_{\rm T \it ;\lambda n}
        +
        \mathcal{E}^{(2)}_{\rm T \it ;\lambda n},
    \end{aligned}
\end{equation}
where the two terms above arise from the  leading order and residual terms of the Dyson substitution, respectively:
\begin{equation}\label{eq: error map (1) time cut-off}
    \begin{aligned}
        &\mathcal{E}^{(1)}_{\rm T \it ;\lambda n}(t,t_0)=
        \\&
        i\IA 
        \nu_d\int_{t_0}^t ds
        \mathcal{U}_{\rm q}(t,s)K^d_{\lambda n}(s)\left[1-\chi_n(s)\right]\mathcal{U}^{\lambda n}_{\rm T}(s,t_0)
        \Az
    \end{aligned}
    \end{equation}
and
\begin{equation}
    \begin{aligned}
        \mathcal{E}^{(2)}_{\rm T \it ;\lambda n}(t,t_0)
        =
        -\IA&
        \nu_d \nu_e \int_{t_0}^t ds \int_{t_0}^s du 
        \\
        &\times
        \mathcal{U}_{\rm q}(t,s)
        \\
        &\times
        K^d_{\lambda n}(s) \left[1-\chi_n(s)\right]
        \mathcal{U}_{\rm T}(s,u) 
        \\
        &\times
        K^e_{\lambda n}(u)
        \chi_n(u)
        \mathcal{U}^{\lambda n}_{\rm T}(u,t_0)
        \Az.
    \end{aligned}
\end{equation}

We bound the error of the maps individually, starting with $\mathcal{E}^{(2)}_{\rm T \it ;\lambda n}$. 
In Eq.~\eqref{eq: tracenorm of partial trace bound} [or \eqref{Bound on trace norm of generic error term}] we saw that $\norm{\Tr_A[X\rho Y]}\leq \norm{X}\norm{Y}\norm{\rho}_{\Tr}$, where $\Tr_A$ denotes the partial trace over the ancilla qubits. Using this result together with the triangle inequality and the sub-multiplicativity of the operator norm, we obtain
\begin{equation}
\begin{aligned}
    \norm{\mathcal{E}^{(2)}_{\rm T \it ;\lambda n}(t,t_0)}
    \leq &
    4\gamma\Delta\xi \int_{t_0}^t ds \int_{t_0}^s du
    \left[1-\chi_n(s)\right]\chi_n(u)
    \\
    \times&
    \norm{K_{\lambda n}(s)}
    \norm{K_{\lambda n}(u)},
\end{aligned}
\end{equation}
where the factor of $4$ comes from counting over the left and right multiplication indices, $d$ and $e$. Recalling that $K_{\lambda n}$ is the restriction of $K_{\lambda n,1}$ to the two lowest Fock states of the oscillator, we use the bound on $K_{\lambda n;1}$ from Eq.~\eqref{eq: Bound on K_qb} and the fact that $\chi_n\leq 1$ to obtain
\begin{equation}\label{eq: bound on error map (2) time cut-off}
\begin{aligned}
    \norm{\mathcal{E}^{(2)}_{\rm T \it ;\lambda n}(t,t_0)}
    \leq &
    4\gamma\Delta\xi \int_{t_0}^t ds \int_{t_0}^s du
    \left[1-\chi_n(s)\right]
    \\
    \times&
    \sum_{\alpha,\alpha'}\abs{g_{\rm f\it;\alpha\lambda }(s-\xi_n)}
    \abs{g_{\rm f\it;\alpha'\lambda}(u-\xi_n)},
\end{aligned}
\end{equation}

We now bound the term $\mathcal{E}^{(1)}_{\rm T \it ;\lambda n}$ in Eq.~\eqref{eq: error map (1) time cut-off}. We first note that the operator to the right of $\mathcal{U}_{\rm q}$ in Eq.~\eqref{eq: error map (1) time cut-off} has its $(\lambda,n)$--support confined to the subspace spanned by $\ketbra{0}{1},\ketbra{1}{0}$. Performing a $K_{\lambda n}$--Dyson substitution on $\mathcal{U}_{\rm q}$ in Eq.~\eqref{eq: error map (1) time cut-off} results in a leading order term $\mathcal{U}^{\rm q}_{\lambda n}$, which preserves the confinement of the $(\lambda,n)$--support. Since this support is purely off-diagonal, the map resulting from the leading order term of the Dyson substitution vanishes in the partial trace over the ancilla qubits. This leaves us with the following map resulting from the residual term of the Dyson substitution
\begin{equation}
    \begin{aligned}
        \mathcal{E}^{(1)}_{\rm T \it ;\lambda n}(t,t_0)
        =
        i\IA&
        \nu_d\nu_e\int_{t_0}^t ds\int_s^t du
        \\
        &\times 
        \mathcal{U}_{\rm q}(t,u)
        \\
        &\times
        K^e_{\lambda n}(u)
        \mathcal{U}_{\rm q}^{\lambda n}(u,s)
        \\
        &\times
        K^d_{\lambda n}\left[1-\chi_n(s)\right](s)\mathcal{U}_{\lambda n}^{\rm T}(s,t_0)
        \Az.
    \end{aligned}
\end{equation}
By the same arguments used above for bounding the norm of $\mathcal{E}^{(2)}_{\rm T \it ;\lambda n}$, we obtain the bound
\begin{equation}\label{eq: bound on error map (1) time cut-off}
\begin{aligned}
    \norm{\mathcal{E}^{(1)}_{\rm T \it ;\lambda n}(t,t_0)}
    \leq &
    4\gamma\Delta\xi \int_{t_0}^t ds \int_s^t du
    \left[1-\chi_n(s)\right]
    \\&
    \times    \sum_{\alpha,\alpha'}\abs{g_{\rm f\it;\alpha\lambda }(s-\xi_n)}
    \abs{g_{\rm f\it;\alpha'\lambda}(u-\xi_n)}.
\end{aligned}
\end{equation}

We next combine the two bounds in Eqs.~\eqref{eq: bound on error map (2) time cut-off} and \eqref{eq: bound on error map (1) time cut-off} to obtain a bound for $\norm{\mathcal{E}_{\rm T \it ;\lambda n}(t,t_0)}$. Using the triangle inequality and the fact that the integration domains in Eq.~\eqref{eq: bound on error map (2) time cut-off} and Eq.~\eqref{eq: bound on error map (1) time cut-off} line up, we find the following bound
\begin{equation}
\begin{aligned}
    \norm{\mathcal{E}_{\rm T \it ;\lambda n}(t,t_0)}
    \leq &
    4\gamma\Delta\xi \int_{t_0}^t ds \int_{t_0}^t du
    \left[1-\chi_n(s)\right]
    \\&    \times\sum_{\alpha,\alpha'}\abs{g_{\rm f\it;\alpha\lambda }(s-\xi_n)}
    \abs{g_{\rm f\it;\alpha'\lambda}(u-\xi_n)}.
\end{aligned}
\end{equation}

We finally obtain a bound for $ \norm{\mathcal{E}_{\rm T }(t,t_0)}$ by summing the above bound over all ancilla qubits
\begin{equation}\label{eq: error time cut double integral}
\begin{aligned}
     \norm{\mathcal{E}_{\rm T }(t,t_0)}
     \leq&\sum_{\lambda,n}4\gamma\Delta\xi \int_{t_0-\xi_n}^{t-\xi_n} ds \int_{t_0-\xi_n}^{t-\xi_n} du \left[1-\chi_0(s)\right]
    \\
    &
     \times
    \sum_{\alpha,\alpha'}
    \abs{g_{\rm f\it;\alpha\lambda }(s)}
    \abs{g_{\rm f\it;\alpha'\lambda}(u)}.
\end{aligned}
\end{equation}
By extending the integration over $s$ to the entire real line and then applying the same argument used below Eq.~\eqref{eqa:f result}, we arrive at the bound
\begin{equation}\label{eq: truncation 1}
\begin{split}
    \norm{\mathcal{E}_{\rm T }(t,t_0)}\leq& 4\gamma(T+\Delta\xi)\int_{-\infty}^\infty du\int_{-\infty}^\infty ds\left[1-\chi_0(s)\right]
    \\
    &  \times \sum_{\lambda,\alpha,\alpha'}   \abs{g_{\rm f \it;\alpha \lambda}(s)}\abs{g_{\rm f \it;\alpha' \lambda}(u)},
\end{split}
\end{equation}
where $T=t-t_0$ is the full simulation time. 
Using the inequality [see Appendix~\ref{app: Matrix g sum bound} for the proof]
\begin{equation}\label{eq: matrix g sum bound}
    \sum_{\lambda,\alpha,\alpha'}\abs{g_{\alpha\lambda}(s)}\abs{g_{\alpha'\lambda}(u)}
    \leq \norm{\bg(-s)}_{2,1}\norm{\bg(-u)}_{2,1},
\end{equation}
we obtain, relabeling integration variables
\begin{equation}
\begin{split}
    \norm{\mathcal{E}_{\rm T }(t,t_0)}\leq& 4\gamma(T+\Delta\xi)\int_{-\infty}^\infty du\int_{-\infty}^\infty ds\left[1-\chi_0(s)\right]
    \\
    \times&
    \norm{\bg_{\rm f} (s)}_{2,1}\norm{\bg_{\rm f}(u)}_{2,1}.
\end{split}
\end{equation}
Recalling the definition of  $\Gamma_{\rm f}$ [see also below Eq.~\eqref{eq: def gamma f and tau f}], 
we find 
\begin{equation}\label{eq: General time cut-off error}
\begin{split}
    \varepsilon_3\leq
    &
    2(T+\Delta\xi) \sqrt{\gamma\Gamma_{\rm f}}
    \\
    &\times\int_{-\infty}^\infty ds \left[1-\chi_0(s)\right]\norm{\bg_{\rm f}(s)}_{2,1}.
\end{split}
\end{equation}
Since $1-\chi_0(t)$ vanishes for $|t|\leq \tau_{\rm c}$, we see that the norm of the error map is small if $\norm{\bg_{\rm f}}_{2,1}$ has negligible weight in the regions where $\abs{t}\geq\tau_{\rm c}$. The above constitutes our best bound for $\varepsilon_3$.

While the above bound is generally valid, we can obtain a simple expression for the right-hand-side in terms of the bath correlation time $\tau_{\rm f}$ and $\Gamma_{\rm f}$. Since the integrand in the above integrand is nonzero only when $\frac{\abs{s}}{\tau_{\rm c}}\geq 1$, we find
\begin{equation}
\begin{aligned}
    \norm{\mathcal{E}_{\rm T }(t,t_0)}\leq & 2\sqrt{\gamma\Gamma_{\rm f}}(T+\Delta\xi) \frac{1}{\tau_{\rm c}}\int_{-\infty}^\infty ds \abs{s} \norm{\bg_{\rm f}(s)}_{2,1}.
\end{aligned}
\end{equation}
Defining $\tau_{\rm f}\coloneq\int_{-\infty}^\infty ds \abs{s} \norm{\bg_{\rm f}(s)}_{2,1}/\int_{-\infty}^\infty ds  \norm{\bg_{\rm f}(s)}_{2,1}$ analogously to $\tau$, we find
\begin{equation}
\begin{aligned}
    \norm{\mathcal{E}_{\rm T }(t,t_0)}\leq 
    \Gamma_{\rm f}(T+\Delta\xi) \frac{\tau_{\rm f}}{\tau_{\rm c}}.
\end{aligned}
\end{equation}
In Appendix~\ref{app:Gamma tau bound proof} 
we show that $\Gamma_{\rm f}\tau_{\rm f}\leq   55.16\Gamma\tau$ with the choice of filtering function $\varphi$ made in Eq.~\eqref{eq: phi bump function}. Thus, we obtain the following bound on the error
\begin{equation}
\begin{aligned}
    \varepsilon_3\leq 
    55.16\Gamma\frac{\tau}{\tau_{\rm c}}(T+\Delta\xi).
\end{aligned}
\end{equation}
This tells us that we can emulate the effects of the bath to precision $\mathcal{O}(\varepsilon\Gamma T)$  by choosing $\tau_{\rm c} \sim \tau/\varepsilon$.

\subsection{Trotter error}\label{app: Trotter error}
In this section, we bound the error from Trotterizing the evolution of the truncated Hamiltonian $
    H_{\rm T}=H_{\rm S}+\sqrt{\gamma} \sum_\alpha S_\alpha\otimes A_{{\rm T};\alpha}(t)$ first defined in Eq.~\eqref{eq: H ATR} with $A_{{\rm T};\alpha}(t)$ given in Eq.~\eqref{eq: multi channel ancilla operators} and $\Delta t$ a finite trotter step. Such a Trotterized evolution is directly implemented in the ATA; thereby this bound completes the establishment of an error bound for the ATA, see e.g. Fig.~\ref{Fig: error flowchart}. 
 For simplicity, we focus on the case where $H_{\rm S}$ is  time-independent.  Trotterization techniques for time-ordered exponentials are analogous to time-independent Trotterization, and both first and second-order formulas exist, see, e.g., Ref.~\cite{Huyghebaert_1990}. The case of time-dependent $H_{\rm S}(t)$ can thus be analyzed with similar techniques as below, but is more cumbersome; we therefore leave such an investigation for future studies. 

Our construction of the ATA  employs 
a second-order Trotterization scheme. Denoting $t_\ell\coloneq t_0+\ell\DT$ {and defining $M\coloneq\operatorname{round}(T/\Delta t)$ as the total number of Trotter steps}, this scheme approximates the time-evolution operator generated by $H_{\rm T}$,
\begin{equation}\label{eq: U_T def}
    U_{\rm T}(t_M,t_0)=\mathcal{T}e^{-i\int_{t_0}^{t_M} dt' H_{\rm T}(t')},
\end{equation}
with the Trotterized system-ancilla ATR evolution operator given by $\prod_{\ell=1}^{M}G_{\ell}$, where \begin{equation}\label{eq: G_ell def}
    G_\ell=e^{-\frac{iH_{\rm S}\DT}{2}}\left[\mathcal{T}e^{-i\int_{t_{\ell-1}}^{t_\ell}dt'\sqrt{\gamma} \sum_{\alpha}S_{\alpha}\otimes A_{{\rm T};\alpha}(t')}\right] e^{-\frac{iH_{\rm S}\DT}{2}},
\end{equation}
as in Eq.~\eqref{eq: Trotter main} in the main text. 
Letting and $\rho_{\rm T}(t_M)$ {denote} the system reduced density matrix generated by the system-ancilla Hamiltonian $H_{\rm T}$ and $\rho_{\rm Trotter}(t_M)$ the system reduced density matrix generated by the Trotterized system-ancilla time-evolution operator, we define the Trotter error:
\begin{equation}
    \varepsilon_4\coloneq \lVert\rho_{\rm T}(t_M)-\rho_{\rm Trotter}(t_M)\rVert_{\Tr}.
\end{equation}
In Appendix~\ref{app: Trotter bound 1} we prove that the sum of the distances between the Trotterized and original time-evolution increments provides an upper bound for the deviation. With the notation $U_\ell\coloneq U_{\rm T}(t_\ell,t_{\ell-1})$, we have the bound \begin{equation}\label{eq: accumulated Trotter bound}
    \norm{\rho_{\rm T}(t_M)-\rho_{\rm Trotter}(t_M)}_{\Tr}\leq 2\sum_{\ell=1}^{M}\lVert U_\ell-G_\ell \rVert.
\end{equation}
 Therefore,
\begin{equation}\label{eq: Trotter bound 1}
    \begin{aligned}
        \varepsilon_4\leq 2\sum_{\ell=1}^{M}\lVert U_\ell - G_\ell \rVert.
    \end{aligned}
\end{equation}

We next bound the terms on the right-hand-side of \eqref{eq: Trotter bound 1}. To this end, for brevity, we introduce the notation $X = H_{\rm S}$ and $Y(t) =\sqrt{\gamma}\sum_{\alpha}S_{\alpha}\otimes A_{{\rm T};\alpha}(t)$. The evolution we wish to Trotterize is of the form $\mathcal{T}e^{-i\int_{s}^{s+z}X(u)+Y(u)du}$; our goal hence is to bound the distance of this operator to  $e^{-\frac{iXz}{2}}\left[\mathcal{T}e^{-i\int_{s}^{s+z}Y(u)du}\right]e^{-\frac{iXz}{2}}$. {{Here} we included an argument, $u$, in $X(u)$, whenever $X$ is part of the time-ordered exponential argument despite the fact that we take $X$ to be time independent, to signify that $X$ \emph{cannot} be pulled outside the integral in the time-ordered exponential.}
Such a bound can be conveniently established using standard Trotterization techniques~\cite{Huyghebaert_1990} by introducing the operator 
\begin{equation}
\begin{aligned}
    &F(z,s)\coloneq\\
   &e^{\frac{iXz}{2}}\left(\mathcal{T}e^{-i\int_{s}^{s+z}Y(u)du}\right)^{-1}\!\!e^{\frac{iXz}{2}}\mathcal{T}e^{-i\int_{s}^{s+z}X(u)+Y(u)du}-1,
\end{aligned}
\end{equation}
which compares the {original} time-evolution {increment} followed by an inverse Trotter{ized increment} to the identity.
{To relate $F(\Delta t,z)$ to the operator difference we wish to bound, we use} $F(0,s)=0$ to write $F(\DT,s)=\int_0^{\DT}dz \frac{d}{dz}F(z,s)$. Using the integral triangle inequality,   we  find
\begin{equation}
    \begin{aligned}
        \lVert F(\Delta t,s)\rVert \leq\int_{0}^{\DT} dz \lVert\frac{d}{dz}F(z,s)\rVert.
    \end{aligned}
\end{equation}
{Exploiting the} invariance of the operator norm under multiplication by unitary operators to multiply both sides by $e^{\frac{-iX\DT}{2}}\left(\mathcal{T}e^{-i\int_{s}^{s+\DT}Y(u)du}\right)e^{\frac{-iX\DT}{2}}$, we thus find
\begin{equation}\label{eq: Trotter error bound 3}
    \begin{aligned}
        \lVert &e^{\frac{-iX\DT}{2}}\left(\mathcal{T}e^{-i\int_{s}^{s+\DT}Y(u)du}\right)e^{\frac{-iX\DT}{2}}
        \\&
        -\mathcal{T}e^{-i\int_{s}^{s+\DT}X(u)+Y(u)du}\rVert \leq\int_{0}^{\DT} dz \lVert\frac{d}{dz}F(z,s)\rVert.
    \end{aligned}
\end{equation}
We note that the a sum over $\ell$ on the left-hand-side with $s=t_{\ell-1}$ gives an upper bound on the Trotter error by Eqs.~\eqref{eq: U_T def}, \eqref{eq: G_ell def}, and \eqref{eq: Trotter bound 1}.

As the next step, we bound $\lVert\frac{d}{dz}F(z,s)\rVert $. Through direct evaluation and use the triangle inequality, one can verify  
\begin{equation}\label{eq: Trotter bound 2}
    \begin{aligned}
        &\lVert\frac{d}{dz}F(z,s)\rVert\\
        &\qquad\leq\int_0^z dw\int_0^w du \Bigg[\frac{1}{2}\norm{[[X,Y(s+z)],Y(s+w)]}\\
        &\qquad\qquad\qquad+\frac{1}{4}\left[ \norm{[X,[X,Y(z+s)]]}\right]\\
        &\qquad\qquad\qquad+\frac{1}{2} \int_{w}^{z} dv\norm{[[X,Y'(v+s)],Y(s+u)]}\\&\qquad\qquad\qquad+\frac{1}{2}\norm{[X,Y'(w+s)]}\Bigg].
    \end{aligned}
\end{equation}
Notice that integrating this equation leads to the well-known Suzuki-Trotter error bound when $Y(t)$ is constant \cite{Suzuki91}.
We now analyze the terms of integrand in \eqref{eq: Trotter bound 2} individually. For brevity, we 
focus on the first term here. The remaining three terms can be bounded by similar arguments.

To bound the first term in Eq.~\eqref{eq: Trotter bound 2}, we first use our definitions of $X$ and $Y$ to obtain
\begin{equation}
\begin{aligned}
     &\norm{[[X,Y(t)],Y(s)]}=\\
        &\gamma\sum_{\alpha,\beta}\norm{[[H_{\rm S},S_\alpha]\otimes A_{{\rm T};\alpha}(t),S_\beta\otimes A_{{\rm T};\beta}(s)]}.
\end{aligned}
\end{equation}
Hence,
\begin{equation}
\begin{aligned}
         &\norm{[[X,Y(s+z)],Y(s+w)]}\leq\\& 2\gamma\sum_{\alpha,\beta}\norm{[H_{\rm S},S_\alpha]}\norm{ A_{{\rm T};\alpha}(s+z)}\norm{S_\beta}\norm{ A_{{\rm T};\beta}(s+w)]}.
\end{aligned}
\end{equation}
{Recall from the definition of $A_{{\rm T};\beta}(t)$ [Eq.~\eqref{eq: multi channel ancilla operators}] that}   $\norm{ A_{{\rm T};\beta}(t)]}\leq \frac{1}{\sqrt{\Delta\xi}}\sum_{n=-\infty}^{\infty}\sum_{\lambda}\abs{g_{{\rm f};\beta\lambda}(t-\xi_n)}\Delta\xi$. Thus, using the inequality $f(y)\leq \frac{1}{\delta}\int_{x_0}^{x_0+\delta} dx f(x)+\int_{x_0}^{x_0+\delta} dx \abs{f'(x)}$ for all $x_0\leq y\leq x_0+\delta$ and $f$ absolutely continuous, we find
\begin{equation}
\begin{aligned}
        \norm{ A_{{\rm T};\beta}(t)]}\leq \frac{1}{\sqrt{\Delta\xi}}\int_{-\infty}^{\infty}ds'\Big[\sum_{\lambda}&\abs{g_{{\rm f};\beta\lambda}(t-s')}\\&+\Delta\xi \abs{g_{{\rm f};\beta\lambda}'(t-s')}\Big]
\end{aligned}
\end{equation}
Defining $\mathbf{s}=(s_1,s_2,...,s_{N_C})$, $\Dot{\mathbf{s}}=(\Dot{s}_1,\Dot{s}_2,...,\Dot{s}_{N_C})$,  with 
\begin{equation}
    \begin{split}
        s_\alpha&= \norm{S_\alpha}=1,\\
        \Dot{s}_\alpha&= \norm{[H_{\rm S},S_\alpha]},\\
    \end{split}
\end{equation}
and using the Cauchy-Schwarz inequality, we thus find
\begin{equation}
\begin{aligned}
        \sum_{\beta}\norm{S_\beta}\norm{ A_{{\rm T};\beta}(t)]}\leq \frac{1}{\sqrt{\Delta\xi}}\norm{\mathbf{s}}_2&\int_{-\infty}^{\infty}\!\!\! ds' \Big[\norm{\bg_{\rm f}(t-s')}_{2,1}\\&+\Delta\xi \norm{\bg_{\rm f}'(t-s')}_{2,1}\Big],
\end{aligned}
\end{equation}
with $\norm{\cdot}_2$ denoting the standard Euclidean norm. 
Defining $C_{\rm f}\coloneq \frac{\sqrt{\gamma} \left(\int_{-\infty}^{\infty}ds\norm{\bg'_{\rm f}(s)}_{2,1}\right)}{\Gamma^{3/2}}$, it follows from the definition of $\Gamma_{\rm f}$ that
\begin{equation}
    \sqrt{\gamma}\sum_{\beta}\norm{S_\beta}\norm{ A_{{\rm T};\beta}(t)]}\leq \norm{\mathbf{s}}_2\frac{1}{\sqrt{\Delta\xi}}\left(\!\frac{\sqrt{\Gamma_{\rm f}}}{2}+\Delta \xi\Gamma^{3/2}C_{\rm f}\right)\!.
\end{equation}
{By a similar line of arguments} we find 
\begin{equation}
  \sqrt{\gamma}\sum_{\alpha}\norm{[H_{\rm S},\!S_\alpha]}\norm{ A_{{\rm T};\alpha}(t)]}\!\!\leq\!\! \sqrt{\gamma \Delta\xi}\norm{\Dot{\mathbf{s}}}\!\!\!\!\sum_{n=-\infty}^{\infty}\!\!\!\norm{\bg_{\rm f}(t-\xi_n\!)}_{2,1}.
\end{equation}
\begin{widetext}
Thus  it now follows that \begin{equation}
    \norm{[[X,Y(z+s)],Y(s+w)]}\leq 2\norm{\Dot{\mathbf{s}}}_2\norm{\mathbf{s}}_2\left(\frac{\sqrt{\Gamma_{\rm f}}}{2}+\Delta \xi\Gamma^{3/2}C_{\rm f}\right)\sqrt{\gamma}\sum_{n=-\infty}^{\infty}\norm{\bg_{\rm f}(z+s-\xi_n)}_{2,1}.
\end{equation}
Using {this,} \eqref{eq: Trotter bound 2}, \eqref{eq: Trotter bound 1}, and \eqref{eq: Trotter error bound 3} we find, for the first error term, 
\begin{equation}
    \begin{aligned}
        &\sum_{\ell=1}^{M}\int_0^{\Delta t} dz \int_0^z dw\int_0^w du \frac{1}{2}\norm{[[X,Y(t_\ell+z)],Y(t_\ell+w)]}\\&\quad \leq \frac{(\Delta t)^2}{2}\norm{\Dot{\mathbf{s}}}_2\norm{\mathbf{s}}_2\left(\frac{\sqrt{\Gamma_{\rm f}}}{2}+\Delta \xi\Gamma^{3/2}C_{\rm f}\right)\sqrt{\gamma}\sum_{n=-\infty}^{\infty}\sum_{\ell=1}^{M}\int_{0}^{\Delta t} dz\norm{\bg_{\rm f}(t_{\ell-1}+z-\xi_n)}_{2,1}.
    \end{aligned}
\end{equation}
We next use that $\sum_{\ell=1}^{M}\int_{0}^{\Delta t} dz\norm{\bg_{\rm f}(t_{\ell-1}+z-\xi_n)}_{2,1}=\int_{t_0}^{M\Delta t} dz\norm{\bg_{\rm f}(t_{\ell-1}+z-\xi_n)}_{2,1}$ and  $    \sum_{n=-\infty}^\infty
    \int_{t_0- \xi_n}^{t-\xi_n}dt f(t)
    \leq
    \left(\frac{T}{\Delta \xi}+1 \right)
    \int_{-\infty}^\infty dt f(t),$ for any measurable, non-negative function $f$; 
see Appendix \ref{app: bound on sums of integrals}, Eq. \eqref{eq: bound on sum of integrals}, where a proof is provided. These facts together imply 
\begin{equation}
    \begin{aligned}\label{eq: Trotter error 1}
        &\sum_{\ell=1}^{M}\int_0^{\Delta t} dz \int_0^z dw\int_0^w du \frac{1}{2}\norm{[[X,Y(t_\ell+z)],Y(t_\ell+w)]} \leq \frac{(\Delta t)^2}{2\Delta\xi}\left(T+\Delta\xi\right)\norm{\Dot{\mathbf{s}}}_2\norm{\mathbf{s}}_2\left(\frac{\sqrt{\Gamma_{\rm f}}}{2}+\Delta \xi\Gamma^{3/2}C_{\rm f}\right)\frac{\sqrt{\Gamma_{\rm f}}}{2}.
    \end{aligned}
\end{equation}
Analyzing the remaining terms of \eqref{eq: Trotter bound 2} in a similar way gives us:
    \begin{align}\label{eq: Trotter error 2}
        \sum_{\ell=1}^{M}\int_0^{\Delta t}\!\! dz \int_0^z\!\! dw\int_0^w\!\! du \frac{1}{4}\norm{[X,[X,Y(z+t_\ell)]]}&\leq  \frac{(\Delta t)^2}{8\Delta\xi}\left(T+\Delta \xi\right)\norm{\Ddot{\mathbf{s}}}_2 \sqrt{\Delta\xi}\frac{\sqrt{\Gamma_{\rm f}}}{2}, \\ \label{eq: Trotter error 3}
        \sum_{\ell=1}^{M}\int_0^{\Delta t}\!\! dz \int_0^z\!\! dw\int_0^w\!\! du\frac{1}{2}\norm{[X,Y'(w+t_\ell)]}&\leq \frac{(\Delta t)^2}{4\Delta\xi}\left(T+\Delta \xi\right)\norm{\Dot{\mathbf{s}}}_2 \sqrt{\Delta\xi} \Gamma^{3/2} C_{\rm f},
    \end{align}
and 
\begin{equation}
    \begin{aligned}\label{eq: Trotter error 4}
        \sum_{\ell=1}^{M}\!\int_0^{\Delta t}\!\!\!\! dz\!\! \int_0^z\!\!\!\! dw\!\!\int_0^w\!\!\!\! du \frac{1}{2}\int_{w}^{z} \! dv&\norm{[[X,Y'(v+t_\ell)],Y(u+t_\ell)]}\! \leq\!  \frac{(\Delta t)^2}{2\Delta\xi}\left(T+\Delta \xi\right)\norm{\Dot{\mathbf{s}}}_2\ \!\!\! \norm{\mathbf{s}}_2\! \left(\frac{\sqrt{\Gamma_{\rm f}}}{2}+\Delta \xi\Gamma^{3/2}C_{\rm f}\right)\! \frac{\Gamma^{3/2}C_{\rm f}\Delta t}{3},
    \end{aligned}
\end{equation}
where $\Ddot{\mathbf{s}}=(\Ddot{s}_1,\Ddot{s}_2,...,\Ddot{s}_{N_C})$,  with 
$
        \Ddot{s}_\alpha= \norm{[H_{\rm S},[H_{\rm S},S_\alpha]]}.$
Hence, by combining Eqs.~\eqref{eq: Trotter error 1} 
-\eqref{eq: Trotter error 4} with \eqref{eq: Trotter bound 1}, \eqref{eq: Trotter error bound 3}, and \eqref{eq: Trotter bound 2}, we obtain the following bound for the total accumulated Trotter error bound for the ATA:
\begin{equation}\label{eq: pre Trotter error}
\begin{split}
    \varepsilon_{4}
    \leq&
    2\frac{(\DT)^2}{\Delta\xi}(T+\Delta\xi) 
     \Bigg[ \sqrt{\Delta\xi}\left(\frac{\sqrt{\Gamma_{\rm f}}\norm{\Ddot{\mathbf{s}}}_2}{8} 
    +\frac{\norm{\Dot{\mathbf{s}}}_2 \Gamma^{3/2}C_{\rm f}}{4}\right) 
    +\frac{\norm{\Dot{\mathbf{s}}}_2 \norm{\mathbf{s}}_2}{2} \left(\frac{\sqrt{\Gamma_{\rm f}}}{2}+\frac{1}{3}\Gamma^{3/2}C_{\rm f}\DT\right)\left(\frac{\sqrt{\Gamma_{\rm f}}}{2}+\Gamma^{3/2}C_{\rm f}\Delta \xi\right)   
    \Bigg].
\end{split}
\end{equation}
Next, we note that $\norm{\mathbf{s}}_2^2=N_C$.
Using this result, along with $\sqrt{\Gamma_{\rm f}}\leq \frac{2^{3/4}4}{\sqrt{\pi}} \sqrt{\Gamma}$, and noticing that $\int_{-\infty}^{\infty}ds\abs{\bg'_f(s)}_{2,1}\leq \frac{2^{3/4}4}{\sqrt{\pi}} \int_{-\infty}^{\infty}ds\abs{\bg'(s)}_{2,1}$ \footnote{This follows by $\bg_{\rm f}'(s)=[\bg'\ast \varphi](s)$ and a similar argument to that given in Appendix~\ref{app:Gamma tau bound proof} in \eqref{eq: Minkowski phibound} -- \eqref{eq: phibound 3}}, we find
\begin{equation}\label{eq: Trotter error}
\begin{split}
    \varepsilon_4
    \leq&
    \frac{2^{3/4}4}{\sqrt{\pi}} \frac{(\DT)^2}{\Delta\xi}(T+\Delta\xi) 
     \Bigg[ \sqrt{\Delta\xi\Gamma}\left(\frac{\norm{\Ddot{\mathbf{s}}}_2}{4} 
    +\frac{\norm{\Dot{\mathbf{s}}}_2 \Gamma C}{2}\right) 
    +\norm{\Dot{\mathbf{s}}}_2 \sqrt{N_C} \Gamma \left(\frac{1}{2}+ \frac13 C\Gamma\DT\right)\left(\frac{1}{2}+ C\Gamma \Delta \xi\right)   
    \Bigg].
\end{split}
\end{equation}
with $C=\sqrt{\gamma}\Gamma^{-3/2} \left(\int_{-\infty}^{\infty}ds\abs{\bg'(s)}_{2,1}\right)$.
We see from the second-order Trotter error in \eqref{eq: Trotter error} how small $\DT$ must be chosen to guarantee the desired accuracy $\varepsilon_{4}=O(\varepsilon\Gamma T)$. Assuming  $\Delta\xi\leq \frac{\varepsilon}{\Gamma}$, we find the following requirement on $\DT$:

\begin{equation}
\begin{split}
    \DT \leq& \min\left[
     \left(\frac{\varepsilon \Delta\xi }{C\Gamma \norm{ \Dot{\mathbf{s}}}_2 \sqrt{N_C}}\right)^{\frac{1}{3}},
    \left(\frac{\varepsilon  }{C^2\Gamma ^2\norm{\Dot{\mathbf{s}}}_2 \sqrt{N_C}}\right)^{\frac{1}{3}},
    \sqrt{\frac{\varepsilon \Delta\xi }{\norm{\Dot{\mathbf{s}}}_2 \sqrt{N_C}}},
    \sqrt{\frac{\varepsilon }{C\Gamma \norm{\Dot{\mathbf{s}}}_2 \sqrt{N_C}}},
    \sqrt{\frac{\varepsilon \sqrt{\Gamma\Delta\xi}}{\norm{\Ddot{\mathbf{s}}}_2}},
    \sqrt{\frac{\varepsilon \sqrt{\Gamma\Delta\xi}}{C\Gamma\norm{\Dot{\mathbf{s}}}_2}}\right].
\end{split}
\end{equation}
\end{widetext}
In the limit where $\varepsilon\to0$, the dominant inequality is
\begin{equation}
    \DT\leq \sqrt{\frac{\varepsilon \Delta\xi }{\norm{\Dot{\mathbf{s}}}_2 \sqrt{N_C}}}.
\end{equation}
This establishes the Trotter error bound used in Eq.~\eqref{eq:resource requirements} of the main text.

\section{Bounds on the filtered bath characteristic scales}
\label{app:Gamma tau bound proof}

In this appendix, we prove that the relevant interaction rate and timescale for the filtered jump-correlator, $\bg_{\rm f}(t)$, can be bounded by those of its unfiltered counterpart, $\bg(t)$. This is important as the errors encountered in the appendices above often depend on these timescales. The results of this section justify our re-expression of all bounds in terms of the corresponding rates and timescales associated with $\bg$; $\Gamma$ and $\tau$ [see e.g. \eqref{eq: Gamma Tau def main text} of the main text].
The characteristic {interaction} rate and correlation time of $\bg_{\rm f}$ are defined as
\begin{equation}\label{eq: def gamma f and tau f}
    \Gamma_{\rm f} \coloneq 4\gamma \left[\int_{-\infty}^\infty dt \norm{\bg_{\rm f}(t)}_{2,1} \right]^2,\  \tau_{\rm f}\coloneq \frac{\int_{-\infty}^\infty dt \norm{\bg_{\rm f}(t)t}_{2,1}}{\int_{-\infty}^\infty dt \norm{\bg_{\rm f}(t)}_{2,1}}.
\end{equation}
In this appendix, we prove that \begin{equation}
\begin{aligned}\label{eq: gamma_f tau_f bound}
        \Gamma_{\rm f}&\leq 14.41\Gamma,\\
    \Gamma_{\rm f}\tau_{\rm f}&\leq 14.41 \Gamma\tau+ 40.75\Gamma/\Omega.
\end{aligned}
\end{equation}
We do not claim optimality of the numerical constants, and they may be further reduced by a different choice of $\tilde{\varphi}(\omega)$. Choosing $\Omega\geq  1/\tau$, we therefore have
\begin{equation}\label{eq: physical Gamma tau vs filtered Gamma tau}
    \Gamma_{\rm f}\tau_{\rm f}=  \mathcal{O}(\Gamma\tau). 
\end{equation}
To establish the bound above, we will, for convenience, use the following notation: 
\begin{equation}
    \begin{aligned}
        \int_{f(t)> c} h &\coloneqq \int_{-\infty}^{\infty} dt \theta(f(t)-c) h(t),\\
        \int_{f(t)< c} h &\coloneqq \int_{-\infty}^{\infty} dt \theta(c-f(t)) h(t),
    \end{aligned}
\end{equation}
for any functions $f(t)$ and $h(t)$, with $\theta(\cdot)$ denoting the Heaviside step function. Similarly, we use the notation 
\begin{equation}
    \int_{\substack{f_1(t)>c_1\\f_2(t)>c_2}} h \coloneqq \int_{-\infty}^{\infty} dt \theta(f_1(t)-c_1)\theta(f_2(t)-c_2) h(t),
\end{equation}
with obvious generalizations to other directions of the inequalities in the integration domain.

To bound $\Gamma_{\rm f}$, we invoke Minkowski's integral inequality, which states \begin{equation}\label{eq: Minkowski phibound}
    \begin{aligned}
         &\sqrt{\sum_\alpha \abs{\int_{-\infty}^\infty ds g_{\alpha\beta}(t-s)\varphi(s)}^2}\\&\qquad \qquad\qquad \leq \int_{-\infty}^\infty ds\abs{\varphi(s)} \sqrt{\sum_\alpha \abs{ g_{\alpha\beta}(t-s)}^2}.
    \end{aligned}
\end{equation}
Squaring both side of this inequality and integrating over $t$, we therefore have
\begin{equation}\label{eq: Gamma physical vs Gamma filtered}
    \Gamma_{\rm f}\leq \norm{\varphi}_1^2\Gamma.
\end{equation} 
We now show that, by our choice of $\varphi$ in Eq.~\eqref{eq: phi bump function}, we have 
\begin{equation}\label{eq: phibound 3}
    \norm{\varphi}_1\leq \frac{4 \ 2^{3/4}}{\sqrt{\pi
   }}\approx 3.8\approx \sqrt{14.41},
\end{equation}
from which the first bound in Eq. \eqref{eq: gamma_f tau_f bound} follows. (Note that the above bound is valid with any matrix-valued function, $\mathbf{h}$, with integrable $\norm{\mathbf{h}}_{2,1}$, in place of $\bg$.)
To prove \eqref{eq: phibound 3}, we first note \begin{equation}\label{eq: phibound}
    \norm{\varphi}_1\leq \int_{\abs{t}<s_0}\abs{\varphi}+\int_{\substack{\abs{t^2\varphi}>s_1\\
\abs{t}>s_0}}\abs{\varphi}+\int_{\substack{\abs{t^2\varphi}<s_1\\
\abs{t}>s_0}}\abs{\varphi},
\end{equation}
{where $s_0$ and $s_1$ are free parameters to be chosen below.}
Using the Cauchy-Schwarz inequality in the first term, we find $\int_{\abs{t}<s_0}\abs{\varphi}\leq \sqrt{2 s_0}\norm{\varphi}_2=\sqrt{s_0/\pi}\norm{\tilde{\varphi}}_2\leq \sqrt{2 s_0\Omega/\pi}$.
For second term, notice that \begin{equation}
\int_{\substack{\abs{t^2\varphi}>s_1\\
\abs{t}>s_0}}\abs{\varphi}\leq \frac1{s_1}\int_{-\infty}^{\infty}dt\abs{t^2\varphi(t)^2}.
\end{equation}
Using $(s_1)^{-1}\int_{-\infty}^{\infty} dt\abs{t^2\varphi(t)^2}=\frac{1}{2s_1\pi}\int_{-\infty}^{\infty} dt \abs{\frac{d}{d\omega}\tilde{\varphi}(\omega)}^2$ and $\frac{1}{2s_1\pi}\int_{-\infty}^{\infty} d\omega \abs{\frac{d}{d\omega}\tilde{\varphi}(\omega)}^2\leq \frac{\Omega}{ 2s_1 \pi}\left(\frac{4}{\Omega}\right)^2$, we find 
\begin{equation}
    \int_{\substack{\abs{t^2\varphi}>s_1\\
\abs{t}>s_0}}\abs{\varphi}\leq \frac{8}{\pi s_1 \Omega}.
\end{equation}
For the third term in \eqref{eq: phibound} we observe that \begin{equation}
    \int_{\substack{\abs{t^2\varphi}<s_0\\
\abs{t}>s_1}}\abs{\varphi}\leq 2s_1 \int_{s_0}^{\infty}dt \frac{1}{t^2} =2\frac{s_1}{s_0}.
\end{equation}
Thus, $\norm{\varphi}_1\leq \sqrt{2 s_0\Omega/\pi}+ \frac{8}{\pi s_1 \Omega}+2\frac{s_1}{s_0}$. {Minimizing} the right-hand-side with respect to $s_0$ and $s_1$, we find Eq.~\eqref{eq: phibound 3} at $s_0=4\sqrt{2}/\Omega$ and $s_1=4\ 2^{1/4}/(\sqrt{\pi}\Omega)$.

We prove the second bound in Eq.~\eqref{eq: gamma_f tau_f bound} by a similar approach{ to the first bound}: first we note that \begin{equation}\label{eq: gamma_f tau_f product}
    \Gamma_{\rm f}\tau_{\rm f}=\gamma\int_{-\infty}^{\infty}dt \norm{\bg_{\rm f}(t) t}_{2,1}\int_{-\infty}^{\infty}dt \norm{\bg_{\rm f}(t)}_{2,1}.
\end{equation}
We identify the last factor  as $\sqrt{\Gamma_{\rm f}/4\gamma}$, which we have bounded above. To bound the first factor, we use that $\int_{-\infty}^{\infty}ds \bg(t-s)\varphi(s) t=\int_{-\infty}^{\infty}ds (t-s)\bg(t-s)\varphi(s)+\int_{-\infty}^{\infty}ds \bg(t-s)\ s \varphi(s) $. A straightforward computation, using Minkowski's integral inequality, as in \eqref{eq: Minkowski phibound}, then shows
\begin{equation}\label{eq: gft bound}
    \begin{aligned}
        &\int_{-\infty}^{\infty}dt \norm{\bg_{\rm f}(t) t}_{2,1}\\
        &\leq \norm{\varphi}_1\int_{-\infty}^{\infty}dt \norm{\bg(t) t}_{2,1}+\norm{t\varphi}_1\int_{-\infty}^{\infty}dt\norm{\bg(t)}_{2,1}.
    \end{aligned}
\end{equation}
With our choice of $\varphi$ in Eq.~\eqref{eq: phi bump function}, we  show below that
\begin{equation}\label{eq: phibound4}
   \norm{t\varphi}_1\leq  \frac{2^{1/4}\  16}{\sqrt{\pi}}\frac{1}{\Omega}\approx 10.74 \frac{1}{\Omega}.
\end{equation}
Eq.~\eqref{eq: gamma_f tau_f bound} follows  by combining this result with \eqref{eq: gamma_f tau_f product}, \eqref{eq: phibound4}, and \eqref{eq: phibound 3}. 

To prove \eqref{eq: phibound4}, observe first that  
\begin{equation}\label{eq: phibound2}
    \norm{t\varphi}_1\leq \int_{\abs{t}<r_0}\abs{t\varphi}+\int_{\substack{\abs{t^3\varphi}>r_1^2\\
\abs{t}>r_0}}\abs{t\varphi}+\int_{\substack{\abs{t^3\varphi}<r_1^2\\
\abs{t}>r_0}}\abs{t\varphi},
\end{equation}
{where $r_0$ and $r_1$ are free parameters to be chosen below.}
For the first term in \eqref{eq: phibound2} we use the Cauchy-Schwarz inequality again to bound
$
\int_{\abs{t}<r_0}\abs{t\varphi}\leq \sqrt{2r_0}\norm{t\varphi}_2.
$
Using {Placherel's Theorem, which states }that the Fourier transform preserves the $L^2$-norm up to a normalization constant, it hence follows that 
\begin{equation}
    \int_{\abs{t}<r_0}\abs{t\varphi}\leq \sqrt{\frac{r_0}{\pi}}\left \lVert\frac{d}{d\omega}\tilde{\varphi}\right \rVert_2\leq \sqrt{\frac{8 r_0}{\pi\Omega}},
\end{equation}
where the latter inequality follows from the explicit form of $\tilde{\varphi}$ given in Eq.~\eqref{eq: phi bump function}.
For the second term in \eqref{eq: phibound2} we use \begin{equation}
    \int_{\substack{\abs{t^3\varphi}>r_1^2\\
\abs{t}>r_0}}\abs{t\varphi}\leq \frac{1}{r_1^2}\int_{-\infty}^{\infty} dt\abs{t^2\varphi(t)}^2.
\end{equation} 
Using again Plancherel's Theorem, $\int_{-\infty}^{\infty}dt\abs{t^2\varphi(t)}^2=\frac{1}{2\pi}\int_{-\infty}^\infty d\omega \abs{\frac{d^2}{d\omega^2}\tilde{\varphi}(\omega)}^2$, and the explicit $\tilde{\varphi}$ from Eq.~\eqref{eq: phi bump function}, it follows that
\begin{equation}
    \begin{aligned}
        \int_{-\infty}^{\infty} dt\abs{t^2\varphi(t)}^2&\leq  \frac{4^4}{2\pi r_1^2 \Omega^3 }.
    \end{aligned}
\end{equation}
For the last term in \eqref{eq: phibound2} we have 
\begin{equation}
    \int_{\substack{\abs{t^3\varphi}<r_1^2\\
\abs{t}>r_0}}\abs{t\varphi}\leq 2(r_1)^{2} \int_{r_0}^{\infty}dt\frac{1}{t^2} =2\frac{r_1^2}{r_0}.
\end{equation}
Combining the results above, we find 
\begin{equation}
   \norm{t\varphi}_1\leq  \sqrt{\frac{8 r_0}{\pi\Omega}}+ \frac{4^4}{2\pi r_1^2 \Omega^3 }+2\frac{r_1^2}{r_0}.
\end{equation}
{Minimizing} the right-hand-side with respect to $r_0$ and $r_1$  (achieved with $r_0=8\sqrt{2}/\Omega$ and $r_1=4\ 2^{3/8}/(\pi^{1/4}\Omega)$) we obtain Eq.~\eqref{eq: phibound4}.

\section{Trotterizing ancilla--system couplings when operators coupled to baths commute}\label{app: LRI for multiple channels}
In this appendix, we describe how the system--ancilla coupling can be Trotterized for multiple quantum noise channels with commuting $\{S_\alpha\}$. 

To recap, for multiple channels, we use the following operator to advance the coupling in the $m$'th Trotter step
\begin{eqnarray} 
    V_{m}& \coloneq&  \mathcal{T}\exp\left[-i\int_{t_{m-1}}^{t_m} ds\sqrt{\gamma} \bS\cdot \bA_{\rm T}(s)\right].
\end{eqnarray}
[See Eq. \eqref{eq: multi channel V} of the main text].
If $\{S_\alpha\}$ mutually commute, we may factorize $V_m$ into unitaries acting on individual ancillas
\begin{equation}
    V_{m}=\prod_{n,\lambda} V_{m;\lambda n},
\end{equation}
where
\begin{equation}
    V_{m;\lambda n}\coloneq
    \mathcal{T}\exp\left[-i\sqrt{\gamma}\int_{I_{mn}} ds \bS\cdot \bA_{{\rm T}:\lambda n}(s)\right],
\end{equation}
and where  the integration domain $I_{mn}$ is the intersection between the $m$th Trotter step interval and the time cut-off interval for ancilla $n$,
$
    I_{mn}\coloneq [t_{m-1},t_m]\cap[\xi_n-\tau_{\rm c},\xi_n+\tau_{\rm c}]
$. Finally, the $\alpha$th entry of $\bA_{{\rm T}:\lambda n}$ is given by
\begin{equation}
    A_{{\rm T}: \alpha \lambda n}(s)\coloneq \sqrt{\Delta\xi}g^*_{{\rm f}:\alpha\lambda}(s-\xi_n)\sigma^+_{\lambda n}+h.c.
\end{equation}
The ATA for multiple commuting quantum noise channels can then be executed with the pseudocode in Fig. \ref{pseudocode1} by compiling $V_m$ as a product of gates acting only on single ancillas.

\section{Derivations of commonly used identities}\label{app: Common Relations and Notation}
In this appendix, we derive the  bounds  in Eqs.~, \eqref{eq: Dyson split expansion in text 1}, \eqref{eq: Dyson split expansion in text 2}, \eqref{eq: tracenorm of partial trace bound}, \eqref{eqa:f result}, \eqref{eq: matrix g sum bound}, and \eqref{eq: accumulated Trotter bound} that are   used throughout Appendices \ref{app: errors} and \ref{app:Gamma tau bound proof}.

\subsection{Dyson substitutions: proof~of~{Eqs.}~\texorpdfstring{\eqref{eq: Dyson split expansion in text 1}}{eqD1}~and~\texorpdfstring{\eqref{eq: Dyson split expansion in text 2}}{eqD2}}\label{app: Dyson Expansion}

In this section, we derive general identities for Dyson substitutions of unitary superoperators given in Eqs. \eqref{eq: Dyson split expansion in text 1} and \eqref{eq: Dyson split expansion in text 2}. Specifically, we show that given a Hamiltonian $H(t)=H_1(t)+H_2(t)$, the unitary superoperator $\mathcal{U}(t,t_0)$ generated by $H(t)$ --- i.e. the time-evolution superoperator --- satisfies
\begin{equation}
    \begin{split}\label{Dyson split expansion}
        \mathcal{U}(t,t_0)=
        \mathcal{U}_1(t,t_0)
        -i\nu_d 
        \int_{t_0}^tds 
        \mathcal{U}_1(t,s)H^d_2(s)\mathcal{U}(s,t_0),
    \end{split}
\end{equation}
and
\begin{equation}
    \begin{split}\label{Dyson split expansion2}
        \mathcal{U}(t,t_0)=
        \mathcal{U}_1(t,t_0)
        -i\nu_d 
        \int_{t_0}^tds 
        \mathcal{U}(t,s)H^d_2(s)\mathcal{U}_1(s,t_0).
    \end{split}
\end{equation}
where $\mathcal{U}_1(s,t_0)$ denotes the unitary superoperator generated by $H_1(t)$. In Eqs.~\eqref{Dyson split expansion} and \eqref{Dyson split expansion2}, we refer to the right-hand-sides as $H_2$-Dyson substitutions of $\mathcal{U}$.

To prove \eqref{Dyson split expansion}, recall that, given a Hamiltonian $H(t)$, the corresponding time-evolution superoperator is
\begin{equation}\label{eq: superoperator time evolution}
    \mathcal{U}(t,t_0)=\mathcal{T}\exp\left(-i\sum_{d\in\{L,R\}}\nu_d\int_{t_0}^t ds H^d(s)\right),
\end{equation}
where $L$ denotes left multiplication and $R$ denotes right multiplication, and $\nu_L=-\nu_R=1$. From now on, we contract indices, so it is implicit that
\begin{equation}
    \nu_d H^d \coloneq \sum_{d\in\{L,R\}}\nu_dH^d.
\end{equation}
The derivative of $\mathcal{U}$ evaluated at time $s$ is (by definition)
\begin{equation}
    \frac{\partial \mathcal{U}(t,t_0)}{\partial t}\bigg\rvert_s=-i\nu_d H^d(s)\mathcal{U}(s,t_0).
\end{equation}
Clearly $\mathcal{U}(t_0,t_0)=\mathcal{I}$, where $\mathcal{I}$ is the identity superoperator. Hence, integrating the above equation, we find
\begin{equation}\label{Simple Dyson}
    \mathcal{U}(t,t_0)=\mathcal{I}-i\nu_d\int_{t_0}^tds H^d(s)\mathcal{U}(s,t_0).
\end{equation}
Likewise, the derivative with respect to $t_0$ yields
\begin{equation}
    \frac{\partial \mathcal{U}(t,t_0)}{\partial t_0}\bigg\rvert_s=i\nu_d\mathcal{U}(t,s) H^d(s).
\end{equation}
Integrating this derivative from $t$ to $t_0$ and exchanging the limits of integration in exchange for a sign change, we find
\begin{equation}\label{Simple Inverse Dyson}
    \mathcal{U}(t,t_0)=\mathcal{I}-i\nu_d\int_{t_0}^tds\mathcal{U}(t,s) H^d(s).
\end{equation}

With the above preliminaries established, we now prove Eqs.~\eqref{Dyson split expansion} and \eqref{Dyson split expansion2}: If the Hamiltonian is written as $H(t)=H_1(t)+H_2(t)$, the superoperator associated with the unitary time evolution operator is then given by Eq \eqref{eq: superoperator time evolution}. On the other hand, consider the superoperator
\begin{equation}
        \tilde{\mathcal{U}}(t,t_0)=
        \mathcal{U}_1(t,t_0)
        -i\nu_d 
        \int_{t_0}^tds 
        \mathcal{U}_1(t,s)H^d_2(s)\mathcal{U}(s,t_0).
\end{equation}
By straightforward differentiation we find 
\begin{equation}
    \frac{\partial}{\partial t}\!\left(\tilde{\mathcal{U}}(t,t_0)-\mathcal{U}(t,t_0)\right)\!\!=-i\nu_d H_1^d(t)\left(\tilde{\mathcal{U}}(t,t_0)-\mathcal{U}(t,t_0)\right)\!.
\end{equation}
It follows that $$\tilde{\mathcal{U}}(t,t_0)-\mathcal{U}(t,t_0)=\mathcal{U}_1(t,t_0)\left(\tilde{\mathcal{U}}(t_0,t_0)-\mathcal{U}(t_0,t_0)\right)=0$$ 
and hence the superoperator $\mathcal{U}(t,t_0)$ is equal to $\tilde{\mathcal{U}}(t,t_0)$, and \eqref{Dyson split expansion} follows. Eq.~\eqref{Dyson split expansion2} follows by a similar argument.

\subsection{Bound on the trace norm of a partial trace: {p}roof~of~Eq.~\texorpdfstring{\eqref{eq: tracenorm of partial trace bound}}{eqTrace}}\label{app: Inequality for the trace norm of a partial trace}
In this section we prove Eq.~\eqref{eq: tracenorm of partial trace bound}, i.e., we provide a bound for the trace norm of a partial trace of a product of bounded operators, one of which is a trace-class operator. 
This bound is used to bound the operator norm of the error maps in Appendices \ref{app: discretization error}, \ref{app: qubit error}, and \ref{app: cut-off error}. Specifically Eq. \eqref{eq: tracenorm of partial trace bound} states that for any trace class operator $\rho$ on a composite Hilbert space $\mathscr{H}_A\otimes\mathscr{H}_B$ and bounded operators $X$ and $Y$ on that same Hilbert space, we have  
\begin{equation}\label{Bound on trace norm of generic error term}
    \norm{\Tr_B\left[X \rho Y\right]}_{\rm tr}\leq \norm{X}\norm{Y}\norm{\rho}_{\Tr},
\end{equation}
where $\Tr_B$ denotes the partial trace over $\mathscr{H}_B$. We will often think of $\rho$ as a quantum state, however, the inequality is generally valid for any trace class operator.

To prove Eq.~\eqref{Bound on trace norm of generic error term}, we use the fact that there exist two {orthonormal} bases for $\mathscr{H}_A$, $\{\ket{\psi_j}\}_{j}$ and $\{\ket{\phi_j}\}_{j}$, such that
\begin{equation}
    \norm{\Tr_B\left[X\rho Y\right]}_{\rm tr}
    =
    \sum_j \bra{\psi_j}\Tr_B\left[X\rho Y\right]\ket{\phi_j}.
\end{equation}
This follows from the singular value decomposition of the operator $\Tr_B\left[X\rho Y\right]$ on $\mathscr{H}_A$.
We next 
express $\langle \psi_j|\Tr_{B}[\mathcal O]|\phi_j\rangle =
\sum_{a}\langle \psi{_j}|\otimes \langle a|\mathcal O|\phi{_j}\rangle\otimes|a\rangle  $, with $\{|a\rangle\}$ any orthonormal basis on $\mathscr H_B$,  and note that
there exist a unitary operator, $U$,
such that $U\ket{\psi_i}=\ket{\phi_i}$. Thus,
\begin{equation}
\begin{split}
    \norm{\Tr_B\left[X\rho Y\right]}_{\rm tr}
    =&
    {\sum_{j,a}\bra{\phi_j}\otimes\bra{a}(U\otimes I) X\rho Y\ket{\phi_j}\otimes\ket{a}}
    \\
    =&
    {\Tr\left[Y (U\otimes I) X\rho \right]},
\end{split}
\end{equation}
where we have used the cyclic property of the trace in the last step. Taking the absolute value on both sides, and using Hölder's inequality for Schatten--$p$--norms, $\abs{\Tr[AB]}\leq \norm{A}\norm{B}_{\rm tr}$, we obtain
\begin{equation}
    \abs{\Tr\left[Y (U\otimes I) X\rho \right]}\leq \norm{Y (U\otimes I) X}\norm{\rho}_{\rm tr}.
\end{equation}
Using sub-multiplicativity of the operator norm along with  $\norm{U\otimes I}=1$ has unit operator norm leads to Eq.~\eqref{Bound on trace norm of generic error term}.

\subsection{Proof of Eq.~\texorpdfstring{\eqref{eqa:f result}}{eqF}} \label{app: bound on sums of integrals}
In this section, we prove the bound in Eq.~\eqref{eqa:f result} for a non-negative function, $f(t)$:
\begin{equation}\label{eq: bound on sum of integrals}
    \sum_{n=-\infty}^\infty
    \int_{t_0- \xi_n}^{t-\xi_n}dt f(t)
    \leq
    \left(\frac{T}{\Delta \xi}+1 \right)
    \int_{-\infty}^\infty dt f(t),
\end{equation}
where $T=t-t_0\geq 0$ and $\xi_n =n\Delta\xi$. 
To show this inequality, 
we introduce the characteristic function on an interval
\begin{equation}
    \chi_{[t_a,t_b]}(t)
    \coloneq
    \begin{cases}
        1, & t\in [t_a,t_b]
        \\
        0, & \text{Else}
        \end{cases}.
\end{equation}
This allows us to write the above integrals as
\begin{equation}
    \sum_{n=-\infty}^\infty \int_{t_0- \xi_n}^{t-\xi_n}dt f(t)
    =
    \int_{-\infty}^\infty dt f(t) \sum_{n=-\infty}^\infty \chi_{[t_0-\xi_n,t-\xi_n]}(t).
\end{equation}
The sum over the indicator functions is uniformly bounded
\begin{equation}
    \sum_{n=-\infty}^\infty \chi_{[t_0-\xi_n,t-\xi_n]}(t)
    \leq \left(\frac{T}{\Delta\xi}+1\right).
\end{equation}
Due to the non-negativity of $f(t)$, we thus arrive at the desired inequality.

\subsection{Proof of Eq.~\texorpdfstring{\eqref{eq: matrix g sum bound}}{eqMatg}}\label{app: Matrix g sum bound}
Here, we prove the bound in Eq.~\eqref{eq: matrix g sum bound}, which states 
\begin{equation}\label{eq: jump correlator Cauchy Schwarz}
    \sum_{\lambda,\alpha,\alpha'}\abs{g_{\alpha\lambda}(s)}\abs{g_{\alpha'\lambda}(u)}
    \leq \norm{\bg(-s)}_{2,1}\norm{\bg(-u)}_{2,1},
\end{equation}

To establish this result, we first use the Cauchy-Schwarz inequality to obtain
\begin{equation}
\begin{aligned}
        \sum_{\lambda,\alpha,\alpha'}&\abs{g_{\alpha\lambda}(s)}\abs{g_{\alpha'\lambda}(u)}\\&\leq \sum_{\alpha,\alpha'}\Big(\sum_\lambda \abs{g_{\alpha\lambda}(s)}^2\Big)^{\frac{1}{2}}
    \Big(\sum_\lambda \abs{g_{\alpha'\lambda}(u)}^2\Big)^{\frac{1}{2}}.
\end{aligned}
\end{equation}
Using that $g_{\alpha'\lambda}(u)=g^*_{\lambda\alpha'}(-u)$ we hence find 
\begin{equation}
\begin{aligned}
        \sum_{\lambda,\alpha,\alpha'}&\abs{g_{\alpha\lambda}(s)}\abs{g_{\alpha'\lambda}(u)}\\&\leq \sum_{\alpha,\alpha'}\Big(\sum_\lambda \abs{g_{\lambda\alpha}(-s)}^2\Big)^{\frac{1}{2}}
    \Big(\sum_\lambda \abs{g_{\lambda\alpha'}(-u)}^2\Big)^{\frac{1}{2}}.
\end{aligned}
\end{equation}
We recognize the above right-hand-side as the product of two $2,1$-norms, $\norm{M}_{2,1}\coloneq \sum_{\lambda}\left(\sum_\alpha \abs{M_{\alpha \lambda}}^2\right)^{1/2}$, of $\bg$, i.e.

\begin{equation}
    \sum_{\lambda,\alpha,\alpha'}\abs{g_{\alpha\lambda}(s)}\abs{g_{\alpha'\lambda}(u)}
    \leq \norm{\bg(-s)}_{2,1}\norm{\bg(-u)}_{2,1},
\end{equation}
which is the desired result.

\subsection{Bound on the accumulated Trotter error: {p}roof~of~Eq.~\texorpdfstring{\eqref{eq: accumulated Trotter bound}}{eqAccumulatedT}}
\label{app: Trotter bound 1}
Here prove the  bound in Eq.~\eqref{eq: accumulated Trotter bound}:
\begin{equation}\label{eq: accumulated Trotter bound appendix}
\begin{aligned}
        \lVert \rho_{\rm T}(t_M)&-\rho_{\rm Trotter}(t_M)\rVert_{\Tr}\leq\\& 2\sum_{\ell=1}^{M}\lVert\mathcal{T}e^{-i\int_{t_{\ell-1}}^{t_{\ell}}dt'H_{{\rm T}}(t')}-e^{-\frac{iH_{\rm S}\DT}{2}}V_{\ell} e^{-\frac{iH_{\rm S}\DT}{2}}\rVert,
\end{aligned}
\end{equation}
with $t_M=t_0+M\Delta t$.

To conveniently prove the bound above   we introduce the following notation: we let 
\begin{equation}
    U_{\rm T}(t,t_0)=\mathcal{T}e^{-i\int_{t_0}^t dt' H_{\rm T}(t')},
\end{equation}
 denote full system-ancillas ATR time-evolution operator, with $H_{\rm T}$ given in Eq.~\eqref{eq: H ATR},
and let  \begin{equation}
    G_m=e^{-\frac{iH_{\rm S}\DT}{2}}V_{m} e^{-\frac{iH_{\rm S}\DT}{2}}
\end{equation}
denote the full system-ancilla ATR Trotter step unitary.
Furthermore, we let \begin{equation}
    \tilde{\rho}_{\rm T}(t)=U_{\rm T}(t,t_0)(\rho(t_0)\otimes\ketbra{\boldsymbol{0}})U_{\rm T}(t,t_0)^\dagger,
\end{equation} 
 denote the evolution of the density matrix of the {\it composite} system-ancilla system generated by $H_{\rm T}(t)$, and likewise let  \begin{equation}
    \tilde{\rho}_{\rm Trotter}(t_\ell)=\left[\prod_{m=1}^{\ell}G_m\right]\rho(t_0)\otimes\ketbra{\boldsymbol{0}}\left[\prod_{{m'=1}}^{{\ell}}G_{m'}\right]^\dagger.
\end{equation} 
denote the corresponding time evolved state of the composite system with the  Trotterized evolution. 
We finally, introduce the shorthand $U_{{\rm T};\ell}\coloneq U_{\rm T}(t_\ell,t_{\ell-1})$.
\begin{widetext}
With the notation above, it clearly follows that $\tilde{\rho}_{\rm T}(t_\ell)=U_{\rm T;\ell}\tilde{\rho}_{\rm T}(t_{\ell-1})U_{{\rm T};\ell}^\dagger$, and $\tilde{\rho}_{\rm Trotter}(t_\ell)=G_\ell\tilde{\rho}_{\rm Trotter}(t_{\ell-1})G_\ell^\dagger$. Therefore,
\begin{equation}
    \tilde{\rho}_{\rm T}(t_\ell)-\tilde{\rho}_{\rm Trotter}(t_\ell)=U_{t;\ell}\tilde{\rho}_{\rm T}(t_{\ell-1})U_{t;\ell}^\dagger-G_\ell \tilde{\rho}_{\rm Trotter}(t_{\ell-1})G_\ell^\dagger.
\end{equation}
Adding and subtracting $U_{t;\ell}\tilde{\rho}_{\rm Trotter}(t_{\ell-1})U_{t;\ell}^\dagger$, taking the trace norm, and using the triangle inequality we find 
\begin{equation}\label{eq: Appendix trotter bound 1}
    \begin{aligned}
        \lVert \tilde{\rho}_{\rm T}(t_\ell)-\tilde{\rho}_{\rm Trotter}(t_\ell)\rVert_{\Tr}\leq& \lVert U_{t;\ell}[\tilde{\rho}_{\rm T}(t_{\ell-1})-\tilde{\rho}_{\rm Trotter}(t_{\ell-1})]U_{t;\ell}^\dagger\rVert_{\Tr}+\lVert U_{t;\ell}\tilde{\rho}_{\rm Trotter}(t_{\ell-1})U_{t;\ell}^\dagger-G_\ell \tilde{\rho}_{\rm Trotter}(t_{\ell-1})G_\ell^\dagger\rVert_{\Tr}.
    \end{aligned}
\end{equation}
As the trace-norm is unitarily invariant, the first term of \eqref{eq: Appendix trotter bound 1} gives
\begin{equation}
    \lVert U_{t;\ell}[\tilde{\rho}_{\rm T}(t_{\ell-1})-\tilde{\rho}_{\rm Trotter}(t_{\ell-1})]U_{t;\ell}^\dagger\rVert_{\Tr}= \lVert \tilde{\rho}_{\rm T}(t_{\ell-1})-\tilde{\rho}_{\rm Trotter}(t_{\ell-1})\rVert_{\Tr}.
\end{equation}
For the second term of \eqref{eq: Appendix trotter bound 1} we add and subtract $G_\ell\tilde{\rho}_{\rm Trotter}(t_{\ell-1})U_{t;\ell}^\dagger$, and use the triangle inequality to find
\begin{equation}
    \begin{aligned}
        \lVert U_{t;\ell}\tilde{\rho}_{\rm Trotter}(t_{\ell-1})U_{t;\ell}^\dagger-G_\ell \tilde{\rho}_{\rm Trotter}(t_{\ell-1})G_\ell^\dagger\rVert_{\Tr}\leq& \lVert [U_{t;\ell}-G_\ell]\tilde{\rho}_{\rm Trotter}(t_{\ell-1})U_{t;\ell}^\dagger\rVert_{\Tr}+\lVert G_\ell\tilde{\rho}_{\rm Trotter}(t_{\ell-1})(U_{t;\ell}-G_\ell)^\dagger\rVert_{\Tr}.
    \end{aligned}
\end{equation}
Using that $\norm{X \tilde{\rho}_{\rm Trotter} Y }_{\rm tr}\leq \norm{X}\norm{Y},$ for $X$ and $Y$ bounded operators [Eq. \eqref{Bound on trace norm of generic error term} with $\mathscr{H}_B=\mathbb{C}$ and $\lVert \tilde{\rho}_{\rm Trotter}\rVert_{\Tr}=1$], we thus find
\begin{equation}
    \begin{aligned}
        \lVert U_{t;\ell}\tilde{\rho}_{\rm Trotter}(t_{\ell-1})U_{t;\ell}^\dagger-G_\ell \tilde{\rho}_{\rm Trotter}(t_{\ell-1})G_\ell^\dagger\rVert_{\Tr}\leq 2\lVert U_{t;\ell}-G_\ell\rVert.
    \end{aligned}
\end{equation}
Thus,
\begin{equation}
    \begin{aligned}
        \lVert \tilde{\rho}_{\rm T}(t_\ell)-\tilde{\rho}_{\rm Trotter}(t_\ell)\rVert_{\Tr}\leq& \lVert \tilde{\rho}_{\rm T}(t_{\ell-1})-\tilde{\rho}_{\rm Trotter}(t_{\ell-1})\rVert_{\Tr}+2\lVert U_{t;\ell}-G_\ell\rVert.
    \end{aligned}
\end{equation}
By induction and by using $\lVert \tilde{\rho}_{\rm T}(t_{0})-\tilde{\rho}_{\rm Trotter}(t_{0})\rVert_{\Tr}=0$, it follows that 
\begin{equation}
    \lVert \tilde{\rho}_{\rm T}(t_M)-\tilde{\rho}_{\rm Trotter}(t_M)\rVert_{\Tr}\leq 2\sum_{\ell=1}^{M}\lVert\mathcal{T}e^{-i\int_{t_{\ell-1}}^{t_{\ell}}dt'H_{{\rm T}}(t')}-e^{-\frac{iH_{\rm S}\DT}{2}}V_{\ell} e^{-\frac{iH_{\rm S}\DT}{2}}\rVert.
\end{equation}
 Eq.~\eqref{eq: accumulated Trotter bound}~[or \eqref{eq: accumulated Trotter bound appendix}] now follows by noting that 
\begin{equation}
    \lVert \rho_{\rm T}(t_M)-\rho_{\rm Trotter}(t_M)\rVert_{\Tr}\leq \lVert \tilde{\rho}_{\rm T}(t_M)-\tilde{\rho}_{\rm Trotter}(t_M)\rVert_{\Tr}
\end{equation}
which follows from the bound $\lVert\Tr_{B}(\tilde{\rho})\rVert_{\Tr}\leq\lVert \tilde{\rho} \rVert_{\Tr} $, valid for any trace-class operator $\tilde{\rho}$ on a composite Hilbert space $\mathscr{H}_A\otimes \mathscr{H}_{B}$ [see Eq.~\eqref{Bound on trace norm of generic error term}, with $\mathscr{H}_B$ being the ancillas Hilbert space, and with $X=Y=I$].
\end{widetext}
\newpage

\bibliography{Reference.bib}

\end{document}